\documentclass[sn-mathphys-ay]{sn-jnl}

\usepackage{graphicx}%
\usepackage{multirow}%
\usepackage{amsmath,amssymb,amsfonts}%
\usepackage{amsthm}%
\usepackage{mathrsfs}%
\usepackage[title]{appendix}%
\usepackage{xcolor}%
\usepackage{textcomp}%
\usepackage{manyfoot}%
\usepackage{booktabs}%
\usepackage{algorithm}%
\usepackage{algorithmicx}%
\usepackage{algpseudocode}%
\usepackage{listings}%
\usepackage{verbatim}
\usepackage{aas_macros}
\usepackage{tikz}
\usepackage{xspace}
\usepackage{hyperref}

\definecolor{mpc}{rgb}{0.0, 0.3, 0.7}
\definecolor{mpc2}{rgb}{0.4, 0.1, 0.5}

\definecolor{ipc}{rgb}{0.7, 0.0, 0.2}
\definecolor{ipc2}{rgb}{0.5, 0.0, 0.1}

\newcommand{\medd}{$\dot{m}_{\rm Edd}$}
\newcommand{\mbh}{M$_{\rm BH}$}
\newcommand{\nxv}{$\sigma^2_{\rm nxv}$}

\begin{document}

\noindent This work is published in \textit{"La Rivista del Nuovo Cimento"}, DOI:\hyperlink{https://doi.org/10.1007/s40766-025-00072-5}{10.1007/s40766-025-00072-5}\\

Cite as: Paolillo, M., Papadakis, I. Continuum optical-UV and X-ray variability of AGN: current results and future challenges. Riv. Nuovo Cim. (2025). https://doi.org/10.1007/s40766-025-00072-5
 
\title[Optical-UV and X-ray variability of AGN]{Continuum optical-UV and X-ray variability of AGN: current results and future challenges}


\author[1,2,3]{\fnm{Maurizio} \sur{Paolillo}\email{maurizio.paolillo@unina.it}}

\author[4,5]{\fnm{Iossif} \sur{Papadakis}\email{jhep@physics.uoc.gr}}

\affil[1]{\orgdiv{Dipartimento di Fisica "Ettore Pancini"}, \orgname{Università degli Studi di Napoli Federico II}, \orgaddress{\street{C.U. Monte S.Angelo, via Cintia}, \city{Napoli}, \postcode{80126}, \country{Italy}}, \textit{Orcid} 0000-0003-4210-7693}

\affil[2]{\orgdiv{INAF--Osservatorio Astronomico di Capodimonte}, \orgaddress{\street{Salita Moiariello 16}, \city{Napoli}, \postcode{80131}, \country{Italy}}}

\affil[3]{\orgdiv{INFN--Sez. di Napoli}, \orgaddress{\street{C.U. Monte S.Angelo, via Cintia}, \city{Napoli}, \postcode{80126}, \country{Italy}}}

\affil[4]{\orgdiv{Department of Physics and Institute of Theoretical and Computational Physics}, \orgname{University of Crete}, \orgaddress{\street{Voutes University Campus}, \city{Heraklion}, \postcode{71003}, \country{Greece}},\textit{Orcid} 0000-0001-6264-140X}

\affil[5]{\orgdiv{Institute of Astrophysics}, \orgname{FORTH}, \orgaddress{\street{Voutes University Campus}, \city{Heraklion}, \postcode{71110}, \country{Greece}}}

\abstract{
Active Galactic Nuclei (AGN) are believed to be powered by accretion of matter onto a supermassive black hole. 
A fundamental ingredient in shaping our understanding of AGN is their variability across the entire electromagnetic spectrum. Variability studies have the potential to help us understand the geometry of the emitting regions (in various energy bands), their causal relations, and the physics of the accretion processes. This review focusses on the observational properties of AGN variability in the optical/UV/X-ray bands (where most of the AGN luminosity is emitted) and their dependence on the AGN physical parameters (i.e. mass, luminosity, accretion rate). We also discuss possible interpretations in the context of accreting compact systems, and we review the use of variability as a tool to discover AGN and trace their properties across cosmic time, using both ground and space facilities. Finally, we discuss the opportunities and challenges provided by current and next-generation optical/X-ray surveys, to use variability as an effective tool to probe the growth of super massive black holes in the Universe.}




\maketitle

\section{Introduction} 
\label{sec:intro}

From the very early discovery of Active Galactic Nuclei (AGN) as radio sources associated with nearby galaxies, variability both in the radio (first), in the optical and X-rays (later) was recognised as one of the defining features of this class of astronomical sources. For instance, C. Hoffmeister initially mis-identified BL Lacertae as a variable star \citep{Hoffmeister1929}. Later, in 1956, A. Deutsch at the Pulkovo Observatory reported that the nucleus of NGC\,5548 varied by 1 mag. In 1958 Antoinette de Vaucouleurs noticed fluctuations in the photoelectric magnitudes of NGC\,3516, NGC\,4051, and NGC\,4151 exceeding the photometric errors, and in 1960 A. Sandage reported the variability of 3C\,48, that is, before the true nature of quasars was understood \citep[][and refereces therein]{Shields1999}. These fast variations implied that the size of the optical source must be small. 
The major breakthrough on the fundamental nature of AGN occurred in 1963, when Maarten Schmidt published the redshift of quasar 3C\,273 \citep{3c27363}.  
Shortly afterward, high redshifts were measured for other star-like radio sources. If indeed extragalactic, these objects should be orders of magnitude more luminous than other radio galaxies that were identified by then. This fact, coupled with the small size of the emitting region (as implied by the detection of fast variations), was very difficult to explain by physical processes that operate in normal galaxies. 

Both observational facts (i.e. large luminosity emanating from a small region) were pivotal to the realisation, in the late 1960s, that AGN are probably powered by accretion of mass onto ``super-massive black holes" (SMBH, i.e. black holes $10^6-10^{10}$ times the solar mass; \citealp{lyndenbell69}). The presence of a SMBH can explain, in principle, the large luminosities emitted by these objects. For example, if a particle of mass $m$ falls into an object of mass $M$, it releases gravitational energy $GMm/R$. The larger $M$, the larger the energy released. At the same time, if this object is a black hole, then the particle can release gravitational energy as it travels all the way to the event horizon. Even if we consider that the particle releases energy up to a distance of $3R_S=6R_G$\footnote{$R_S=2GM/c^2$ is the Schwartschild radius of the black hole, while $R_G=GM/c^2$ is the gravitational radius.}, i.e. the innermost stable orbit of a particle around a non-rotating black hole (BH), then the amount of gravitational energy released is 1/6\,m$c^2$\,! Thus the presence of a SMBH not only increases $M$, but also decreases the distance $R$ a particle can reach (for a given $M$), hence maximising the energy release. At the same time, this model can explain the constraints that variability observations impose on the size of the central engine of AGN. 

Supermassive black holes are now believed to exist in the center of most if not all massive galaxies, since the mass of the black hole correlates well with the velocity dispersion of the galactic bulge (the M–sigma relation) or with bulge luminosity \cite[][and references therein]{Kormendy2013,Heckman2014,Donofrio2021}. However, not all galaxies host an active nucleus. The nucleus becomes active whenever a supply of material for accretion comes within the sphere of influence of the central black hole and there is a mechanism which removes angular momentum from the gas, so that it will start accreting towards the black hole.  

According to the standard model of AGN, material accretes into the black hole in the form of an accretion disc. Viscous processes in the accretion disc can transport matter inwards and angular momentum outwards. As the gas accretes, it releases gravitational energy that can heat the disc. To the first order, the spectrum of the accretion disc is the sum of black-body spectra, the temperature of which increases inward \citep[see e.g.][]{SS1973,novikovthorne}. In the case of AGN, where the BH mass is large, the spectrum is expected to peak in the optical-ultraviolet waveband, where most of the bolometric luminosity of an AGN is emitted, thus explaining the so-called ``Big Blue Bump'' in their spectral energy distribution. 

Active galactic nuclei are also strong X-ray emitters,
with the 2--10 keV luminosity being up to 15\% of the bolometric luminosity \cite[e.g.][]{lusso12,gupta24}. The disc temperature in AGN is not high enough to explain the X-ray emission we observe from these objects. It is generally believed that the X-rays are produced by Inverse-Compton scattering of the disc photons by hot, energetic electrons (with $kT_e\sim 50-100$ keV). The electrons are located in a region called the X-ray ``corona" (in analogy with the respective active region in the Sun). The X-ray corona should be located near the SMBH, where most of the power is released. This is also suggested by the strong (i.e. large-amplitude and very fast) variations we observe in this spectral band. However, we still do not know the corona geometry (extended or compact, flat or sperical), and its location with respect to the accretion disc (i.e., whether it is in a slab-like configuration on top of the inner disc, or within the disc, or located on top of the BH, at a certain height, $h$). 

The inner region of AGN, where almost all of the gravitationally energy is released and where most of it luminosity is emitted, is very small in size, and cannot be resolved by current telescopes, either ground-based or space-borne\footnote{This has been strictly true until the observations of Sgr\,A$^\star$ and M87 by the {\it Event Horizon Telescope} collaboration, but this recent accomplishment is still limited to two sources in the very nearby Universe and only in the radio band.}. In fact, it will not be possible to resolve the inner region of the AGN in the near and probably even distant future.  On the other hand, it is generally believed that variability studies can help us probe the physical processes that operate in the inner region of AGN. It is mainly because of this reason that variability studies in the optical/UV/X-ray bands (where most of the luminosity is emitted) have grown steadily in the last few decades.

A comprehensive review of all manifestations of AGN variability is complicated and beyond the scope of this paper. In this work, we review the status of the continuum flux variability of AGN observed mainly in the optical/near-UV and X-ray parts of the electromagnetic spectrum, reflecting the properties of the inner accretion flow and of the X-ray corona. We point out that here have been several reviews in the past that also address some of the topics that we discuss below \citep[e.g.][]{mushotzky93, Ulrich1997,Gaskell2003,Czerny2006}.

We focus on radio-quiet sources since the variability of radio-loud AGN is intrinsically different in nature, being associated with the physics of relativistic jets. 
We will also not discuss the optical/UV emission line variability, changing-state AGN, transient accretion events such as those caused by Tidal Disruption Events, or variations caused by SMBH binaries. Regarding the X-ray part, we will not discuss disc reverberation in the X-ray domain, spectral variability, or variations due to Quasi-Periodic Eruptions. We refer the reader interested in these topics to the many reviews in the literature that discuss these aspects of AGN variability  \cite[e.g.][]{Peterson2001a, Uttley2014, Graham2015a, Graham2015b, Cackett2021, Dai2021, Popovic2021, Ricci2023,   Komossa2025, DOrazio_Charisi24, kara25}. 

Our aim is three-fold. First, to shed some clarity on the methods commonly employed to study variability in astrophysical sources characterised by stochastic aperiodic red-noise variability. Then, to review our current understanding of the variability properties of AGN; we discuss how these trace the fundamental properties of accreting supermassive black holes, i.e. mass and accretion rate, and how variability can assist in detecting and characterising AGN populations. Finally, our work is intended to contribute to the exploitation of next generation timing studies of AGN, which will be made possible thanks to facilities such as the \textit{Vera C. Rubin} Observatory, eROSITA, \textit{Euclid} and so forth, which are expected to revolutionise the field in the next decade.

\section{Optical variability}
 
\subsection{Measuring optical/UV variability}\label{sec:optvar_methods}

The optical/UV variability of AGN has been studied with a variety of methods, starting from simple statistical quantities tracing the amplitude and timescales of the observed variability to more sophisticated approaches in both the time and frequency space.
\citet{Sánchez-Sáez2021} and \citet{DeCicco2021} discuss a number of these statistics (with references to the original papers) and their effectiveness in measuring the properties of lightcurves and in identifying AGN among other classes of sources. In general, however, all these statistics intend to measure the absolute or fractional standard/mean deviation or other robust estimators of the lightcurve \textit{variance} over some typical timescale $\Delta t$, or express the degree of correlation between different timescales or photometric bands (\citealp{DeCicco2025}, also see \citealp{Ulrich1997} and references therein). 

A more comprehensive approach, trying to estimate both the amplitude and characteristic timescales of variability makes use of the so-called ``Structure Function'' (SF), defined as the average magnitude variation as a function of timescale. Formally, for a lightcurve measured at different epochs $i$ and $j$ separated in time by $\Delta t \equiv\tau=t_j-t_i$, the SF can be defined as (\citealt{Kozlowski2016}, also see \citealt{Paltani1999,Graham2014} as well as \citealt{Simonetti1985} for the original statistical description):
\begin{equation}
    \mbox{SF}_{obs}(\Delta t)=\sqrt{2\,\sigma_s^2+2\,\sigma_n^2-2\,cov(s_i,s_j)}
\end{equation}
where $\sigma_s^2$ is the variance of the signal, $\sigma_n^2$ is the variance of the noise (assumed to be uncorrelated with itself and with the signal) and $cov(s_i,s_j)$ is the \textit{auto-covariance} of the signal between epochs $i$ and $j$\footnote{The auto-covariance function, $cov(s_i,s_j)$, is sometimes also expressed as $R(\tau)$, where $\tau\equiv\Delta t=t_j-t_i$.}. Thus, if we intend to retrieve the intrinsic SF of the signal:
\begin{align}
    \mbox{SF}_{true}(\Delta t)=\sqrt{[\mbox{SF}_{obs}(\Delta t)]^2-2\,\sigma_n^2}=\sqrt{2\,\sigma_n^2(1-\mbox{ACF}(\Delta t))}= \\ \nonumber
    =\mbox{SF}_\infty\sqrt{1-\mbox{ACF}(\Delta t)}
\end{align}
where $\mbox{ACF}(\Delta t)\equiv cov(s_i,s_j)/\sigma_s^2$ is the \textit{autocorrelation function} and $\mbox{SF}_\infty=\sqrt{2}\,\sigma_s$ is the asymptotic value when $\Delta(t)\rightarrow \infty$. In practice the SF is typically measured as:
\begin{equation}
     \mbox{SF}_{obs}(\Delta t)=\mbox{rms}[s_i-s_j]\equiv \mbox{rms}[s(t)-s(t+\Delta t)].
     \label{rmsSF}
\end{equation} 
In essence, the SF represents a non-parametric measure of the ACF of the signal, and is thus linked to the Power Spectral Density (PSD), as described in Sect.\,\ref{sec:x-rayvar_psd}.

\begin{figure}[]
    \centering
    \includegraphics[width=1.0\textwidth]{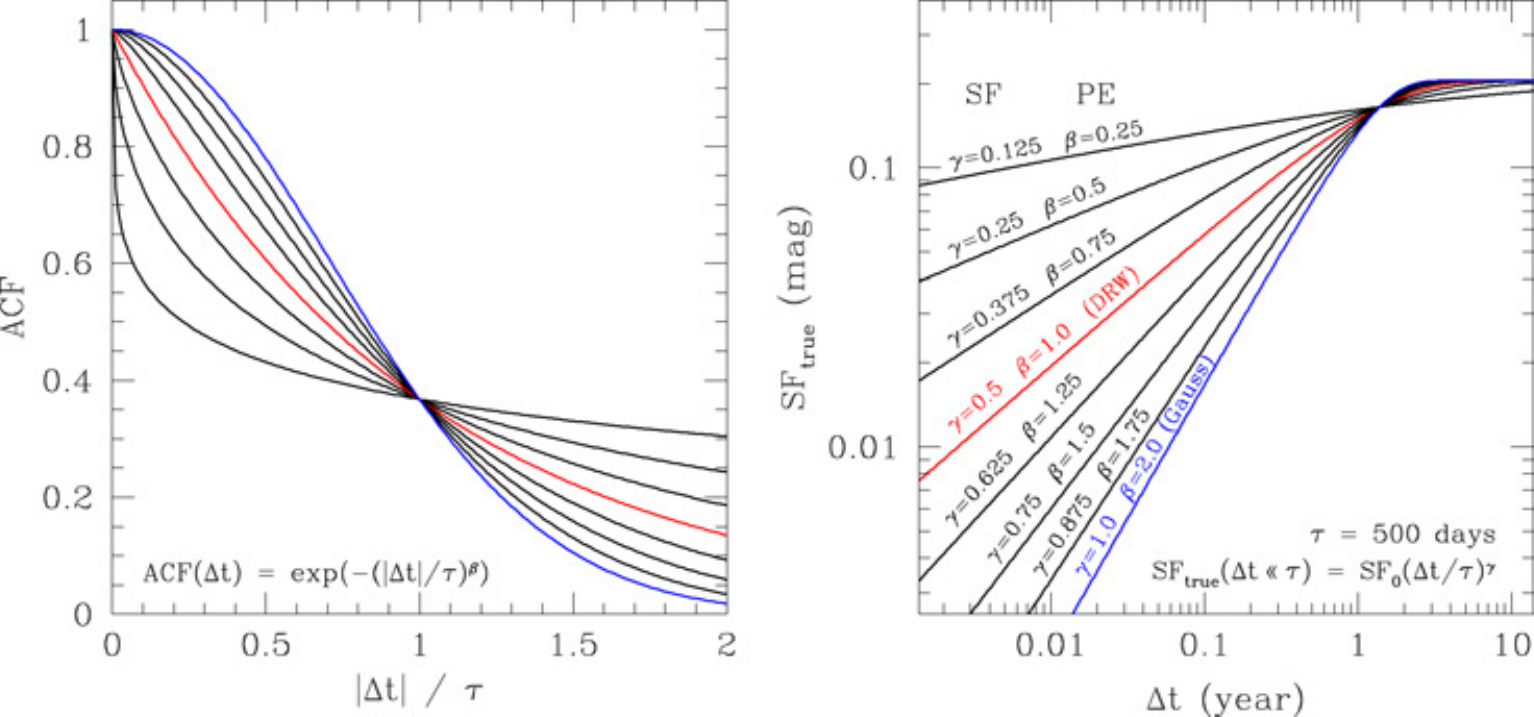}
    \caption{The connection between exponential ACFs ({\it left panel}) and the corresponding SFs ({\it right panel}). The DRW model, marked in red, has $\beta = 1$ and produces a SF of slope $\gamma = 0.5$ for $\Delta t<<  \tau$. Adapted from Fig.\,2 from \cite{Kozlowski2016}}
    \label{figSF}
\end{figure}

SF analysis must be approached with caution since it is often misused, leading to problems in the interpretation of the results, as well as in the comparison with physical and empirical models. First of all, in optical astronomy the signal $s$ in the SF is often computed in terms of \textit{magnitudes} instead of \textit{fluxes}; as shown later in Sect.\,\ref{sec:optvar_phen} the effect on the measured SF is negligible when dealing with small/moderate fluctuations typical of AGN on relatively short timescales (up to years) but becomes important if the amplitude fluctuations are larger. This suggests extreme caution in using SF computed in magnitudes when analysing extreme events such as flares or TDE. Also, the actual measurement of $\mbox{rms}[s(t)-s(t+\Delta t)]$ in eq.\,(\ref{rmsSF}) is often expressed in terms of more robust statistical quantities (sigma-clipped, median average etc.) in order to remove outlier points in the lightcurves. In such a case, comparing the different definitions requires an a priori knowledge of the statistical distribution of the signal. Furthermore,  \citet{Kozlowski2016} points out that the noise contribution is often incorrectly subtracted, leading to an erroneous measurement of the shape/amplitude of the SF. Finally, as discussed by \cite{Emmanoulopoulos2010} in detail, 
SF operates in the time domain (as opposed to the frequency domain probed by the PSD), so that its points are (heavily) correlated (because the light curve points themselves are heavily correlated). This has important consequences. First of all, it prevents proper fitting of models to the observed SF with methods such as chi-square minimisation. Then, and perhaps more importantly, the correlation between the SF points may lead to the appearance of spurious breaks in the observed SF, which depend on the sampling pattern and the length of the lightcurves. Furthermore, the SF amplitude on the longest timescales is also underestimated due to poor sampling related to the finite length of the light curves \cite[see, e.g., Appendix of][]{Bauer2009}.

An alternative approach is to directly fit the observed light curves with some pre-defined empirical model. For example, Damped Random Walk (DRW) has long been proposed as a possible realistic description of AGN variability, and was shown to be in good agreement with the data \cite[e.g.][]{Kelly2009,MacLeod2010,MacLeod2012,Zu2013}. A DRW behaviour would produce a SF with a power-law index of 1/2 below the decorrelation timescale and a flat white-noise behavior on longer timescales (Fig.\,\ref{figSF}). Deviations from this model have also been observed (see Sect.\,\ref{sec:optvar_phen}), and for this reason a more general approach consists in modeling AGN lightcurves as a Continuous-time AutoRegressive Moving-Average (CARMA) processes \citep[e.g.][]{Scargle1981,Priestley1982}. This general statistical model includes DRW as the first-order CARMA (0,1), and the Damped Harmonic Oscillator (DHO) as CARMA(2,1). It also allows modelling of processes with deviations from simple power-law power spectra and multiple characteristic timescales \citep{Kelly2014,Simm2016,Moreno2019,Yu2022}. However, such flexibility comes at the cost of much larger uncertainties and some arbitrariness in deciding up to which order to limit the complexity of the model (i.e. the number of free parameters). In fact it was pointed out very early in the literature that the estimate of the best parameters describing the variability model can be a degenerate problem, and modeling the light curves in this way does not guarantee a correct  determination of the underlying intrinsic physical process (see the Discussion in \citealt{Scargle1981}, also see  \citealt{Kozowski2016b}).
\citet{Graham2014} also proposed to use the ``Slepian wavelet variance'' (SWV) that yields comparable or better performance than SF analysis or DRW fitting but with fewer assumptions and also allows one to derive characteristic timescales. A summary of these methods, applied in the context of the Catalina Real-time Transient Survey (CRTS) to detect both stochastic and extreme variability, can be found in \citet{Graham2017}. More recently, methods based on Gaussian processes have also been tested to address the problems introduced by irregularly sampled time series \citep[e.g.][]{Lefkir2025}.

In the last decade a different approach is emerging, in order to extract information from time series. This approach relies on the use of non-parametric models based on machine learning (ML) algorithms to model and forecast AGN variability. For instance \citet{Tachibana2020} used an autoencoder (AE), which is a type of unsupervised ML algorithm, to study the properties of quasar lightcurves observed by the CRTS over timescales of about a decade. In essence AE algorithms compress the data, representing their properties in a reduced-dimensionality space. They show that AEs can model and predict the future behavior of the lightcurves without any prior assumption, better than a DRW model, suggesting that the latter does not fully describe the quasar variability properties. ML methods have the advantage that they can extract information that go beyond first or second order statistics encoded by the traditional methods described previously (SF, PSD, forward modeling). In fact the CRTS data suggest that quasar lightcurves are intrinsically asymmetric, presenting different brightening and fading timescales, similar to what found previously by e.g., \citet{Giveon1999, Hawkins2002}. Finally, the use of a multi-layer perceptron (MLP) allows to link the most significant features identified by the AE with the physical parameters of the sources (redshift, mass, accretion rate).

\subsection{Observational evidence and properties of optical/UV AGN variability}
\label{sec:optvar_phen}
The first attempts to characterise the optical variability of AGN date back to the second half of the twentieth century. In fact, given the wealth of variability studies conducted over the last 70 years, it would be challenging (and to some extent counterproductive) to try to mention all the past literature here. Our approach will instead be to cite some of the most relevant and representative works and refer the reader to their introduction to find previous pivotal papers.

Variability is an ubiquitous property of AGN at every wavelength. In the optical/UV part of the spectrum, variability is observed on timescales ranging from hours (minutes in the most extreme cases) to decades. If we focus on the \textit{intrinsic and persistent} variability, 
such variability appears stochastic and nonperiodic, with some exceptions where external factors may influence the accretion process (e.g. Quasi-Periodic Eruptions or QPE, Tidal Distruption Events or TDE, binary SMBH). 

\begin{figure}
    \centering
    \includegraphics[width=0.9\linewidth]{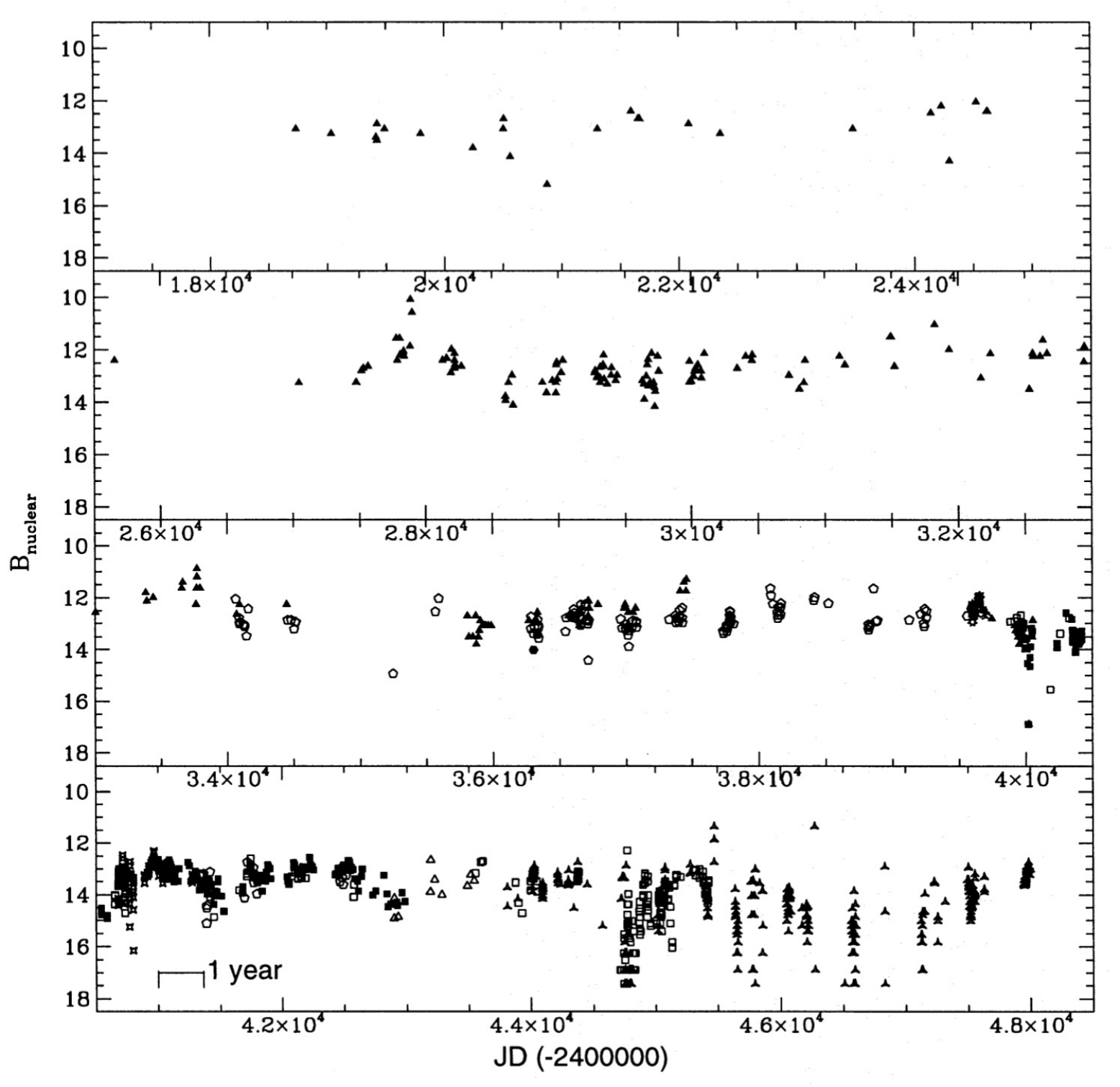}
    \caption{Nuclear lightcurve of NGC4151 in B band, over the period 1910-1991. From Fig.\,1 of \cite{Longo1996}.}
    \label{fig:Longo91}
\end{figure}
 
Originally, variability studies employed both extensive historical records as well as long monitoring campaigns \footnote{These campaigns were often designed for Reverberation Mapping studies aimed at measuring the BH mass and the structure of the Broad Line Region (BLR).} of nearby galaxies, where the nucleus is clearly separable from the rest of the galaxy. The targets of such campaigns have been, for example, NGC 4151 where historical photometric records allowed to derive light curves that span more than half a century (Fig.\,\ref{fig:Longo91}, \citealp{Longo1996}, also see \citealt{Edelson2017}) or NGC\,5548, which soon revealed that variability at different wavelengths is correlated \citep[see e.g.][]{Clavel1991}. The UV/optical study of 12 nearby AGN (namely NGC\,3783, NGC\,4152, NGC\,5548, NGC\,7469, Akn\,120, Mrk\,79, Mrk\,110, Mrk\,335, Mrk\,509, Mrk\,590, Mrk\,817, 3C\,390.3), using data from the {\it International Ultraviolet Explorer/IUE} and the ``AGN Watch" campaigns, suggested that 
the variability amplitude increases as a power law as a function of the light curve length, up to a characteristic timescale of $5-100$ days, that depends on the BH mass, and then it flattens in resemblance to what is observed in X-rays \citep{Collier2001}. A precise determination of the characteristic timescale was, however, hampered by large uncertainties and the authors could not discard the presence of multiple characteristic timescales in some sources (also see Sect.\,\ref{sec:optvar_scaling}).

\begin{figure}
    \centering
    \includegraphics[width=0.8\linewidth]{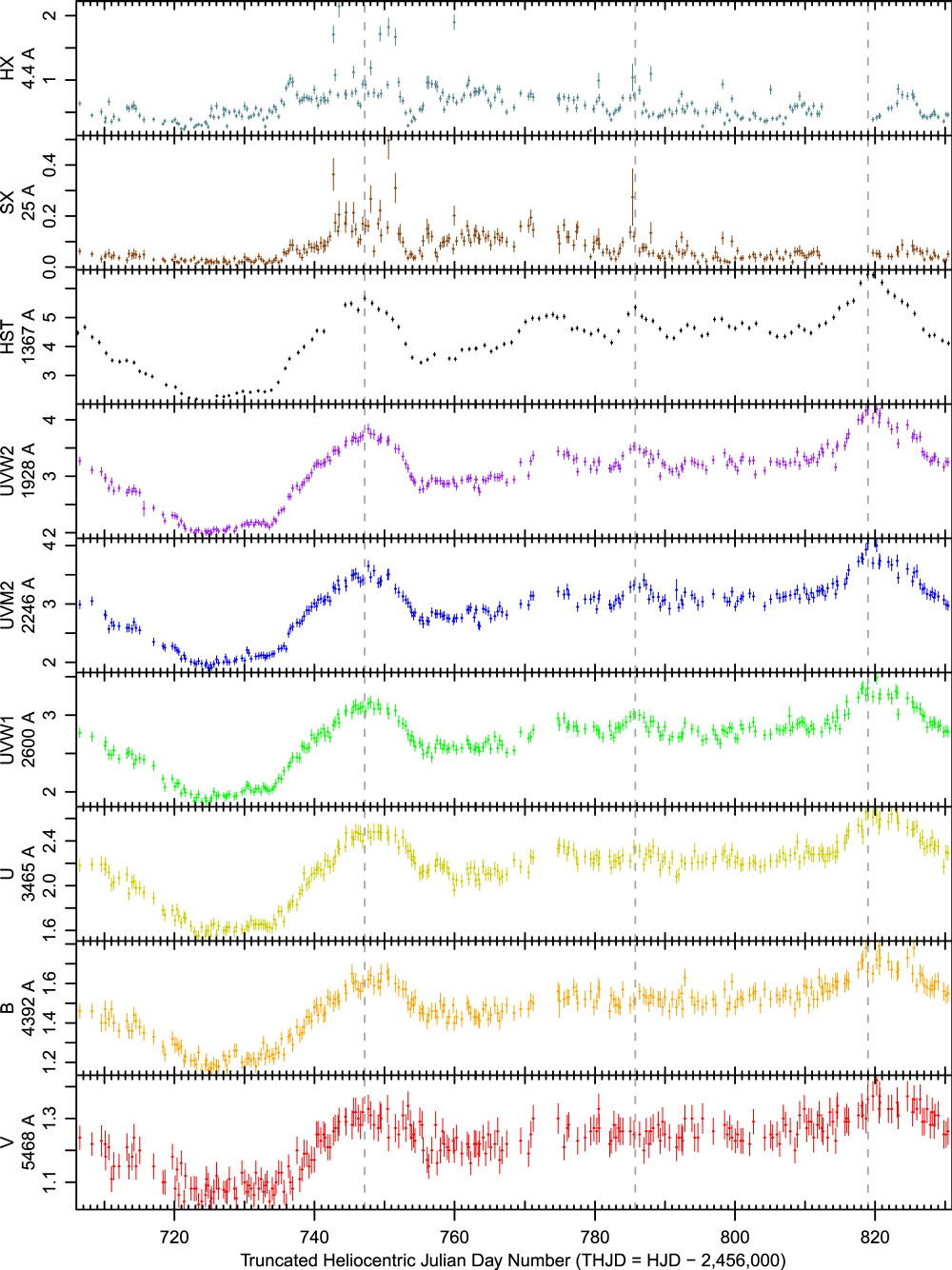}
    \caption{X-ray (top two panels) and UVW2, UVM2, UVW1, U, B, and V light curves from the ``AGN STORM" campaign of NGC\,5548 with {\it Swift}. The third panel from top shows far UV light curve from {\it HST} observations. The vertical grey dashed lines indicate three local maxima in the HST light curve. From Fig.\,3 of \citet{edelson15}.}
    \label{fig:Fausnaugh16}
\end{figure}

\begin{figure}
    \centering
    \includegraphics[width=0.45\linewidth,height=5cm]{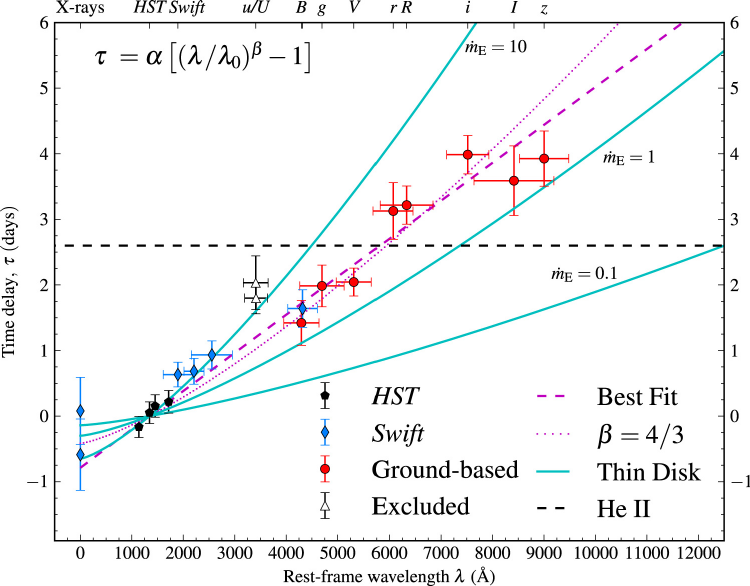}
    \includegraphics[width=0.45\linewidth,height=5cm]{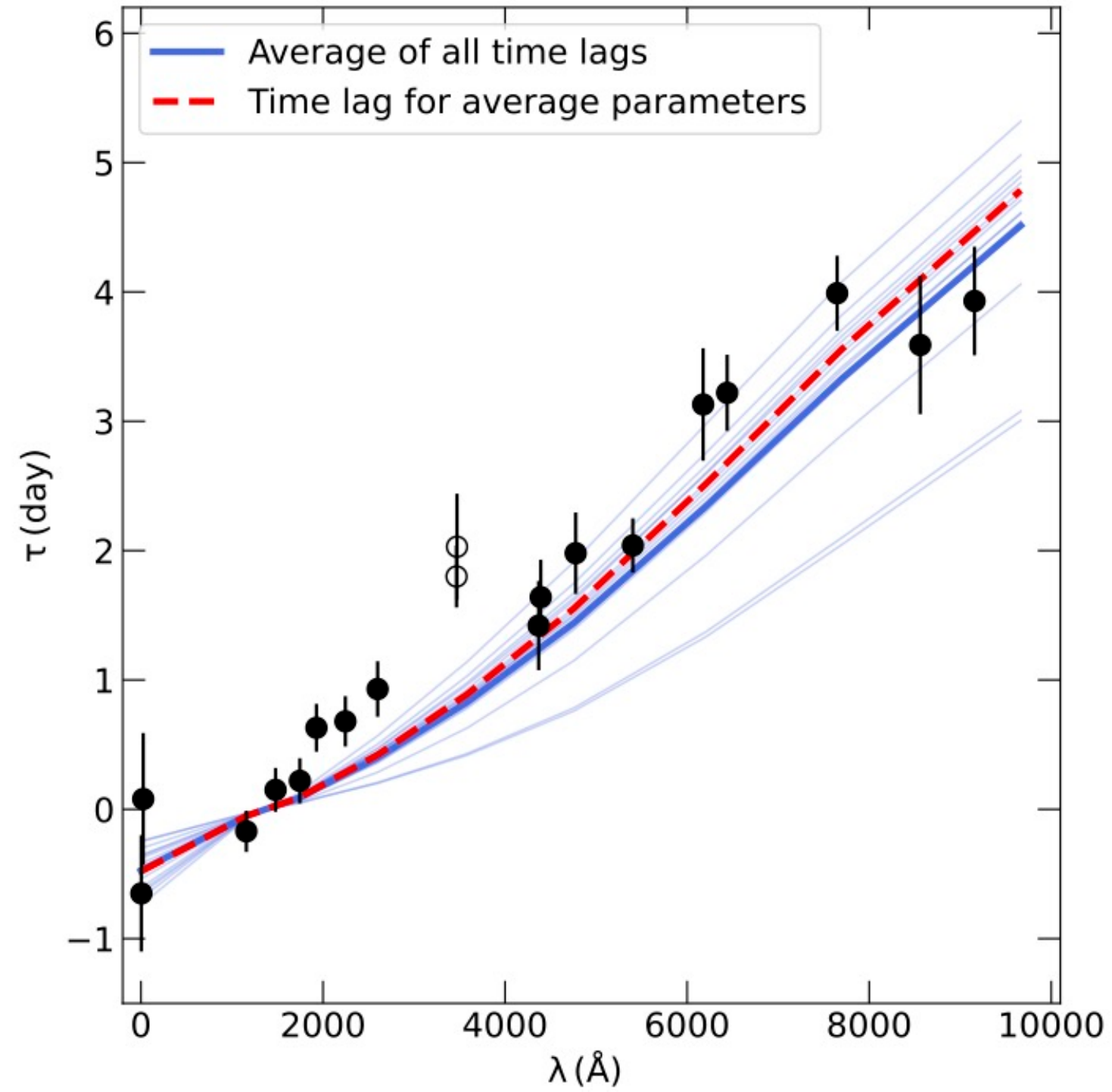}
    \caption{{\it Left panel:} Lag ($\tau$) vs. wavelength for NGC 5548, from the AGN STORM campaign. The magenta dashed line shows the best-fit to the data, and the solid cyan lines show the predicted time lags in the case of X-ray reverberation, based on a rather simple model, as defined by eq.\,12 in \cite{Fausnaugh2016}. From Fig.\,5 of \cite{Fausnaugh2016}. {\it Right panel:} The same time lags for NGC 5548. The dashed red and solid blue lines show the expected time lags in the case of X-ray reverberation and an accretion rate of 0.05 of the Eddington limit, using the more sophisticated models of \cite{Kammoun23}. The observed time-lags are broadly consistent with the X-ray reverberation model, even if the accretion rate of the object is as low as 0.05 of the Eddington limit, contrary to earlier suggestions. An excess of time-delays on top of the model predicted time-lags is clearly visible between  2000-4000 \AA. This could be due to delays due to the continuum emission from gas in the BLR (e.g. \citealp{korista19}). From Fig.\,9 of \citet{Kammoun24}.}
    \label{fig:Cakett2020}
\end{figure}

Intensive multi-wavelength monitoring campaigns have shown that the erratic behavior of AGN lightcurves is also prevalent on 
the shortest sampled time-scales, in all bands. 
Furthermore, the variations at longer wavelengths follow the variations in the far-UV, with a lag that increases with wavelength, as expected by the reprocessing of incident X-ray radiation (\citealp{Cackett2021} and references therein). A clear example can be seen in the ``AGN STORM'' \citep{DeRosa2015} observations of NGC\,5548
(Fig.\,\ref{fig:Fausnaugh16}).  Originally, the observed time delays appeared to challenge the simple optically thick and geometrically thin disk scenario due to the fact that the delays appeared to be larger than expected (Fig.\,\ref{fig:Cakett2020}) although more careful modeling showed that this may not be the case (see further discussion in \ref{sec:optvar_model}).

The other well-studied class of AGN are quasars, where the emission of the host galaxy is negligible in comparison to the nuclear source. Quasar samples also have the advantage of allowing one to explore a larger part of the parameter space, in terms of luminosities, masses, and redshift, with respect to nearby AGN, which probe mainly the low-end of the luminosity/mass functions. Many historical studies exist in the literature, which we will explore further in Sect.\,\ref{sec:optvar_discover}.

There is a general consensus on the fact that the structure functions/power spectra derived from optical lightcurves of local AGN and quasars have a red-noise behaviour, i.e. the variability amplitude (or power) increases toward longer timescales (lower temporal frequencies). This behaviour was observed early on, based on ensamble studies of QSO samples in specific sky regions \cite[SA57, SA94, SGP;][and references therein]{Hook1994, Trevese1994, Cristiani1996}. 

\begin{figure}
    \centering
    \raisebox{0.2cm}{\includegraphics[width=0.49\linewidth,height=0.42\linewidth]{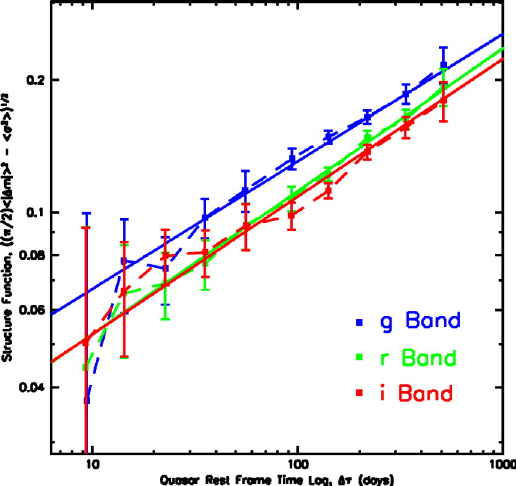}}
    \includegraphics[width=0.50\linewidth,height=0.44\linewidth]{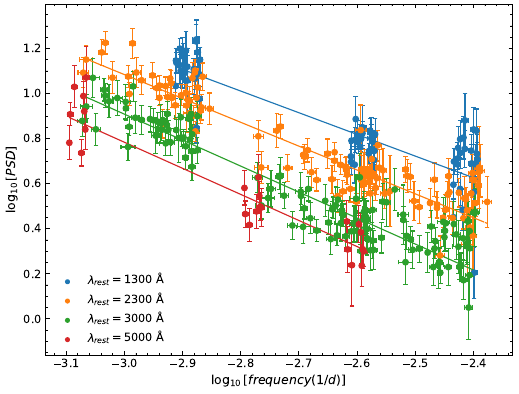}
    \caption{{\it Left panel}: Quasar structure functions for $g, r, i$ passbands, color-coded by band. No accounting has been made for any other variability dependencies, such as luminosity, wavelength, or redshift. Single–power-law fits to the data are also shown. From Fig.\,8 of \cite{VandenBerk2004}. {\it Right panel}: The same result is confirmed from the ensemble PSDs of quasars with different $\lambda_{rest}$ (indicated with different colours) after correcting for the dependencies on \mbh~ and \medd. From Fig.\,16 of \cite{Petrecca2024}.}
    \label{fig:VandenBerk04}
\end{figure}

The effectiveness of ensemble variability studies was clearly demonstrated by \cite{VandenBerk2004} who, using only two epochs from the Sloan Digital Sky Survey (SDSS) for 25\,000 spectroscopically confirmed quasars, found that optical/UV variability depends on rest-frame time lag, luminosity, rest frame wavelength, redshift, the presence of radio and X-ray emission, and the presence of broad absorption lines. The dependence of the variability amplitude on the wavelength is shown in Fig.\,\ref{fig:VandenBerk04} ({\it left panel}). \cite{DeVries2005} used archival data from the Palomar Optical Sky Survey \cite[POSS,][]{Reid1991}, the Second-Generation Guide Star Catalog \cite[GSC2,][]{mclean1998book} and early SDSS \citep{Abazajian2004} to constrain QSO variability over more than 40 years. They found that the increase in variability with increasing time lags is monotonic and constant, with no sign turnover in the SF up to timescales of 40 yr. The SF is modeled as a power law with a slope $0.30\pm0.01$ between 1 and 20 years, but its amplitude is a function of wavelength (larger toward the blue). They also found that high-luminosity quasars vary less than low-luminosity ones. They concluded that this argues against microlensing and starburst models, favoring disk instabilities (see Sect.\,\ref{sec:optvar_model}). 
\citet[also see \citealt{Hawkins2002}]{Hawkins2007} also measured the quasar SF over a period of 26 years, using plates collected by the UK 1.2 m Schmidt telescope at Siding Spring Observatory in Australia over the ESO/SERC field 287. They concluded that if variability is assumed to be due to microlensing this constrains the size of the lensing objects to $> 0.4 M_\odot$, while if variability is intrinsic to the quasar the observed turnover timescale constrains the disk size to $<10^{-2}$ pc, i.e. $\sim 10$ light days.

\begin{figure}
    \centering
    \includegraphics[width=0.7\linewidth]{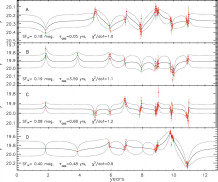}
    \caption{Example of quasar light curves from the SDSS {\it Stripe 82}. Solid and dashed lines show the weighted average and uncertainties of the DRW model light curves that are consistent with the data. From Fig.\,4 of \cite{MacLeod2010}.}
    \label{fig:Stripe82_ltc}
\end{figure}

In addition to single-epoch data, SDSS also conducted a deep survey by repeatedly imaging a $\sim 300$ deg$^2$ area (named {\it Stripe 82}) 70–90 times between 1998 and 2004 \citep{Adelman-McCarthy07,Annis2014}. To date, the {\it Stripe 82} dataset represents a golden standard for quasar variability studies due to the availability of lightcurves in multiple bands imaged almost simultaneously for many, spectroscopically confirmed, quasars. This allows us to build data sets where it is possible to disentangle the multiple dependencies on redshift, luminosity, mass, wavelength, and to compare the results with empirical and physical models. For example, taking advantage of this dataset, \citet{MacLeod2010,MacLeod2012} claimed that quasar lightcurves are described by a DRW model with an accuracy of a few percent (Fig.\,\ref{fig:Stripe82_ltc}). They determined the dependence of the characteristic timescale, $\tau$, and of the asymptotic variability amplitude, SF$_\infty$, on rest-frame wavelength, showing that SF$_\infty$ decreases with increasing wavelength, while $\tau$ increases. They also explored the dependence on luminosity, BH mass and accretion rate, as described further below. Interestingly, they found that radio-loud sources have on average a 30\% larger variability amplitude, while they do not find any systematic difference when comparing ({\it ROSAT}) detected X-ray sources vs. undetected sources.  
Contrary to most previous studies, which study variability in the time domain, \cite{Petrecca2024} used power spectrum analysis to reassess the dependence of variability on the underlying physical parameters. Using bins of fixed mass/luminosity/redshift, the dependence on wavelength (Fig.\,\ref{fig:VandenBerk04}, right panel),  becomes much more evident, strengthening previous results. These types of studies are instrumental in showing the impact that next-generation surveys will have on the field (Sect.\,\ref{sec:nextgen_opt}).

\begin{figure}
    \centering
    \includegraphics[width=0.8\linewidth]{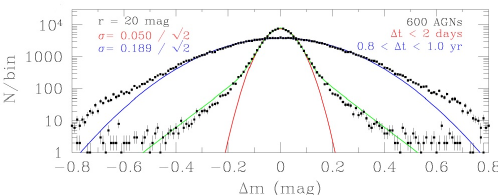}
    \caption{Distributions of magnitude differences for two time lags, $\Delta t < 2$ days (points with red or green fits) and $0.8 <\Delta t < 1$ year (points with blue fits). While the distribution on the shorter timescales is dominated the photometric errors, the distribution of the magnitude differences at the longer lag is due to the intrinsic quasar variability. It is weakly non-Gaussian (blue Gaussian fits), and is better described as a sum of a Gaussian and an exponential function (green fits). Adapted from Fig.\,3 of \cite{Kozlowski2016}.}
    \label{fig:Kozlowski16_PDF}
\end{figure}

\begin{figure}
    \centering
    \includegraphics[width=0.9\linewidth]{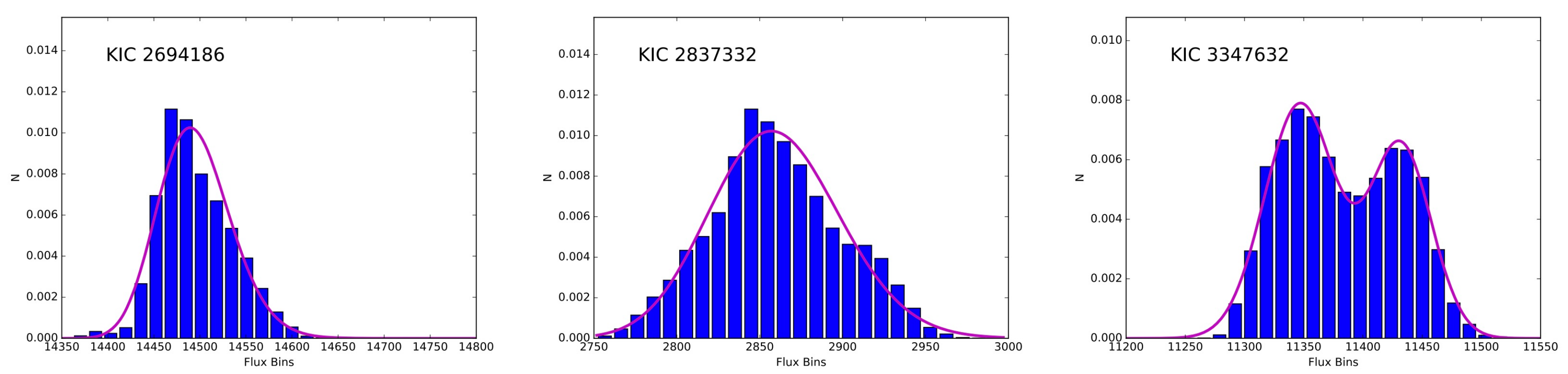}
    \caption{Example of the flux distribution of \textit{Kepler} lightcurves and their best-fit model (magenta line). Adapted from Fig.\,12 of \citet{Smith2018}.}
    \label{fig:Kepler_flux_distro}
\end{figure}

\begin{figure}
    \centering
    \includegraphics[width=0.48\linewidth]{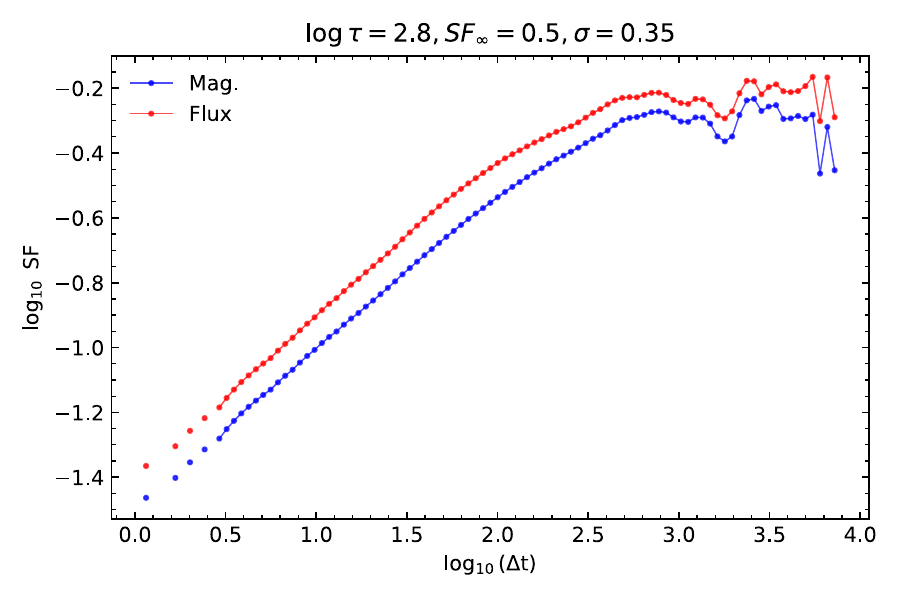}
     \includegraphics[width=0.5\linewidth]{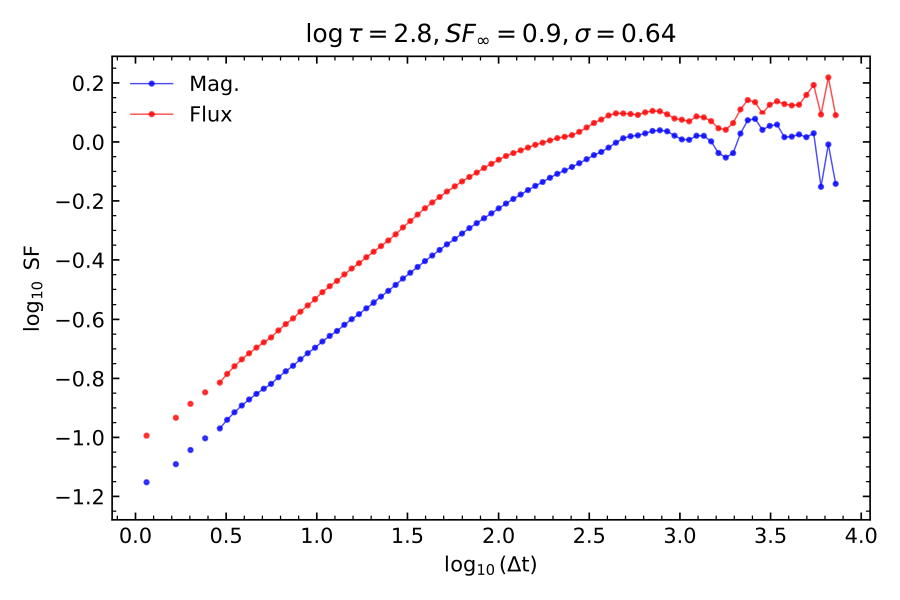}
    \caption{Structure Function derived from a DRW simulation of an AGN lightcurve, using fluxes or magnitudes; the normalization of the SF has been arbitrarily shifted to allow a comparison. As long as the lightcurve variance remains low, shapes of the two SFs remain very similar ({\it left panel}) but as the variance increases the differences become apparent ({\it right panel}). Courtesy of V. Petrecca.}
    \label{fig:SF_flux_vs_mag}
\end{figure}

Several studies (see, e.g., \citealt{MacLeod2012,Kozlowski2016}) suggest that 
the distribution of the amplitude of magnitude differences on various timescales follows a quasinormal probability density distribution in  magnitude space, that is, log-normal in flux space (Fig.,\ref{fig:Kozlowski16_PDF}), 
as predicted by the "propagating fluctuation" model (Sect.\,\ref{sec:optvar_model}). 
However, this result is not confirmed by the \textit{Kepler} lightcurves which display a variety of flux distributions: lognormal, Gaussian and bimodal (Fig.\,\ref{fig:Kepler_flux_distro}). 
It is worth noticing that, as we already discussed in Sect.\,\ref{sec:optvar_methods}, the use of magnitudes in the study of optical lightcurves, instead of fluxes (as originally done by \citealt{Simonetti1985}), often makes the analysis and interpretation more challenging. 
As an example, we show below the SF derived from fluxes and magnitudes of two simulated AGN lightcurves, using a DRW model\footnote{Although the validity of DRW models for representing AGN lightcurves is currently dabated, 
it is appropriate for this simple demonstration.} (Fig.\,\ref{fig:SF_flux_vs_mag}): as long as the variability amplitude is small (e.g. a few tenths of a magnitude) the differences btewwn the two SFs are negligible, but for larger amplitudes the two SFs start to differ significantly, and thus the estimatesq of the slope of the SF and the position of the breaks may be incorrect.

The validity of the DRW model has been challenged both on short- and long-term timescales. At short timescales (i.e. high frequencies), the results obtained from \textit{Kepler} data \citep{Mushotzky2011,Smith2018} suggest a steep power spectrum with an average power-law slope of $\gtrsim -2.5$, indicating a sharp decrease in variability at high frequencies. This result is not consistent with the expected slope of $-2$ predicted by a  DRW model. 
\textit{Kepler} probes very high frequencies inaccessible from the ground and that the range of masses and luminosities covered by \textit{Kepler} observations are lower than the SDSS quasar samples discussed above. 
However, studies based on Palomar Transient Factory (PTF) PanSTARRS, Dark Energy Survey (DES), also confirm that many quasar light curves are steeper/more complex than simple DRW predictions \citep{Caplar2017,Stone2022}. The transition from the DRW regime (defined by a PSD of slope of $-2$) to a slope of $\sim -3$ may occur on a characteristic timescale of $\sim 54$ days, as suggested by \cite{Graham2014} based on a combined sample of CRTS, SDSS and MACHO quasars.

On long timescales, the presence of flattening in the SF has also been challenged. Long monitoring campaigns, which are 3-5 times longer than the expected break timescale, appear to be necessary to properly constrain the presence of the break. 
In fact, shorter light curves often lead to the detection of spurious breaks very close to the longest monitored baseline \citep{Kozlowski2017, Suberlak2021,Stone2022}. Furthermore, a simple DRW model may be inconsistent with the scaling relations discussed in the next section, if the observed lightcurves are different realisations of a universal power spectrum, which depends on the physical parameters of the AGN (see Sect.\,\ref{sec:optvar_scaling}).

Other claimed properties of the AGN lightcurves in the optical/UV bands include the trend of becoming bluer when brighter \cite[e.g.][]{Giveon1999, Trevese2001, VandenBerk2004} and displaying a correlation between brightness and variability (the so-called ``rms-flux" relation), similarly to what is observed in X-rays \citep[see, e.g.,][]{uttley-rms}. However, these findings are also challenged by \textit{Kepler} results that do not indicate such trends at high frequencies.

\subsection{Scaling relations}
\label{sec:optvar_scaling}
In addition to the wavelength, discussed above, the dependence of the variability on other observables was detected as soon as sizable samples of sources with different properties were studied in a systematic way. The most obvious of such trends are the ones with redshift and intrinsic source luminosity. The former is understandable in terms of wavelength dependence and time dilation, because at higher redshift we are observing bluer rest-frame wavelengths (if using a fixed photometric band) and shorter rest-frame timescales. However, it took time to realise and disentangle these effects 
since it requires simultaneous multi-band observations on multiple timescales. 

The anticorrelation between optical variability and luminosity, in the sense that most luminous quasars are less variable than fainter ones, was discovered very early on \citep{Bonoli1979}. This anticorrelation was later confirmed by studies based on tens to hundreds of quasars \citep[e.g.][]{Cristiani1990,Hook1994,Trevese1994,Cristiani1996,Cristiani1997,Giveon1999}, optical \citep{Kelly2009, Rakshit2017, DeCicco2022} and X-ray selected AGN \citep{Simm2016}, as well as ensemble studies of thousands of quasars from the QUEST1, SDSS, Palomar-Quest, Pan-STARRS1, DECaLS, CRTS and PTF surveys \citep[e.g.][]{VandenBerk2004, DeVries2005, Wold2007, Wilhite2008, Bauer2009, MacLeod2010, Zuo2012, Morganson2014, Kozlowski2016, Caplar2017, Li2018, Sun2018, Laurenti2020, Petrecca2024}. 
The anticorrelation can be interpreted as a consequence of the fact that the most luminous AGN are likely to host larger mass black holes 
(if all quasars accrete at similar Eddington rates). 
In this case, their variability should be diluted by the larger emitting region. 

To test this scenario many studies tried to disentangle the dependence of vartiability amplitude on luminosity, BH mass and accretion rate. Unfortunately, most of the time only two of these three parameters can be estimated independently, even if the accretion efficiency is assumed to be known. In particular \citet{Giveon1999, Wold2007, Wilhite2008, Bauer2009, Kozlowski2016} reported a positive correlation between variability and black-hole mass or H$\beta$ equivalent width. On the other hand, \cite{Yu2022} found an anticorrelation with \mbh, as does \cite{Smith2018}, but only for sources of low accretion rate (\medd$<0.1$). \citet{Simm2016,DeCicco2022} do not find evidence of correlation with \mbh\, while \citet{Ai2010, Zuo2012,Sánchez-Sáez2018} claim that the observed dependence of variability amplitude on BH mass is the result of the correlation with luminosity and/or accretion rate.

Regarding the accretion rate, most works suggest the presence of an anticorrelation between the variability amplitude and the accretion rate \citep[e.g.][]{Bauer2009, Ai2010, MacLeod2010, Simm2016, Wilhite2008, Rakshit2017, Li2018, Laurenti2020, DeCicco2022, Yu2022, Arevalo2024, Petrecca2024}, but it is often the case that this result is degenerate with the luminosity dependence.  

\begin{figure}
    \centering
    \includegraphics[width=0.6\linewidth]{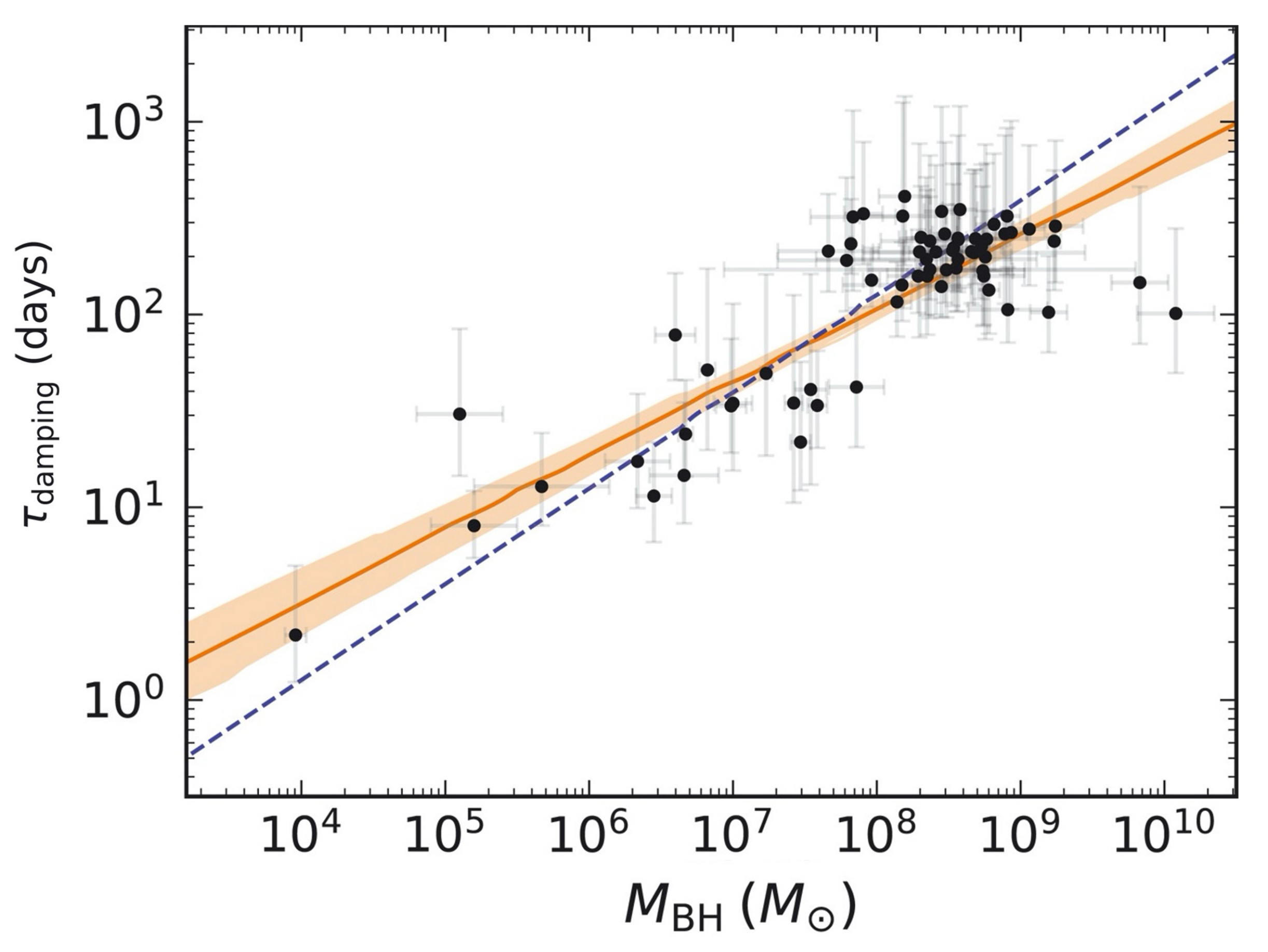}
    \caption{Restframe DRW damping timescale as a function of SMBH mass. The orange solid line is the best-fitting model, which appears close to the fit including also accreting White Dwarfs (dashed line). Adapted from Fig.\,1 of \cite{Burke2021}.}
    \label{fig:Burke2021_f1}
\end{figure}

The above scaling relations generally apply to the variability amplitude, but when the available data allow to constrain them, the characteristic variability timescales also appear to be correlated with AGN physical parameters. For example \cite{Collier2001, Kelly2009, MacLeod2010, Kozlowski2016, Li2018, Smith2018, Sun2018, Burke2020, Burke2021, Suberlak2021, Arevalo2024, Tarrant2025} report a positive correlation with \mbh\ and/or with \medd, derived from DRW/DHO fitting, SF or PSD analysis (Fig.\,\ref{fig:Burke2021_f1}). \citet{Graham2014, Yu2022} also observe an anticorrelation with bolometric luminosity, while other works \citep[e.g.][]{Ai2010, Simm2016} do not observe a significant correlation between the characteristic timescale and the other parameters of the system. Notice, however, that 
what is often claimed to be a characteristic timescale is actually defined at fixed amplitude, thus preventing a direct and meaningful comparison between various works.
\citet{Kozlowski2016} also found a mild dependence of the SF slope on luminosity and a lack of dependence on \mbh, while \citet{Li2018} find a positive correlation with both.
We caution, however, that such results may depend on the specific approach used to model the light curves (e.g. DRW, DHO, etc.) or on the maximum temporal baseline probed by each sample, as discussed previously. A summary of some of these contrasting results is reported in \citet{Suberlak2021} and \citet{DeCicco2022}, and is revised and expanded here in Table \ref{tab:correlation_summary}.

\begin{figure}
    \centering
    \includegraphics[width=0.55\linewidth]{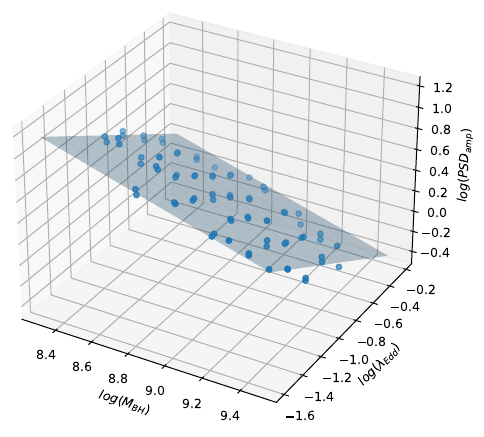}
    \caption{Dependence of PSD amplitudes on \mbh\ and luminosity/accretion rate ($\lambda_{Edd}\simeq$\,\medd) for {\it Stripe 82} quasars. The independent anti-correlation with both quantities is evident, and it defines a narrow plane. Adapted from Fig.\, 15 of \citet{Petrecca2024}.}
    \label{fig:Petrecca_fig15}
\end{figure}

Clearly, much of the confusion arises from the use of different variability estimators, different samples, and the inability to remove the degeneracies among physical properties, intrinsic or observational, due to the limited parameter space accessible to most studies so far. For instance, the SDSS {\it Stripe 82} study of \cite[][see also \citealp{MacLeod2010, Yu2022} and references therein]{Petrecca2024} clearly shows how the availability of long lightcurves in multiple bands and the relatively large range of redshifts, masses and luminosities of the quasars in the sample, allows one to properly account for all such effects. Coupled with a homogeneous analysis of the quasar power spectra, it can reveal clear trends in both mass and accretion rate (or luminosity), as shown in Fig.\,\ref{fig:Petrecca_fig15}. Most notably, the dependence on BH mass and accretion rate appear to be independent from each other, although, strictly speaking, this result is limited to the range of temporal frequencies sampled by the {\it Stripe 82} sample.

\begin{figure}
    \centering
    \includegraphics[width=0.49\linewidth]{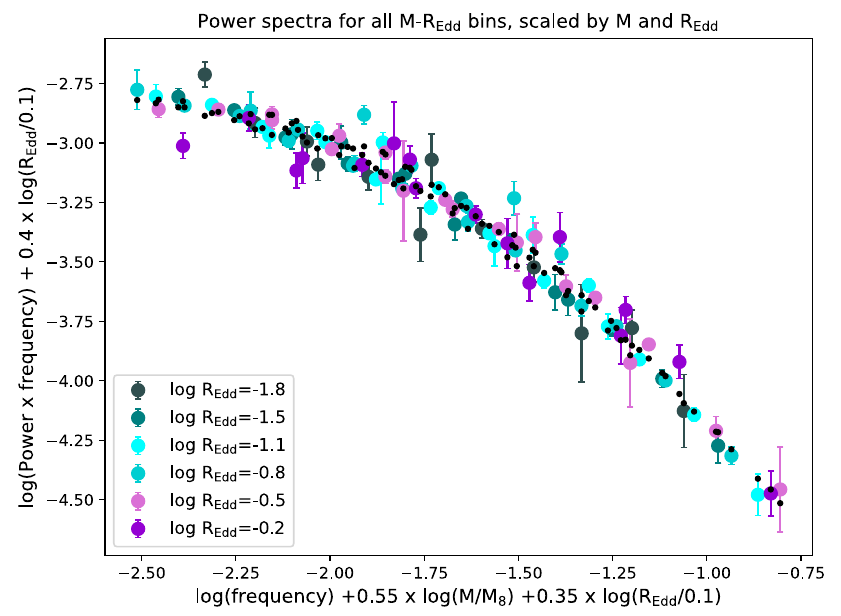}
    \includegraphics[width=0.49\linewidth, height=0.35\linewidth]{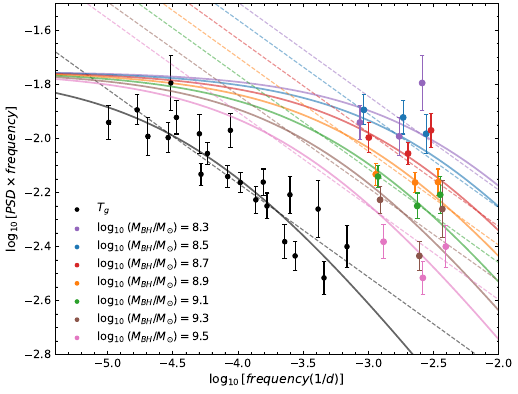}
    \caption{{\it Left panel}: the large dots, color-coded by \medd\ represent the power spectra of the ZTF data, after the re-scaling frequencies and amplitudes according to the best-fitting scaling relations with BH mass and \medd. The black dots represent a `folded' model constructed from a bending powerlaw model with low and high frequency slopes $\alpha_L=-1$ and $\alpha_H=-3$. From Fig.\,5 of \cite{Arevalo2024}. {\it Right panel}: ensemble PSDs of quasars with $\lambda_{rest} = 3000$ \AA, $\log(\lambda_{Edd}) = -1.0$ and various MBH. Black points indicate the combined PSD of quasars when light curves are normalized with the gravitational timescale $T_g$. Dashed and solid lines show the best-fit single and bending power law models. From Fig.\,19 of \cite{Petrecca2024}.}
    \label{fig:Arevalo_Petrecca24_Fig5_19}
\end{figure}

These results have led several authors to speculate that the observed variability properties of quasars (and possibly also lower luminosity AGN) can be explained if there is a universal PSD shape in all AGN and that the characteristic timescales depend on the physical properties of the source (i.e. BH mass and accretion rate). 
\citet[][also see \citealp{Tang2024}]{Tang2023} suggest that the quasar SF in the UV band from the ATLAS survey is consistent with a universal random
walk if time is expressed in units of the orbital or thermal timescales of the emitting material. On the other hand, \cite{Arevalo2024} used the Mexican-hat method of \cite{Arevalo12} to compute power spectra of quasars using ZTF light curves. Combined with the scaling relations discussed above, they built a universal power spectrum which can be modelled by a bending power law and appears steeper than DRW predictions, showing no evidence of flattening at the lowest probed frequencies (Fig.,\ref{fig:Arevalo_Petrecca24_Fig5_19}, left). \cite{Petrecca2024} also proposed that the ensemble PSD of SDSS quasars is consistent with a universal bending power-law shape, when frequencies are rescaled using the gravitational timescale (Fig.\,\ref{fig:Arevalo_Petrecca24_Fig5_19}, right). 

Works based on non-parametric approaches using ML algorithms (Sect.\,\ref{sec:optvar_methods}) also suggest that additional information about the physical properties of the accreting system may be embedded in the detailed shape of the lightcurves themselves, instead of in the global variability amplitude and timescale. Asymmetries in the  brightening and fading phases may in fact depend on the AGN mass and luminosity \citep[][and references therein]{Tachibana2020}.

The results so far suggest that the the same processes governing AGN accretion are shared among sources with different intrinsic properties (mass, accretion rate, spin). Any model trying to explain AGN physics should thus be able to account for the observed correlations between variability and AGN properties that have been observed so far.

\begin{table}[]
    \centering
    \begin{tabular}{l|c|c c c|c c c}
        \hline
        \hline
        Publication & Method & \multicolumn{3}{c|}{Amplitude} & \multicolumn{3}{c}{Timescale}\\
        &  &  \mbh & \medd & $Lum$ & \mbh & \medd & $Lum$\\
        \hline
        \cite{Collier2001} & SF & $\dots$ & $\dots$ & $\dots$ & $+$ & $\dots$ & $\dots$ \\
         \cite{Wilhite2008} & SF & $+$ & $-$ & $-$ & $\dots$ & $\dots$ & $\dots$ \\
         \cite{Kelly2009} & DRW & $--$ & $-$ & $--$ & $+$ & $\circ$ & $+$ \\
         \cite{Ai2010} & EV & $\circ$ & $-$ & $\circ$ & $\dots$ & $\dots$ & $\dots$ \\
         \cite{MacLeod2010} & DRW & $+$ & $--$ & $--$ & $+$ & $\circ$ & $\circ$ \\
         \cite{Graham2014} & SWV & $\dots$ & $\dots$ & $\dots$ & $\dots$ & $\dots$ & $-$ \\
         \cite{Morganson2014} & SF & $\dots$ & $\dots$ & $-$ & $\dots$ & $\dots$ & $\dots$ \\         \cite{Kozlowski2016} & DRW & $+$ & $--$ & $-$ & $+$ & $-$ & $\circ$ \\
         \cite{Simm2016} & EV+PSD & $\circ$ & $-$ & $-$ & $\circ$ & $\circ$ & $\circ$ \\
         \cite{Caplar2017} & SF & $\circ$ & $\dots$ & $--$ & $\dots$ & $\dots$ & $\dots$ \\
         \cite{Rakshit2017} & DRW & $+$ & $-$ & $\sim -$ & $\dots$ & $\dots$ & $\dots$ \\
         \cite{Li2018} & SF & $\circ$ & $-$ & $-$ & $\dots$ & $\dots$ & $\dots$ \\
         \cite{Sánchez-Sáez2018} & SF & $\circ$ & $--$ & $-$ & $\dots$ & $\dots$ & $\dots$ \\
         \cite{Sun2018} & SF & $\dots$ & $\sim -$ & $--$ & $\dots$ & $\circ$ & $+$ \\
         \cite{Laurenti2020} &  SF & $\sim +$ & $-$ & $--$ & $\dots$ & $\dots$ & $\dots$ \\
         \cite{Burke2021} &  DRW & $\dots$ & $\dots$ & $\dots$ & $++$ & $\dots$ & $\dots$ \\
         \cite{Suberlak2021} &  DRW & $+$ & $--$ & $+$ & $+$ & $\dots$ & $\sim +$ \\
          \cite{Yu2022} &  DHO & $-$ & $--$ & $\dots$ & $\circ$ & $\circ$ & $-$ \\        \cite{DeCicco2022} &  SF & $\circ$ & $-$ & $-$ & $\dots$ & $\dots$ & $\dots$ \\
         \cite{Arevalo2024} & PSD & $-$ & $--$ & $\dots$ & $++$ & $+$ & $\dots$\\
         \cite{Petrecca2024} & PSD & $--$ & $--$ & $-$ & $\dots$ & $\dots$ & $\dots$ \\
         \cite{Tarrant2025} & DRW & $\circ$ & $\circ$ & $\dots$ & $++$ & $\circ$ & $\dots$ \\
         \hline
    \end{tabular}
    \caption{Comparison of literature results about the dependence of AGN optical variability on \mbh, \medd, and $L_{bol}$ ($+/-$ = correlation/anticorrelation, $++/--$ = strong correlation/anticorrelation, $\sim +/\sim -$ = weak correlation/anticorrelation, $\circ$ = no significant/conclusive relation found, $\dots$ = dependence not investigated). Method acronyms are: DRW = Damped Random Walk, DHO = Damped Harmonic Oscillator, EV = Excess Variance, PSD = Power Spectral Density, SF = Structure Function, SWV = Slepian wavelet variance. Note that when scaling relations on timescales and amplitudes are degenerate (e.g. for simple power-law SF or PSD) we only considered the latter.}
    \label{tab:correlation_summary}
\end{table}


\subsection{Interpretation of AGN optical variability}\label{sec:optvar_model}

As discussed in Sect.\,\ref{sec:intro}, variability is expected to provide information both on the geometry and the physics of the emitting regions of accreting supermassive black holes. The fastest variability allowed from a coherent, isotropically emitting region is the light-crossing time:
\begin{equation}
    t_{lc}\simeq 0.011 \left ( \frac{M}{10^7 M_\odot} \right )\left ( \frac{R}{10 R_S} \right ) \mbox{days},
\end{equation} 
where $R$ is the radial distance and $M$ is the mass of the central BH. This timescale is small ($\sim$ a few days) even for the largest SMBH and if we consider the disc regions up to $\sim 1000$ R$_S$.
According to the standard Shakura-Sunyaev \citep{SS1973} accretion disk model (which is still debated for AGN), the temperature profile of a geometrically thin Keplerian disc scales as $T(R) \propto R^{-3/4}$. The disc emission in a particular waveband will originate from a broad range of radii, and could be affected by various time-scales that determine the flux variations from accretion discs \citep[see, e.g.][]{Frank_King_Raine_2002}. 
For example, the emitted flux can be modulated on the disc orbital timescale: 
\begin{equation}
    t_{orb}\simeq 0.33 \left ( \frac{M}{10^7 M_\odot} \right )\left ( \frac{R}{10 R_S} \right )^{3/2} \mbox{days}
    \label{eq:tdynamical}
\end{equation} 
which ranges from a few hours to a month for BH in the range $10^{7-9}\,M_\odot$ (at $R=10$\,R$_S$). This timescale is 
equal to the orbital period at radius $R$. The timescale to achieve hydrostatic equilibrium in the disc, and the sound crossing timescale in the disc in the vertical direction are of the same order. 

The thermal timescale of the disc is defined by the ratio of the thermal energy content of the disc per
unit of surface area over the local heating rate by viscous
dissipation, and is given by:
\begin{equation}
    t_{th}\simeq 0.53 \left (\alpha\over 0.1\right )^{-1} \left ( \frac{M}{10^7 M_\odot} \right )\left ( \frac{R}{10 R_S} \right )^{3/2}~\mbox{days}
    \label{eq:thermal}
\end{equation}
where $\alpha$ is the viscosity of the disk. Thus for a fixed viscosity, the thermal timescale is proportional to the orbital timescale. For typical $10^{7-9} M_\odot$ BHs and $\alpha=0.1$ then $t_{th}\sim$ days-years at $R=10\,R_S$. These values appear to be consistent 
claims of breaks (i.e. characteritic timescales) detected in the PSD/SF of AGN \citep[e.g.][]{Collier2001,Smith2018}. 

Works based on DRW lightcurve fitting or SF analysis also propose that the measured characteristic timescales of quasars are in agreement with the disc dynamical or thermal timescales, assuming a viscosity parameter of $\alpha\sim 10^{-3}$ \citep[e.g.][]{Kelly2009,Burke2021,Tang2023}. The thermal hypothesis is further supported by the fact that there are no clear periodicities, as could be expected from Keplerian motions, and by the trend of the observed emission to become bluer when brighter (Sect.\,\ref{sec:optvar_phen}). 
In any case, both timescales intrinsically include a dependence on the mass, which may be imprinted in the correlations discussed in Sect.\,\ref{sec:optvar_scaling}. However, they do not depend on the accretion rate of the disc thus failing to explain in full the observed dependencies.



The emission of the accretion disc can also vary on other timescales, such as the timescale of the propagation of sound waves in the radial direction, $t_{sound-
R}$, and the viscous timescale, $t_{visc}$, which determines the timescales on which variations occur on the local surface density. They are defined as follows: 

\begin{equation}
    t_{sound-R}\simeq 0.53~\left (R\over {10H} \right ) \left ( \frac{M}{10^7 M_\odot} \right )\left ( \frac{R}{10 R_S} \right )^{3/2}~\mbox{days}\\
    \label{eq:tsound}
\end{equation}

\noindent and

\begin{equation}
    t_{visc}\simeq 53 \left (R\over 10 H \right )^2\left (\alpha\over 0.1\right )^{-1} \left ( \frac{M}{10^7 M_\odot} \right )\left ( \frac{R}{10 R_S} \right )^{3/2}~\mbox{days}\\
    \label{eq:tvisc}
\end{equation}


\noindent According to the standard theory of accretion discs, 
\begin{equation}
    \frac{H}{R} \simeq 12.5\, \left(\frac{R}{R_S}\right)^{-1}~\dot{m}_{Edd}\,\left(1-\sqrt{\frac{3R_S}{R}}\right).
\label{eq:hover-mdot}
\end{equation}
Thus the viscous and sound timescales may explain the dependence of the characteristic timescales on accretion rate, as it is claimed by several studies. In general if \medd~of an object is close to 1, $(H/R)$ remains relatively small above $\sim 10\,R_S$ but for super-Eddington ﬂows, or for a hot, optically-thin plasma at the virial temperature this ratio is close to 1, and the viscous timescale is equal to the thermal timescale.
There are additional possible timescales allowed by current accretion models, such as the ``Cold Disk Removal'' timescale, due to the radius where the disk changes from optically-thick to optically-thin \citep{Czerny2006}, the free-fall timescale that may characterise Advection-Dominated Accretion flows \citep[ADAF,][]{Smith2018}.

\begin{figure}
    \centering
    \includegraphics[width=0.65\linewidth]{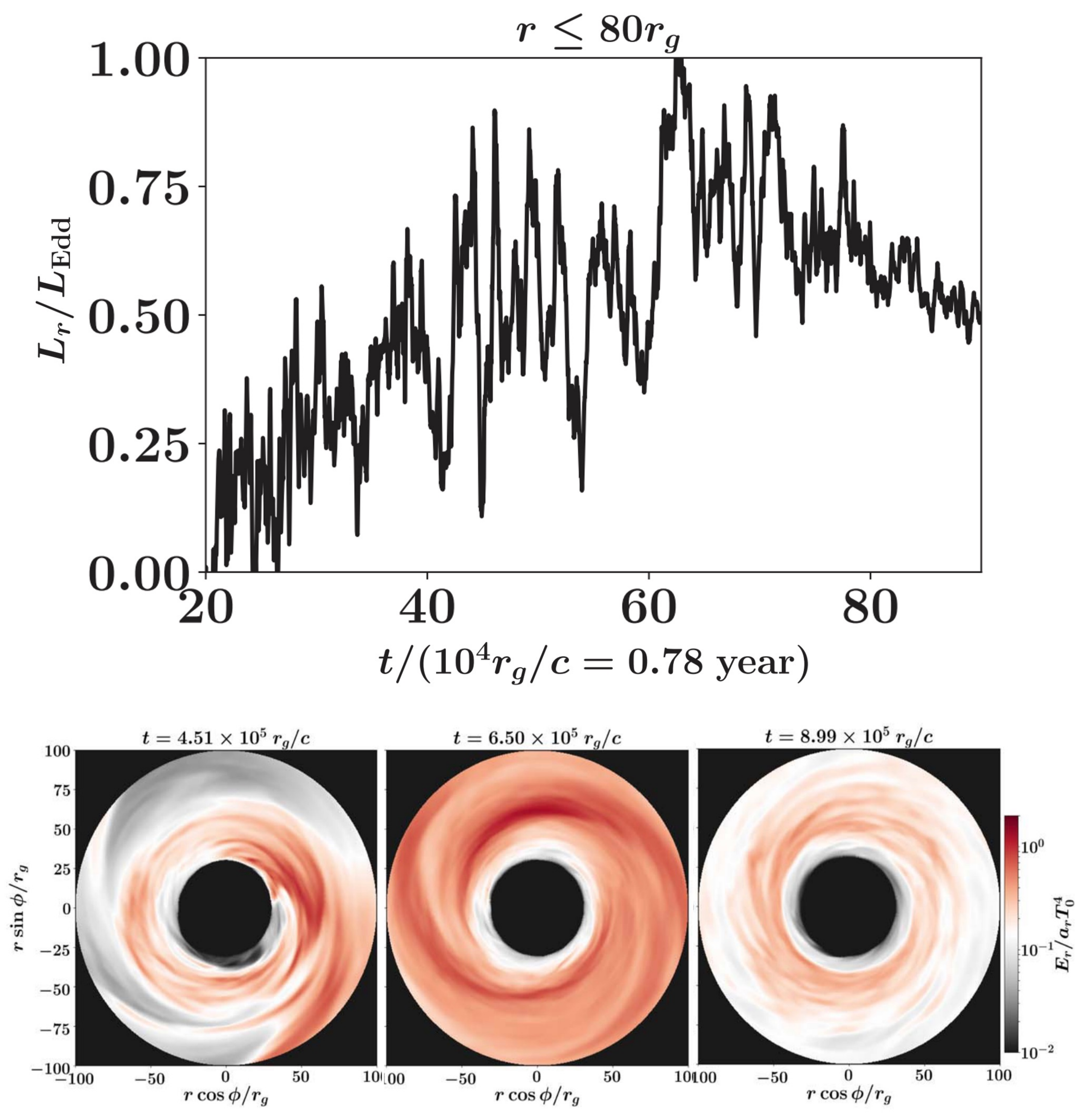}
    \caption{\textit{Top panel}: Total luminosity in units of Eddington luminosity originating in the disk, as a function of time, from MHD simulations. \textit{Bottom panel}: Snapshots of the radiation energy density distributions in the equatorial plane of the disk out to $r=100 r_g$ at three representative epochs. The energy density is scaled with $a_r T_0^4$ with $T_0=2\times 10^5$ K. Adapted from Fig.\,3 and 14 of \cite{Jiang2020}}
    \label{fig:Jiang_Blaes_2020}
\end{figure}

While these basic physical considerations provide a useful framework for understanding AGN variability, it is likely that the actual scenario is more complicated, and 3D Magneto-Hydrodynamical (MHD) simulations are needed to fully understand the disk structure. Although this is a complex topic that we cannot discuss here in detail (see \citealt{Davis2020} for a recent review), we point out that Magneto-Rotational Instability (MRI) is believed to play a major role in generating the disk viscosity and in producing local fluctuations in dissipation and accretion rate, as well as disk clumpiness and vertical stratification. MHD simulations predict that the formation of large magnetic loops takes more than one dynamical timescale, so that the magnetic field affects the X-ray corona and the disk evolution on timescales between the dynamical and thermal one \citep[e.g.][]{Miller2000}. More recent work also shows that when the metal opacity is properly accounted for, the disk may become convectively unstable. This produces strong fluctuations in surface density and disk heating resulting in significant luminosity variations, which may explain the observed AGN variability over a timescale of years to decades (Fig.\,\ref{fig:Jiang_Blaes_2020}, \citealt[][and references therein]{Jiang2020}).

In any case, since the observed emission at every wavelength is due to a superposition of contributions from different parts of the disk, in reality it is questionable if we should expect to observe a dominant timescale in AGN lightcurves (although we do expect the SF/PSD to flatten on long timescales/low frequencies to prevent a diverging total power). In fact, despite several claims (Sect.\,\ref{sec:optvar_phen}), current observations still did not definitively prove the presence of a unique timescale probing the physical properties of the disk (differently from X-rays, see Sect.\,\ref{sec:x-rayvar}) while the case for a correlation of the variability amplitude with physical parameters appears stronger. Clearly a superposition of different effects cannot be excluded and is actually favoured by many studies. In particular, the plasma may respond to non-local perturbations. These include irradiation of the outer disk by radiation generated in the inner parts of the disk itself or the optically thin plasma in the central regions (the ``corona''), or mechanical transport of energy by convection or wind/coronal outﬂows \citep[see][and references therein]{Czerny2006}. 

Microlensing effects caused by intervening bodies within the host galaxy or in the galactic halo may also contribute to the observed variability on short timescales. This form of variability enables mapping of quasar structure down to $< 10^{-6}$ arcsec scales and can be best constrained in lensed quasars, and has suggested that the accretion disk sizes are larger than expected for a thin-disc, requiring to consider extensions/modifications of the standard Shakura-Sunyaev model or the contribution of additional diffuse emission originating from, e.g., the Broad-Line Region (\citealp{Vernardos2024} for a recent review on the topic). However when proper relativistic effects are taken into account, as further discussed below in the context of X-ray reverberation, this tension can be reconciled with the standard disk model.

The ``propagating fluctuation'' model \citep{Lyubarskii1997}
assumes that local fluctuations of the accretion rate in the outer disk propagate inward towards the central regions. While moving inwards, they are further modulated by the local viscous timescale, which becomes shorter with decreasing radius, and thus imprints additional modulations onto the initial flux variability. This model has been invoked to justify the correlations between optical/UV/X-ray variability amplitudes \citep{Middei2017,Laurenti2020}, the linear rms-flux relation, and the lognormal flux distribution often (but not always) observed in AGN lightcurves (Sect.\,\ref{sec:optvar_phen}). However, if indeed fluctuations were propagating inward, then we would expect the variations first to appear in the near-IR and then propagate towards the optical and UV bands with a delay that should be representative of the speed of the fluctuations propagation. However, this is the opposite of what is observed.

In fact, in the last decade, there have been many multi-wavelength monitoring campaigns with the goal of studying the correlation between X-ray, UV, and optical variations in AGN. The targets are mainly near bright Seyferts, which are monitored mainly by the {\it Neil Gherels Swift Observatory}, {\it AstroSat} and {\it NICER} (to a lesser extent) as well as ground-based telescopes in some cases  \citep[e.g.][]{McHardy2014,edelson15,edelson19,Edelson2017,Fausnaugh2016,Cackett2018,Cackett2020,Hernandez19,vincentelli21,kara21,kara23,donnan23,miller23,Kumari23,Kumari24,prince25}. Figure \ref{fig:Fausnaugh16} shows the resulting light curves in the case of the NGC 5548 STORM 1, multi-wavelength monitoring campaign. Similar quality light curves (in terms of sampling rate and duration) have resulted from most of the other campaigns as well. 
In all cases, the UV/optical variations are well correlated,
and the variations propagate from the shorter wavelength to the longer wavelength. Well-correlated variations propagating from the UV to the optical/IR bands are also observed in AGN which are at higher redshifts and brighter than their nearby counterparts \citep[e.g.][]{jiang2017detection,mudd18,homayouni19,homayouni22,yu20,guo2022active,jha2022accretion}.
This is an important observational result that puts strong constraints on theoretical models. Any model for the UV/optical variability of AGN should be able to explain these correlations which appear to hold whenever variable AGN are monitored in many wavebands.  

These results are consistent with the hypothesis of X-ray reverberation in AGN. According to this model, part of the X-rays emitted by the corona in AGN are directed towards the accretion disc. A fraction is `reflected’ by the disc in the X-ray range (this is the so-called ``X-ray reflection component" in AGN). The rest is absorbed, increasing the temperature of the disc and consequently the emitted UV/optical flux as well. Since X-rays are highly variable, the additional UV/optical emission of the disc will also be variable, and the optical/UV variations will be delayed with respect to the X-rays. Since the disc emitting area increases with increasing wavelength, we expect the time lags to also increase with wavelength (as observed). This phenomenon is referred to as the ``disc X-ray thermal reverberation" \citep[e.g.][]{kazanas2001,Cackett2007}.

In the case of a standard \cite{SS1973} accretion disc, we expect the X-ray/UV/optical time delays, $\tau(\lambda)$, to be wavelength dependent following a relation of the form $\tau(\lambda) \propto \lambda^{4/3}$ \citep[see][]{Cackett2007}. The cross-correlation results from the first monitoring campaigns confirmed this prediction. However, it appeared that the amplitude of this relation underestimated the observed lags by a factor of $\sim$ 3-4 (see, e.g., the left panel of Fig.\,\ref{fig:Cakett2020}). \cite{Kammoun21a}, however, studied X-ray reverberation in detail and showed that this is not the case \citep[see also][]{Dovciak22}. 
They assumed an X-ray point-like source that illuminates a standard Novikov-Thorne accretion disc \citep{novikovthorne}. They considered in detail all special and general relativity effects in the light propagation from the X-ray corona to the disc and from the disc to the observer, and they took into account the disc ionisation state when computing the disc reflection flux. As a result, they were able to compute the exact  amount of X-rays that are absorbed by the disc in each radius. This was a major advantage compared to previous works, which simply assumed equal power for the X-ray absorbed flux and the internally generated heat in the disc at each radius. Using the \cite{Kammoun21a} relations, \cite{Kammoun21b} showed that the observed time lags in the nearby Seyferts are fully consistent with X-ray reverberation (see, for example, the right panel in Fig.\,\ref{fig:Cakett2020} for the case of NGC 5548). The same is true for the observed time lags of the more distant AGN \citep{langis24}. 

\begin{figure}
    \centering
    \includegraphics[width=0.49\linewidth, height=0.4\linewidth]{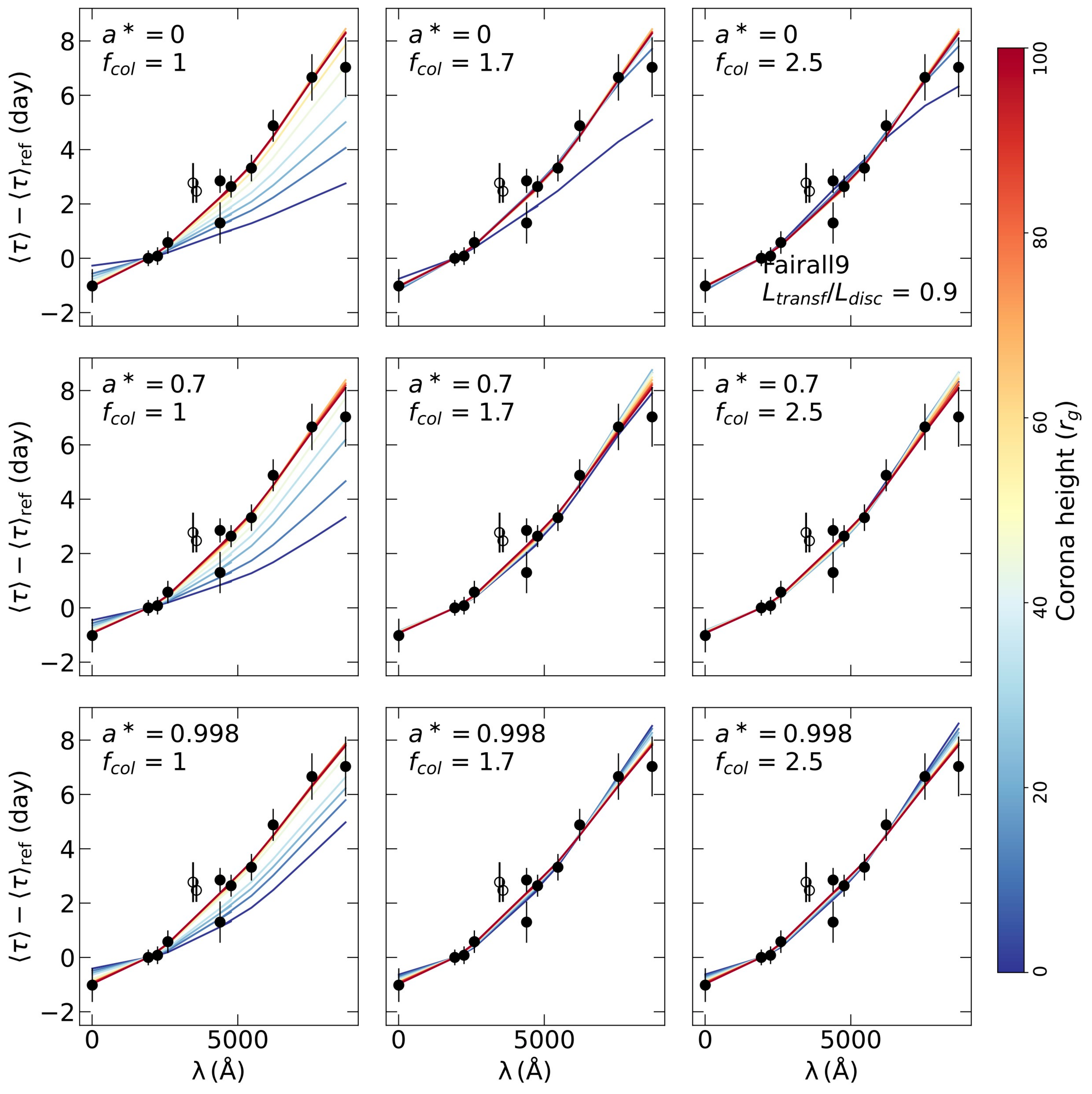}
    \includegraphics[width=0.49\linewidth, height=0.4\linewidth]{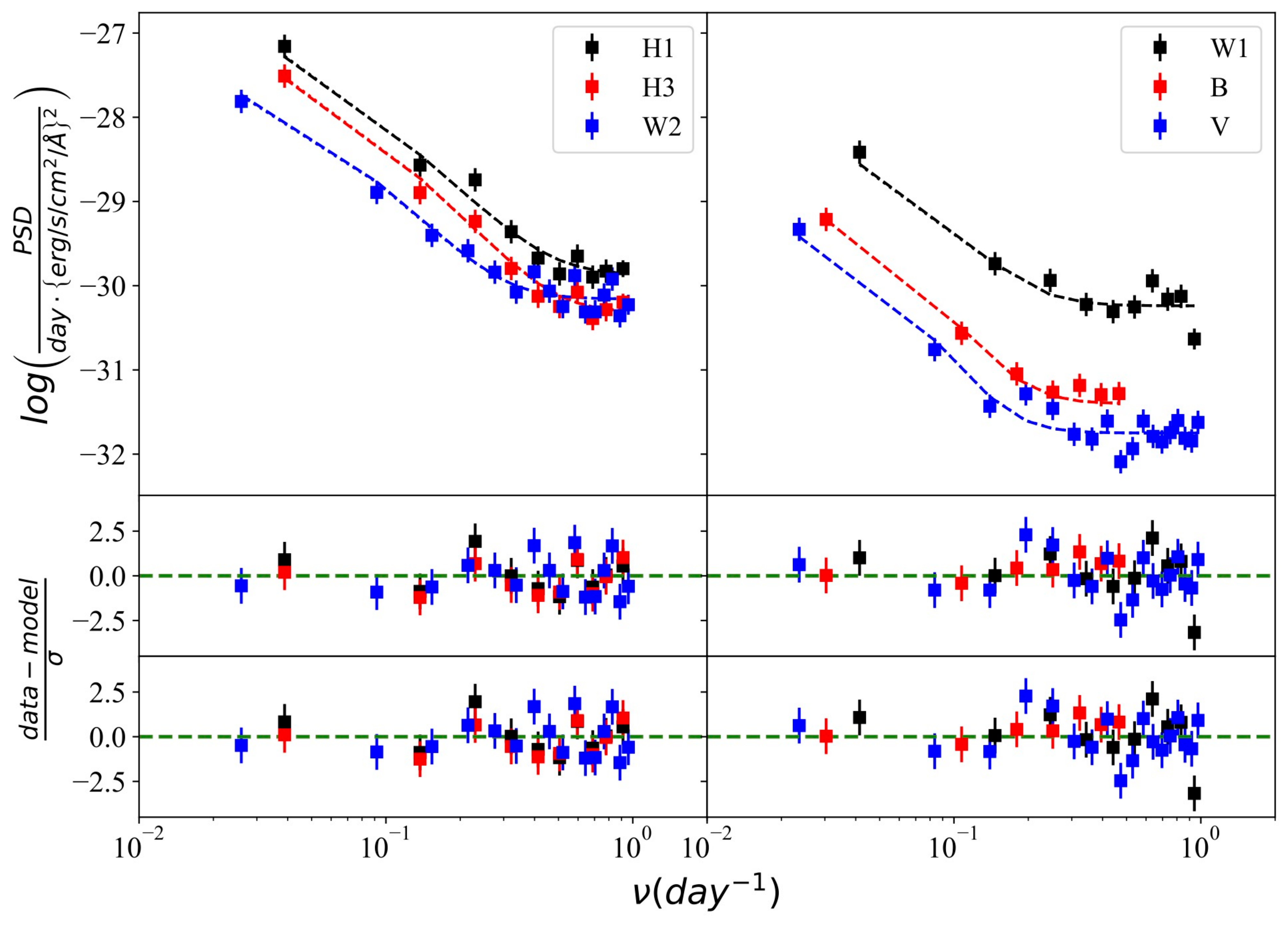}
    \caption{\textit{Left panel}: Fitting the time-lag spectra of Fairall 9 for different values of X-ray luminosity, BH spin, and colour correction factors, $f_{col}$. The colour bar on the right correspond to the different values of the X-ray corona height. Open circles show time-lag measurements that were not used when fitting the data, as they may be affected by emission from BLR. From Figure A1 of \cite{Kammoun23}. {\it Right panel}: UV/optical power spectra of NGC 5548, using the STORM 1 light curves. The dashed lines show the best-fit PSDs assuming X-ray disc thermal reverberation, the observed X-ray PSD, and the same disc response functions that can also fit the observed time lags. From Figure 3 of \cite{Panagiotou22-psds}.}
    \label{fig:reverb}
\end{figure}

\cite{Kammoun23} provided a code that can compute X-ray reverberation time lags for a large parameter space and fitted the observed optical/UV time lags in many Seyferts (left panel, Fig.\,\ref{fig:reverb}). X-ray reverberation models can also fit the wavelength dependence of the observed variance in nearby AGN \citep{Papoutsis24}, the apparent lack of significant cross-correlation between X-ray and optical/UV \citep{Panagiotou22-correlation}, the frequency-dependent time lags in the optical/UV band \citep{Panagiotou25}, the optical/UV spectral variability observed in NGC 5548 \citep{Kammoun24}, and the accretion disc size problems discussed above, inferred from microlensing observations \citep{papadakis22}. The hypothesis of X-ray disc thermal reverberation can also fit well the optical/UV power spectra of NGC 5548 \citep{Panagiotou22-psds} (right panel, Fig.\,\ref{fig:reverb}). At least in this source, (almost) all of the observed UV and optical variations could be due to X-rays illuminating the disc. This possibility alleviates the ``problem" of the fast optical/UV variations that are observed in many AGN, and cannot be explained by the standard accretion disc timescales. 

Disc X-ray thermal reverberation is not the only explanation for the delays between the observed variations in various optical/UV bands. For example, \cite{beard25} studied the X-ray and optical ($g'$-band) variability of NGC 4395, and found that the bending frequency in the optical PSD is much lower than expected in the case of X-ray disc reverberation. They concluded that reprocessing of X-rays by the accretion disc is consistent with part of the optical variability of this source, but it cannot explain all of the variability. At least one other source of variations is needed, particularly at low frequencies.  The authors commend that this extra source of longer variations could be contributed as a result of the BLR diffuse emission and/or propagating disc accretion rate variations.
Also, in this source, \cite{McHardy2023} noted deviations of the time lags from the expectations of simple reprocessing on the longest wavelengths which could be interpreted as disk truncation (allowing the first detection of the outer edge of an accretion disk) or outer disk shielding from some sort of wind.

\begin{figure}
    \centering
    \includegraphics[width=0.49\linewidth, height=0.4\linewidth]{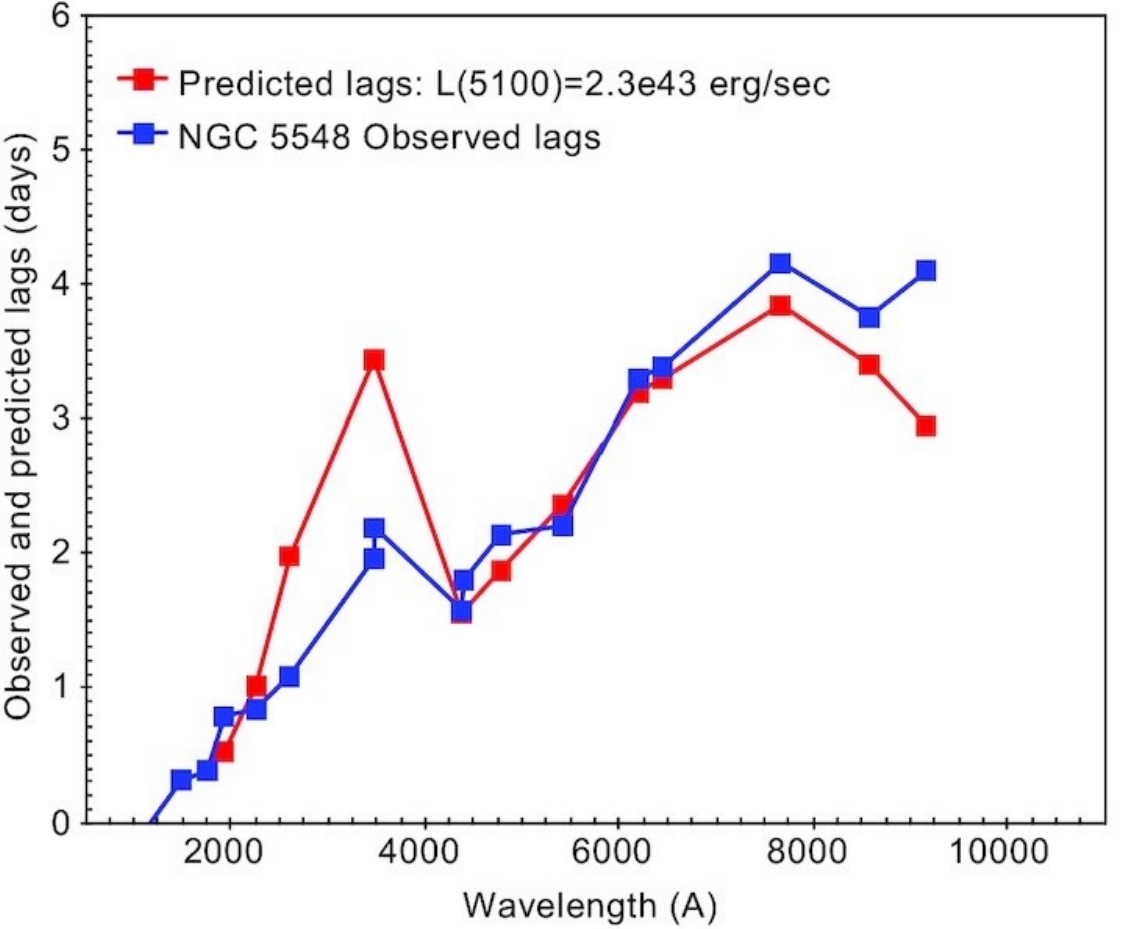}
    \includegraphics[width=0.49\linewidth, height=0.4\linewidth]{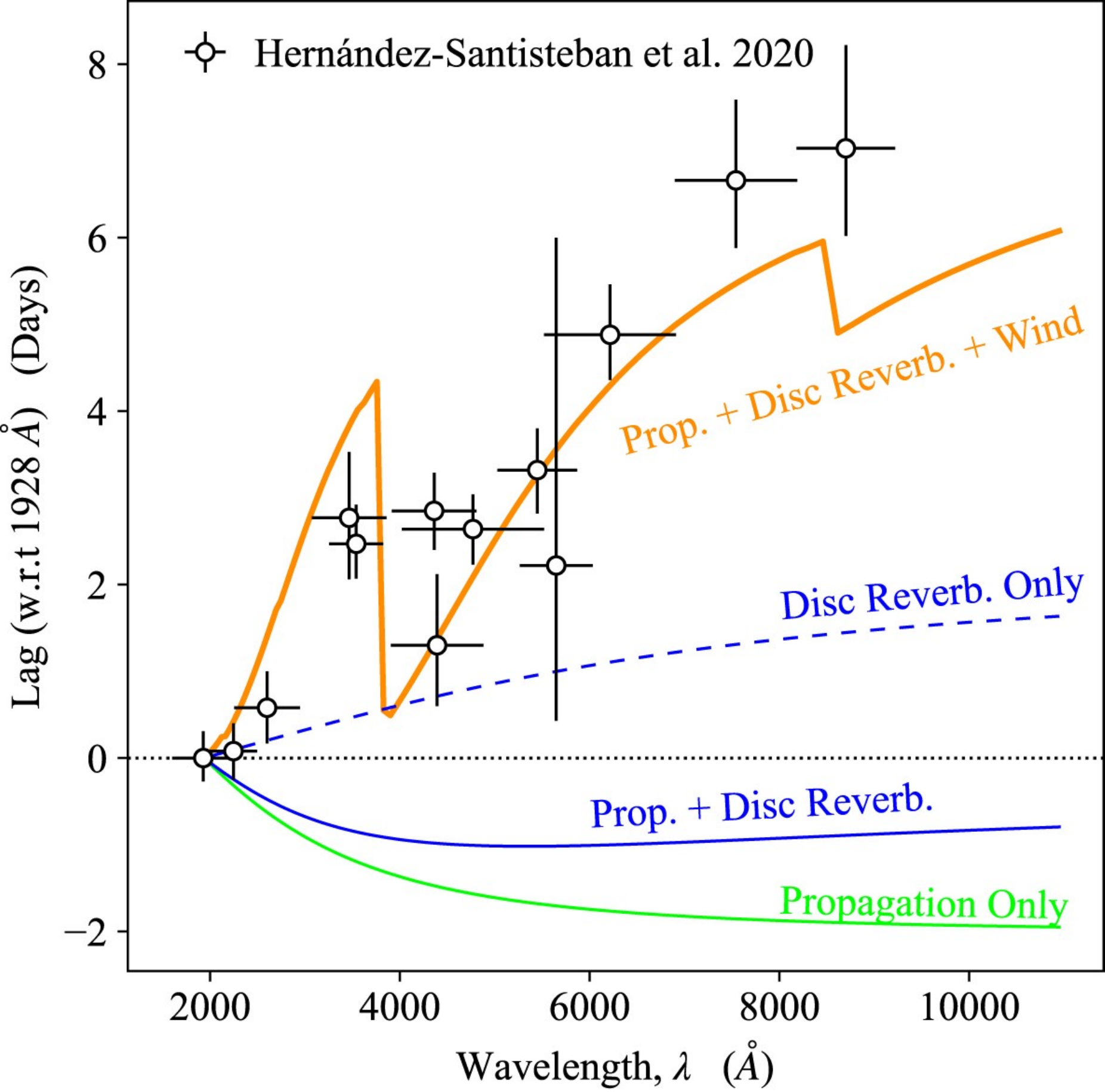}
    \caption{\textit{Left panel}: Observed and calculated broad-band lag-spectra for NGC 5548 (blue and red squares, respectively), in the case of a radiation pressure confined cloud, situated 17$\times L^{1/2}_{5400,44}$ light days from the BH, assuming a covering factor of 0.2. From Figure 3 of \cite{Netzer22}. {\it Right panel}: 
Broad-band time-lags for Fairall 9 \citep{Hernandez19} and the model time lags (orange solid line) including: UV (and X-ray) reverberation from a wind, as well as time-lags due to propagation of accretion rate fluctuations within the disc (gren solid line), and X-ray disc thermal reverberation (dashed blue line), but assuming a small corona height, and not using the \cite{Kammoun23} code. From Figure 10 of \cite{hagen24}.}
    \label{fig:reverbBLR}
\end{figure}

Optical/UV time lags similar to what we observe can be produced by line and continuum emission from gas in the BLR which responds to the continuum extrem UV variations \citep[e.g.][]{korista19,Netzer22}. In fact, \cite{Netzer22} shows that the optical/UV time-lag spectra for many AGN are consistent with the response of diffuse emission from radiation pressure-supported clouds in the BLR with a covering factor of about 0.2, without any need for disc X-ray thermal reverberation (left panel of Fig.\,\ref{fig:reverbBLR}). However, this is not the case with the Fairall 9 time-lags \citep{edelson24}. A complete lack of disc X-ray reverberation would be achieved only if the X-ray source is completely shielded from the disc, by an unknown structure, or in a geometry where the X-ray corona is located within the disc and its height is very small. In this case, the extreme UV part of the disc must be variable to be able to drive the variations of the BLR clouds. 

A similar approach is also presented by \cite{hagen24}. 
They considered an intrinsically variable accretion disc, where slow mass accretion rate fluctuations are generated in the accretion disc and are responsible for the optical/UV disc emission variations. These fluctuations in mass accretion rate also modulate faster X-ray variability from the central region. The model predicts that the intrinsic variability has optical/UV leading the X-rays; however, the model also assumes the presence of a large-scale height wind on the inner edge of the BLR. When it includes reverberation of the variable EUV from the inner wind, then the resulting lagged bound-free continuum matches the observed UV-optical lags. 

It seems that, at the moment, the main driver of the observed time-lags is thought to be due to some kind of high energy photon reverberation; either of the X-ray photons to the disc or of the extreme UV photons by the BLR gas/wind. However, in the second case, the disc must be intrinsically variable (on quite short timescales) to be able to drive the observed variations. Simultaneous modelling and fitting of the observed cross-correlation functions, the power spectral density spectra, and of the time-variable broad-band SEDs will be necessary to determine the mechanism that is responsible for the observed optical/UV and X-ray variability in AGN and, hence, of the X-ray/disc/BLR geometry and physical conditions in AGN.

\subsection{Discovering AGN through optical variability}\label{sec:optvar_discover}

Given the fact that variability is a prominent feature of AGN emission, it is not surprising that it has been widely used to identify active nuclei in galaxies (see \citealt{Mushotzky2004}, for an introductory review on how to find AGN). This approach was adopted not long after the actual discovery of QSO \citep{vandenBergh1973} but became increasingly effective as modern automated photometric techniques were developed at the end of the last century. In fact, the high incidence of variability among quasars observed by \cite{Bonoli1979, Kron_Chiu1981} led several authors to exploit this property to derive homogeneous samples of sources \cite[e.g.][]{Hawkins1983}. Focusing on selected sky regions with deep observations and many photometric standard stars, such as the so-called "Selected Area" (SA) n.57 and n.68, and using observations obtained with the Mayall telescope at Kitty Peak National Observatory, \cite{Koo1986} were able to retrieve samples of quasars down to faint magnitudes (e.g. $B>22$ mag) based on a combination of colours, proper motion, and variability measured between two epochs one year apart. Further studies significantly extended the temporal baseline to more than a decade, using Mayall plates, probing down to $B = 22.6$ mag \citep{Trevese1989,Trevese1994}.

\begin{figure}
    \centering
    \includegraphics[width=0.65\linewidth, angle=0]{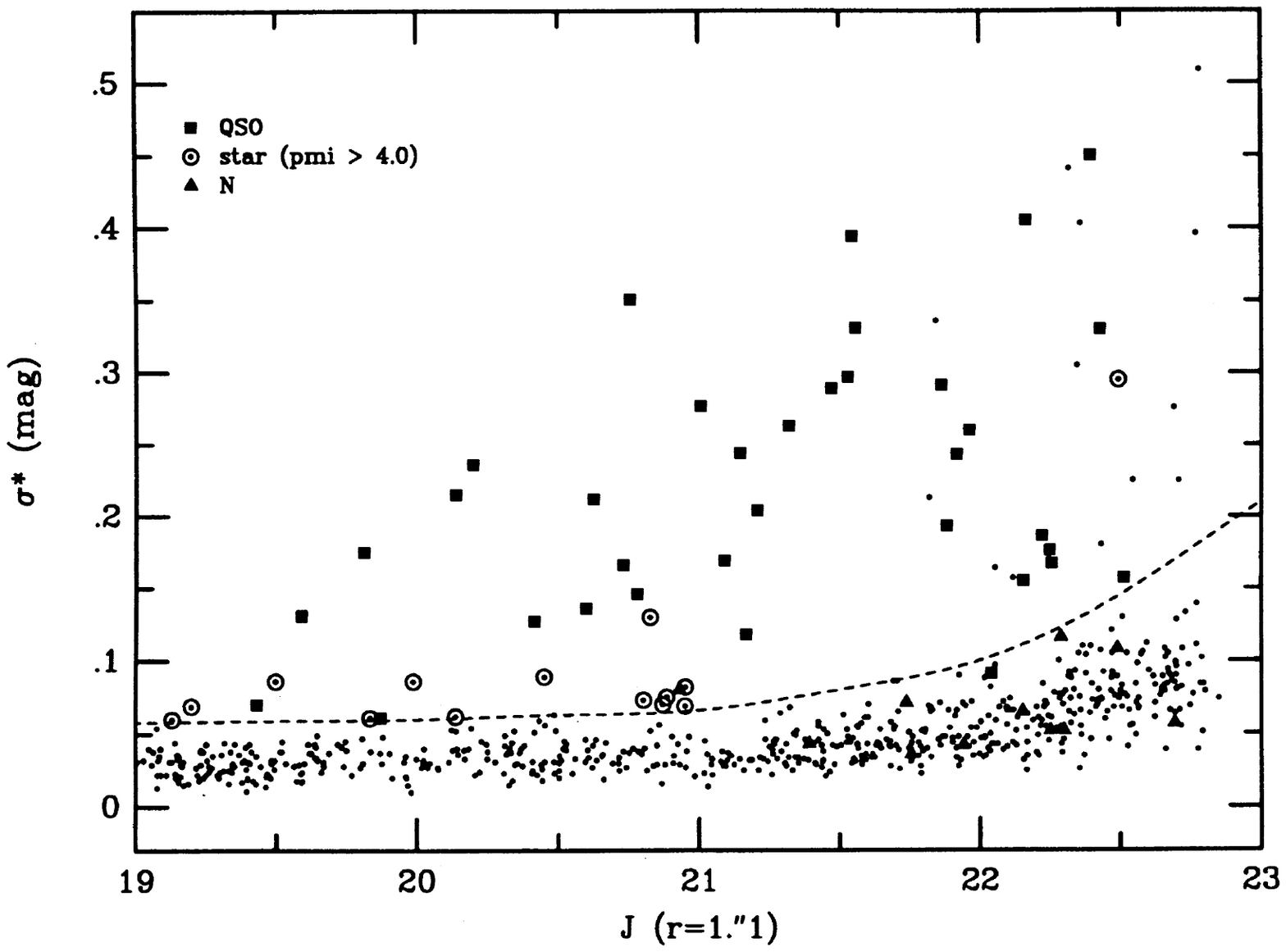}
    \caption{The rms deviation of the $J$ magnitude differences across epochs as a function of the average source magnitude over a 15 yr baseline.  The dotted line shows the adopted variability detection threshold. From Fig.\,1 of \cite{Trevese1994}.}
    \label{fig:Trevese_Fig1}
\end{figure}

The approach introduced by these early works, and followed by many others up to the present, is to compute the rms deviation of the source lightcurves and compare it with the value expected from photometric and calibration uncertainties, as a function of the magnitude of the source (Fig.\,\ref{fig:Trevese_Fig1}); this allows one to identify objects that vary more than a given threshold as potential AGN candidates. The main issue with this method is to properly estimate the uncertainties: errors on magnitudes are often underestimated, and calibration offsets further add to the problem. An effective solution is to use the sources detected in the same field to estimate the actual photometric uncertainties, assuming that the vast majority of them are galaxies or non-variable stars. 
Using this method \cite{Trevese1989} were able to detect QSOs based on variability with $\sim 70\%$ completeness and $\sim 80\%$ reliability with respect to colour-selected samples.
However, it must be noticed that these results are limited to bright quasars, which can be identified as variable pointlike sources, so the efficiency of the method depends on the spatial resolution of the data.

Given the early realisation that lower luminosity AGN seemed to have higher variability than brighter sources (see Sect.\,\ref{sec:optvar_scaling}), \cite{Bershady1998} tried to extend this approach to Seyfert galaxies (e.g. lower luminosity AGN in extended sources). They identified 14 variability-selected galaxies, within the 0.284 deg$^2$ of SA\,57, which appear bluer and more compact than non-variable galaxies. Spectroscopic confirmation was possible for one-third of their sample, indicating that the majority had Seyfert 1 characteristics, and one was consistent with a Seyfert 2 galaxy. They conclude that extended sources make up $\sim 33\%$ of the total AGN population down to luminosities $B_J<22$. Later follow-up studies \citep{trevese2008a} showed that variable sources present optical spectra ranging from typical AGN to XBONG, NELG and LINERs\footnote{XBONG: X-ray Bright Optically Normal Galaxy, NELG: Narrow Emission Line Galaxy, LINER: Low-ionization Nuclear Emission-line Region}.

\begin{figure}
    \centering
    \includegraphics[width=0.45\linewidth, angle=0]{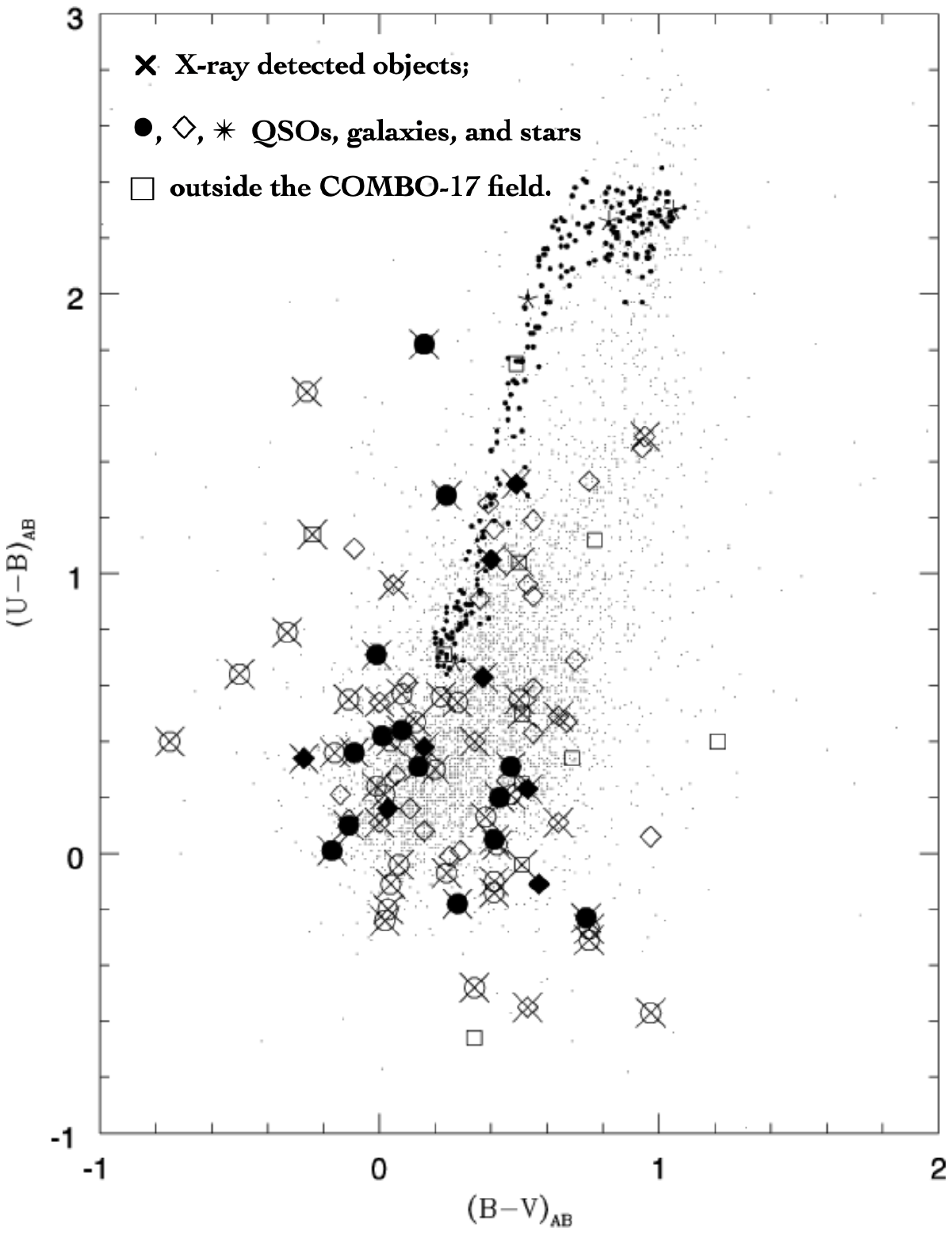}
    \caption{$(U-B)_{AB}$ versus $(B-V)_{AB}$ colours for objects
with EIS photometry in the CDF-S ﬁeld and measured at least at
5 epochs. Non-variable galaxies: small dots; non-variable stars: larger
dots. Variable objects: ﬁlled symbols, objects with spectroscopic redshift; empty symbols, objects without spectroscopic redshift. Adapted from Fig.\,4 of \cite{Trevese2008b}.}
    \label{fig:Trevese_Fig4}
\end{figure}

Similar work was done in the Chandra Deep Field South region by \cite{Trevese2008b}, using 8 epochs from the STRESS SN survey \citep{Botticella2008} spread over 2 years, with the 2.2 m ESO/MPI telescope. They were able to identify 132 variable sources, of which 60\% was confirmed to be AGN based on X-ray, photometric and spectroscopic diagnostics retrieved from the many surveys targeting the same region (COMBO-17, CDF-S, GOODS; Fig.\,\ref{fig:Trevese_Fig4}). 
Spectroscopic follow-up by \cite{Boutsia2009} confirmed that more than half of the unidentified candidates are broad line AGN, while several variable sources are either extended objects which would have escaped the colour selection or objects of very low X-ray to optical ratio, in a few cases without any X-ray detection at all. This implies that variability selection identifies sources that would otherwise escape detection by traditional methods.
On the other hand, the completeness of the variability selection was found to be $\sim 40\%$ but the authors clarified that the efficiency of the method was expected to increase as deeper and better sampled light curves became available in future surveys, as confirmed by later studies discussed below. 

\begin{figure}
    \centering
    \includegraphics[width=0.55\linewidth, angle=0]{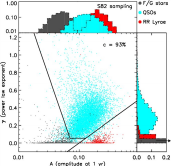}
    \caption{Distribution of the variability Structure Function amplitude $A$ and slope $\gamma$ for objects in {\it Stripe 82}. Spectroscopically confirmed quasars are shown as light blue points; confirmed RR Lyrae and color-selected F/G stars are shown in red and gray, respectively.  The  solid lines define the quasar selection wedge. From Fig.\,5 of \cite{Schmidt2010}.}
    \label{fig:Schmidt2010_Fig5}
\end{figure}

\cite{Schmidt2010} proposed a different approach to select quasars in large surveys, based on the amplitude $A$ and slope $\gamma$ of the Structure Function. Using $\sim 60$ epochs of imaging data, taken from the SDSS {\it Stripe 82} survey over 5 years, they showed that quasars can be effectively separated from other non-variable and variable sources by their location in the $A–\gamma$ plane (Fig.\,\ref{fig:Schmidt2010_Fig5}). They noted that variability performs just as well as colour selection in the identification of quasars with a completeness of 90\% and a purity of 95\%, also in redshift ranges where color selection is known to be problematic. Even with much sparser time sampling, e.g. just six
epochs over 3 years, variability is an efficient quasar classifier reaching $>90\%$ purity and $44\%$ completeness. Similarly, \cite{Palanque-Delabrouille2011} combined $\chi^2$ and SF parameters derived from {\it Stripe 82} lightcurves but used a supervised Neural Network trained on the BOSS spectroscopic sample, to distinguish stars from quasars. In line with previous findings, they were able to retrieve samples with $>84\%$ completeness and about $>70\%$ purity, somewhat dependent on the adopted colour cuts. They also observed that Broad Absorption Line (BAL) quasars are preferentially selected by variability than by color criteria. In \citet{Palanque-Delabrouille2013, Palanque-Delabrouille2016} they implemented this approach to study the quasar luminosity function.
To further improve this methodology, \cite{MacLeod2011} proposed using a damped random walk model parametrised by the damping timescale $\tau$ and the asymptotic amplitude SF$_\infty$. In particular, the inclusion
of the damping timescale information increased the purity from 60\% to 75\% while maintaining a highly complete sample (98\%) even in the
absence of color information. This approach may be limited by the ability of DRW models to reproduce actual AGN lightcurves (see Sect.\,\ref{sec:optvar_model}) and by the fact that $\tau$ is poorly constrained if the data do not probe timescales significantly longer than the damping one (see Sect.\,\ref{sec:optvar_phen}). Nevertheless, this methodology looks promising for current and next-generation surveys spanning several  years (see Sect.\,\ref{sec:nextgen_opt}).

Variability selection has proven to be effective even in very crowded fields. For instance \citet{Kozlowski2010, Kozlowski2013} searched for quasar candidates behind the Magellanic clouds using the Optical Gravitational Lensing Experiment (OGLE) identifying regions of the DRW and photometric parameter space, containing $\sim 60\%$ of all confirmed AGN, and discovering $> 700$ new quasars. A combination of colours, variability, and spectroscopy is also proving effective in identifying quasars close to the Galactic plane ($|b|<20^\circ$), as discussed by, e.g., \citet{Huo2025}.

Despite all these efforts, probing the faint end of the AGN luminosity function by variability selection has proven to be challenging, because of the contamination of the host galaxy, which strongly dilutes the emission of faint, obscured, and/or high-redshift sources when using ground-based observations. 
\citet[][also see \citealt{Sarajedini2000}]{Sarajedini2003} compared two observations of the Hubble Deep Field, separated by a gap of 5 years, with the goal of investigating the population of active galactic nuclei to $z\sim 1$. They found evidence of variability in $8\%$ of galaxies down to $V_{nuc}\sim 27.5$. About $44\%$ of their AGN candidates are X-ray confirmed, while spectroscopic observations reveal that their sample is mainly composed of type 2 AGN. Furthermore, when stacking the X-ray undetected sources, it is usually found that there is a significant excess of hard X-ray emission (N. Brandt, private communication). On the other hand $\sim 40\%$ of the X-ray sources in the field are optically variable, supporting their interpretation as low-luminosity AGN (LLAGN). \cite{Cohen2006} also studied variable AGN in the Hubble Ultra Deep Field, in the $i$-band, down to 28 mag. They concluded that about 1\% of all field objects in the range of $0.1\leq z \leq 4.5$ show variability. They discussed the possibility of tracing AGN activity induced by merger events through variability, although with somewhat inconclusive results.  

The study of faint AGN using observations from space was further extended by \cite{Klesman2007} who used ﬁve HST V-band epochs separated by 45-day intervals in the GOODS-south region. They  focused on the optical variability properties of AGN detected in other bands, namely mid-IR and X-rays, finding that 26\% of them are variable. This fraction increases to $>50\%$ considering ``soft'' X-ray sources, indicating that optical variability preferentially selects Type 1 unobscured AGN. Although most of the known AGN did not reveal significant variability, they concluded that this could change using longer temporal baselines to also probe the obscured population.

\begin{figure}
    \centering
    \includegraphics[width=0.46\linewidth, angle=0]{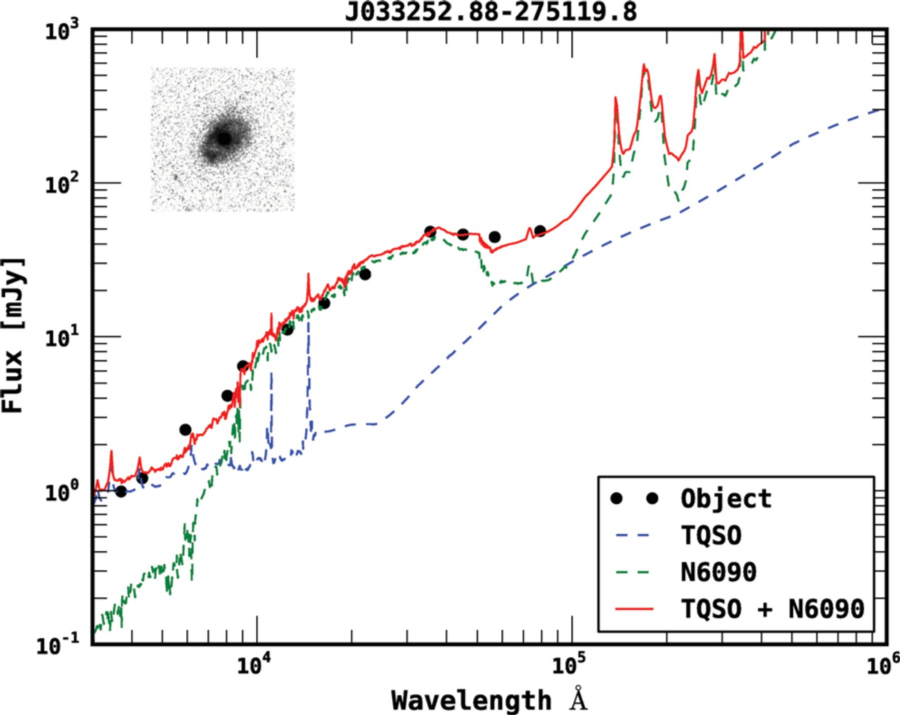}
    \includegraphics[width=0.48\linewidth, angle=0]{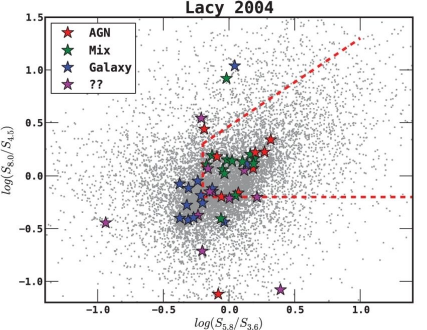}
    \caption{{\it Left panel}: Spectral energy distribution of one variable LLAGN candidate, revealing the presence of both an AGN and  a normal galaxy component. {\it Right panel}: colour–colour plot for AGN selection. The dashed lines show the \cite{Lacy2004} AGN ‘wedge’. Small black dots are all objects with detection in IRAC bands. Stars show variability-selected AGN, colour coded by AGN contribution, based on SED fitting. Variability selection effectively identifies AGN missed by other diagnostics. From Fig.\,8 and 14 of \cite{Villforth2012}.}
    \label{fig:Villforth_Fig8_14}
\end{figure}

\citet{Villforth2010}, used the 5 epochs, $z$-band catalogue of the GOODS field,  spanning $\sim 1$ year, to identify AGN using different statistics. They showed that among $\chi^2$, $F$ and $C$ statistics, the first has the highest retrieval power, especially when the photometric errors are not well constrained and the samples are small. In total, they detected 155 variable sources that represent 1.3\% of the total catalogue, among which 139 AGN candidates down to $z<25.5$ mag. The contamination of their variable AGN sample was estimated to be $\sim 8\%$, but it is strongly dependent on the adopted variability selection threshold and could reach $>100\%$ for lower thresholds. This highlights the main problem of variability selection studies using only a few epochs where, to obtain complete samples, the fraction of contaminants can be very large due to the larger photometric errors at faint magnitudes.
The follow-up study presented in \citet{Villforth2012} confirmed that variability selection reliably identifies AGN (Fig.\,\ref{fig:Villforth_Fig8_14}), predominantly of low
luminosity, often missed by other methods. Their host galaxies span a wide range in the level of ongoing star formation, and massive starbursts are only present in the hosts of the most luminous AGN.
Using \textit{TESS} lightcurves \cite{Treiber2023} additionally showed that space observatories are particularly efficient in selecting low-mass AGN ($M_{BH}\lesssim 10^6 M_\odot$) which represent $\sim 20\%$ of the variability-selected sources.

Although space observations are indeed superior in terms of spatial resolution, until recently they lacked the depth and sky coverage allowed by ground facilities. In order to bridge the gap in the ability to detect low-luminosity AGN, \cite{Choi2014} applied the image-difference technique common to SN searches to {\it Stripe 82} data, in combination with light curve statistics. The method excels in identifying variable sources embedded in complex or blended emission regions such as Type 2 and low-luminosity AGN, reaching 93\% completeness and 71\% purity. Interestingly, 1/3 (two out of six) of their Type 2 AGN candidates show long-term variability like typical Type 1 AGN, a result which could challenge the prediction of the AGN unification model.

\begin{figure}
    \centering
    \includegraphics[width=0.85
    \linewidth, height=0.3\textheight, angle=0]{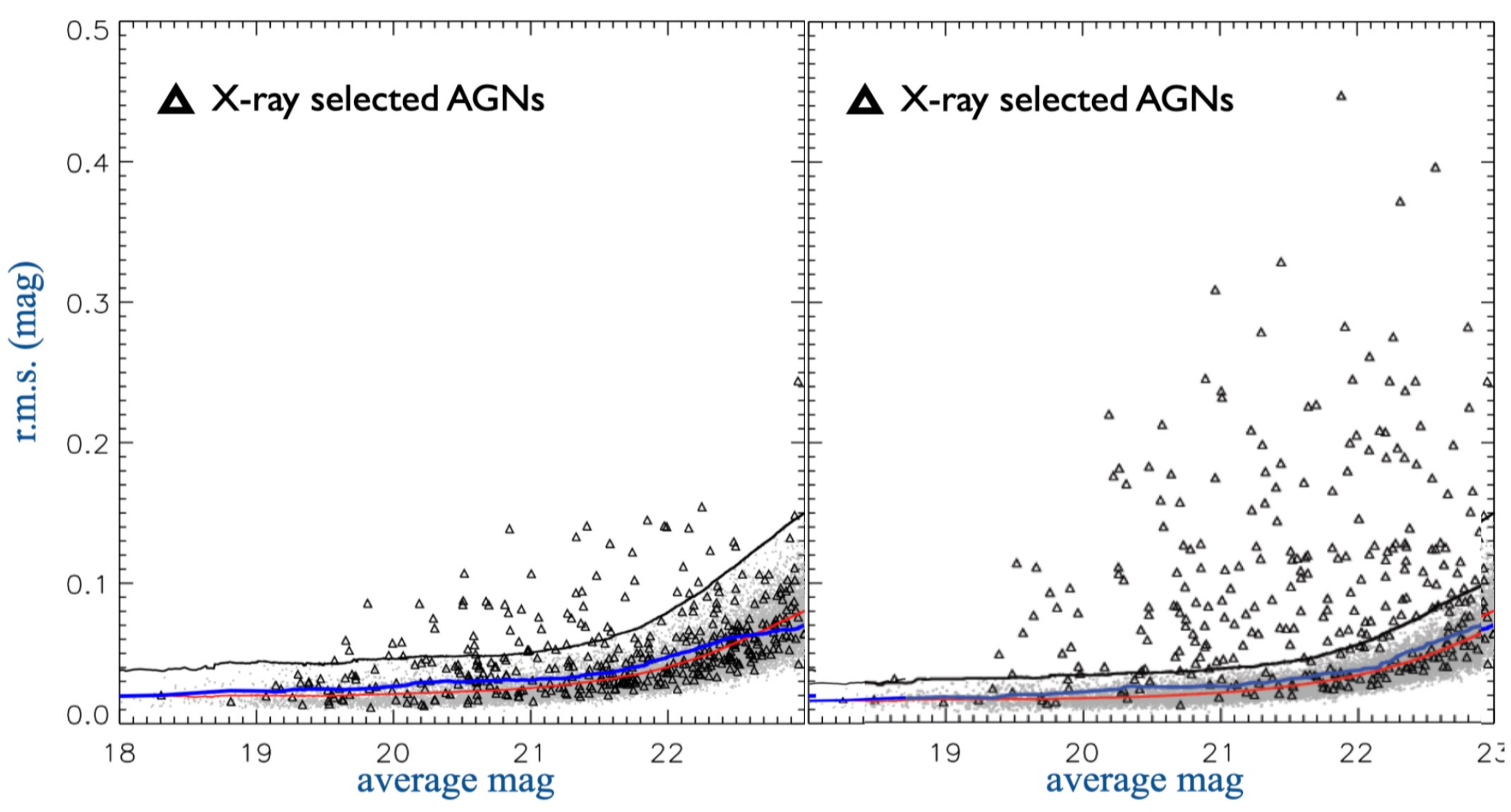}
    \caption{Light curve r.m.s. as a function of the average magnitude for all the non-variable sources in the COSMOS sample (small dots) and for those with an X-ray counterpart and that are spectroscopically confirmed to be AGN (triangles), from the initial 5 months \citep[][{\it left panel}]{DeCicco2015} and the 3 year analysis \citep[][\textit{right panel}]{DeCicco2019}. The red and blue curves represent the running average of the r.m.s. deviation for the two subsamples of sources, respectively. In the 3 year work  59\% of the X-ray sources above the variability threshold (black line), while they were only 15\% after the first 5 months. Adapted from Fig.\,8 of \cite{DeCicco2019}.}
    \label{fig:DeCicco19_Fig8}
\end{figure}

To predict the capabilities of next generation synoptic surveys, the efficiency of variability selection should be compared with other methods based on multiwavelength and morphological information, probing down to the luminosity limits expected from future facilities. To this end \cite{DeCicco2015, Falocco2015, Poulain2020} studied regions that would become the LSST {\it Deep Drilling Fields}. They used wide-field data acquired by the VLT Survey Telescope (VST) within the SUDARE \citep{Botticella2013} and VOICE \citep{Vaccari2016} surveys, to build samples of variability selected AGN using 22-29 epochs covering a baseline of $\sim 6$ months down to $r_{AB}\simeq 23$ mag. This cadence and depth match the one envisioned by the future LSST \citep{Brandt2018}, albeit over a smaller area, and proved useful in estimating the performance expected by the upcoming survey. In fact, due to the wealth of photometric and spectroscopic ancillary data available on these fields, they showed that they can achieve $> 80\%$ purity, considering that the majority of contaminants are SNe which can be easily identified and removed with longer observations or light curve inspection. Among the confirmed active galaxies, Type 1 AGN are prevalent (89\%) w.r.t. Type 2 (11\%). More interestingly, variability selection proved to be able to retrieve only $\sim 15\%$ of all AGN in the field identified by means of spectroscopic or X-ray classification, and is roughly consistent with the infrared selection proposed by \cite{Lacy2004} or \cite{Stern2005}. Although the fraction of retrieved X-ray AGN appears low, the authors noticed that the average variability of the rest of the X-ray sources which fall below the variability selection threshold, is indeed higher than the normal galaxy population, suggesting that longer monitoring campaigns would significantly increase their number.
In fact, the follow-up work of \cite{DeCicco2019} extending the campaign on the COSMOS field up to 3 years, showed that using 55 visits and a 6-times longer baseline it is possible to retrieve 3.5 times more AGN due to the red-noise behaviour of their lightcurves. Furthermore, the fraction of X-ray spectroscopically confirmed AGN retrieved by variability selection increases to $\sim 60\%$ (Fig.\,\ref{fig:DeCicco19_Fig8}) although the sample is still strongly biased towards Type 1 sources.

\begin{figure}
    \centering
    \includegraphics[width=0.4
    \linewidth]{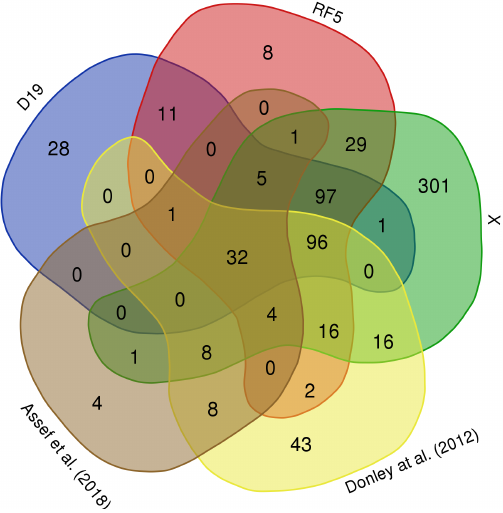}
    \caption{Venn diagram showing the overlap among samples of AGN candidates returned by the random-forest classifier (RF), AGN candidates selected from variability with standard methods in \citet{DeCicco2019}, X-ray AGN selected from X/O diagnostics (X); AGN selected from MIR diagnostics  of \cite{Donley2012} and the sources classified as AGN based on the IR criterion of \cite{Assef2018}. From Fig.\,2 of \cite{DeCicco2021}.}
    \label{fig:DeCicco21_Fig9}
\end{figure}

As astronomy embraced the potential of neural networks and machine learning methods, AGN selection methods also improved. \citet{Graham2017} proposed the use of a stacked ensemble classiﬁer to combine WISE colors with
variability features from structure functions, autoregressive models, and Slepian wavelet
variance measured from CRTS data, obtaining completeness and purities of $\sim 99\%$ although for relatively bright quasars ($V\sim 19.5$ mag).
\cite{Sánchez-Sez2019} explored 70 deg$^2$ with the QUEST-La Silla AGN variability survey down to $m_r\lesssim 21$ through a random forest classifier using a combination of variability parameters and photometric colors. They were able to detect both typical blue sources usually found in previous works and red sources with a significant contribution from the host galaxy. A random forest classifier was also used by \citet{DeCicco2021} to improve the selection of faint and/or Type 2 AGN in the COSMOS field and to forecast the performance of the \textit{Vera C. Rubin} Observatory. Combining variability statistics and optical/near-IR colours, they were able to archive up to $>90\%$ completeness depending on the set of characteristics used in the study, although the result for Type 2 is at best 36\% (Fig.\ref{fig:DeCicco21_Fig9}). \cite{DeCicco2025} further refined this approach, identifying the best features to separate normal galaxies from those containing Type 2 AGN, and reaching a completeness of $\sim 68\%$ and an F1 score\footnote{The F1 metric is defined as the harmonic mean of the \textit{purity} and \textit{completeness}, representing a global estimator of the performance of the classifier.} of 86\%.
\cite{Nakoneczny2025} processed the ZTF lightcurves with a transformer artificial neural network 
showing that the addition of variability features significantly improves the classification of quasar samples with respect to static magnitudes and colors, reaching similar performances as the Gaia or WISE catalogs. They further show that a temporal temporal baseline $\gtrsim 900$ days is needed to obtain a F1 score of 90\% if the sampling cadence is of $\sim 3$ days (as for ZTF), but extending the baseline the cadence can be significantly reduced.

All these efforts demonstrated that AGN selection through optical variability is a powerful method to retrieve large, pure, and reasonably complete samples of AGN, partly complementary to other selection techniques, although Type 2 AGN are still challenging to identify. The availability of deep, extended, multi-band observations expected from ongoing and upcoming surveys are expected to help in overcoming these limitations, as discussed in next section.

\subsection{Prospects for current and next generation optical surveys.}
\label{sec:nextgen_opt}

\begin{figure}
    \centering
    \includegraphics[width=0.85\linewidth]{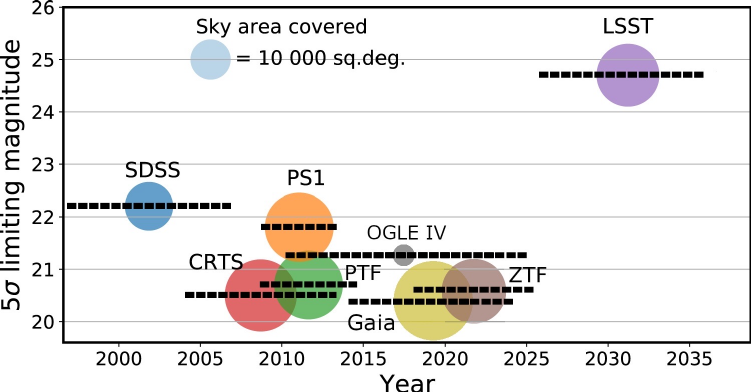}
    \caption{Temporal baseline, sky area covered, and depth of some of the main current and upcoming optical time-domain surveys. The thick dashed lines show the approximate extent of each survey. The y-axis shows the $5\sigma$ limiting magnitude (SDSS-DR15 r, PS1 r, PTF R, ZTF r, LSST r, CRTS V). The circle size is proportional to the  survey area. Modified and updated from Fig.\,2 of \cite{Suberlak2021}.}
    \label{fig:Suberlak_surveys}
\end{figure}

As discussed in the previous sections, despite more than 60 years of studies of AGN variability, we are still missing a full understanding of the underlying physical process and of the connection with the properties of the whole AGN population. Several factors contribute to these limitations. First, previous studies were often based on samples that covered a narrow range of the AGN parameter space. 
Several surveys have been designed and are currently operating for the study of time variable phenomena (PTF, ZTF, CRTS, SDSS-{\it Stripe 82}, OGLE, Pan-STARRS etc., see Fig.\,\ref{fig:Suberlak_surveys}). However, they also suffer from limitations in photometric quality, temporal baseline, wavelength, and area coverage. Furthermore, all ground-based surveys share the problem of angular resolution limits due to the atmosphere and of seasonal gaps in the light curves. More importantly, a proper understanding of AGN variability requires spectroscopic and multiwavelength campaigns in order to derive the physical properties of AGN like mass, accretion rate, obscuration, and inclination.

\begin{figure}[!t]
    \centering
    \includegraphics[width=0.65
    \linewidth]{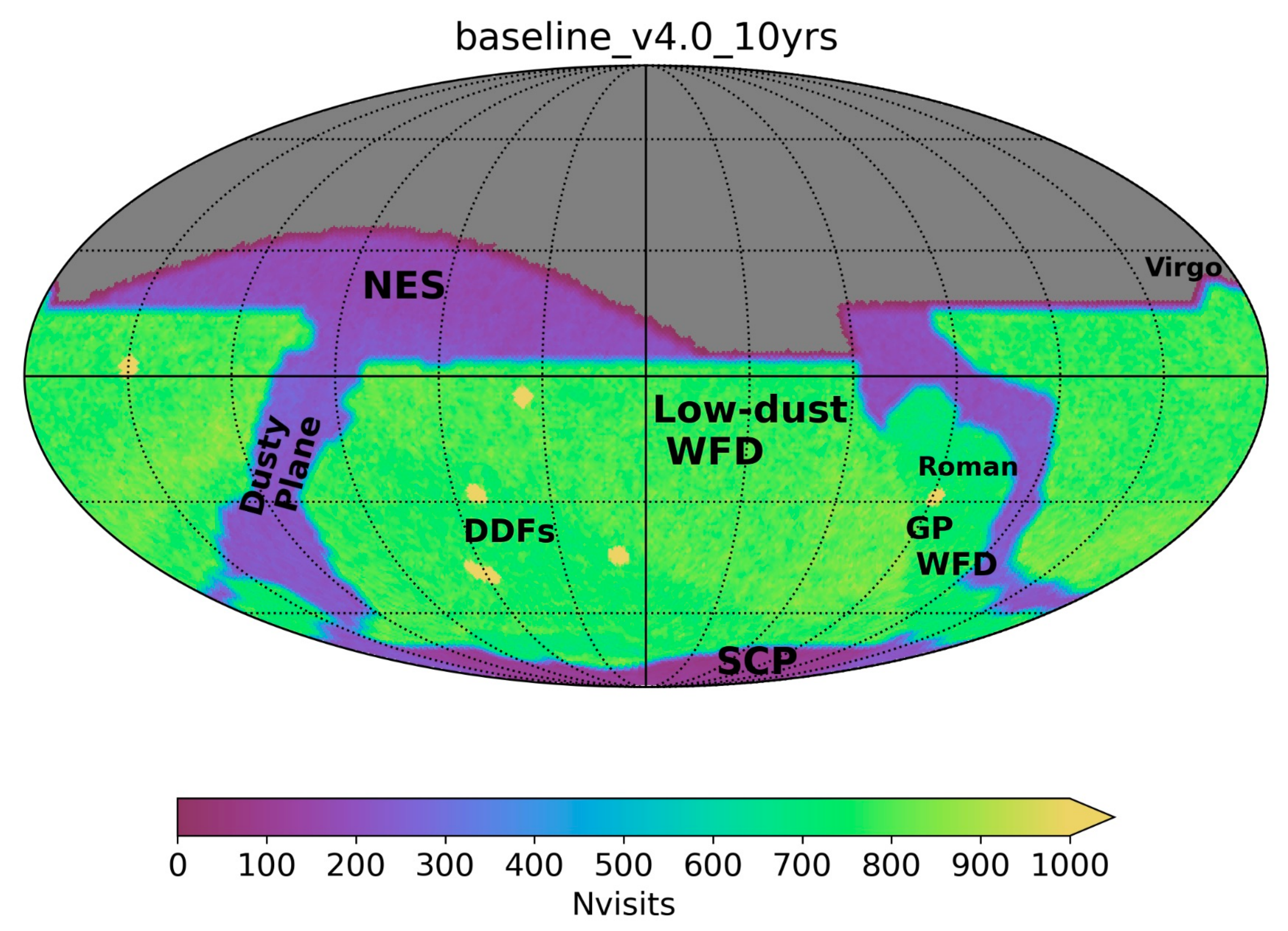}
    \caption{The LSST survey strategy, over the full 10 years duration of the survey, as proposed in the latest recommendations of the Survey Cadence Optimization Committee (see footnote \ref{foot:strategy}).}
    \label{fig:LSST_strategy}
\end{figure}

This decade will see the beginning of operations of the 8.4 m \textit{Vera C. Rubin} Observatory\footnote{\href{https://rubinobservatory.org}{https://rubinobservatory.org}}, which will conduct the Legacy Survey of Space and Time \citep[LSST,][]{Ivezic2019} through the 3.2 gigapixel focal array camera. Although the details of the observing strategy are still being finalised \citep[][also see the LSST website\footnote{\label{foot:strategy}\href{https://www.lsst.org/content/charge-survey-cadence-optimization-committee-scoc}{https://www.lsst.org/content/charge-survey-cadence-optimization-committee-scoc}}]{Bianco2022}, the LSST will conduct a deep-wide-fast survey that will observe a $\sim 18,000$ deg$^2$ region about 800 times (i.e. every few nights in the optical/near-infrared $ugrizy$ bands, Fig.\,\ref{fig:LSST_strategy}) during the anticipated 10 yr of operations. It will also yield co-added images that will probe astronomical sources down to $r\sim 27.5$. These data will include observations of 20 billion galaxies and a similar number of stars, as well as tens of millions of AGN \citep{lsstsciencebook}. In addition to the main survey, LSST will also target a number of \textit{Deep Drilling Fields} (DDF), namely ELAIS-S1, XMM-LSS, CDF-S, COSMOS and \textit{Euclid}-DFS, which will benefit from a faster cadence resulting in significantly deeper final co-adds \citep{Brandt2018}. LSST will also take advantage of a number of ancillary surveys, both photometric and spectroscopic, intended to increase the cadence and length of the lightcurves, the wavelength coverage and the spectroscopic completeness \citep[e.g.][also see the LSST website\footnote{\href{https://www.lsst.org/scientists/in-kind-program}{https://www.lsst.org/scientists/in-kind-program}}]{Swann2019}, .

Considering the number of studies and advances triggered by the SDSS {\it Stripe 82} light curves, 
it is reasonable to assume that LSST will change the field. Preliminary studies have already tried to assess the ability of LSST to detect and characterise AGN through a combination of colours, variability, and proper motions, using simulations and current survey data \citep[e.g.][]{Sartori2019,Suberlak2021, Kovacevic2022, Czerny2023, Fatovic2023, McLaughlin2024}, as well as machine learning algorithms \cite[e.g.][]{Doorenbos2022, Kovacevic2023, Fagin2024}. Among these efforts, \emph{LSST AGN data challenge} \citep{Savic2023} aimed to validate both classical and machine learning approaches for AGN selection and characterisation, using SDSS {\it Stripe 82} lightcurves as a test ground. The performance of supervised (support vector machine, random forest, extreme gradient boosting, artificial neural network, convolutional neural network) and unsupervised models (deep embedding clustering) obtained a classification accuracy of 97.5\% for supervised models and a clustering accuracy of 96.0\% for unsupervised ones on a blinded data set. The experiment confirmed that variability features significantly improve the accuracy of the trained models, and correlation analysis among different bands enables a fast and inexpensive first-order selection of quasar candidates. Such attempts only provide a rough estimate of the LSST potential, since they rely on pre-selected, confirmed quasar samples and do not properly probe the average AGN population accessible to the survey. For this reason, other studies, some mentioned before, tested the ability to detect normal AGN either by using image-difference analysis or extended lightcurves probing some of the DDFs down to single-epoch depths comparable to LSST \citep[e.g,][]{Choi2014, DeCicco2019, DeCicco2022, DeCicco2015}.

Eventually, even LSST will be hampered by seasonal gaps and seeing limitations and by an expected temporal span of 10 years, which has been proven to be insufficient (although not at the same depth and cadence) to provide a full understanding of optical variability in previous studies (see discussion in Sect.\,\ref{sec:optvar_phen} and \ref{sec:optvar_discover}). For this reason it is essential that the astronomical community embarks in an effort to merge all existing and upcoming data to increase the temporal coverage of many thousand to millions of AGN up to several decades, hopefully allowing the complegte characterisation of the AGN variations 
and disentangle the processes dominating different temporal frequency domains. In addition, new methods are being developed to deal with sparsely sampled lightcurves and more flexible empirical models are being developed which will help to deal with some of the limitations of current and future datasets \cite[e.g.][]{Lefkir2025}.

Still, obscured and low-luminosity AGN may remain beyond reach with ground observations even in the next decade. Space observatories can indeed provide access to such populations, but current and upcoming facilities are not designed to efficiently probe the time domain. For instance, the \textit{Euclid} mission \cite{Euclid2024} is surveying the whole sky at unprecedented depth and with Hubble-like angular resolution ($\sim 0.2''$); however, even ignoring the problems due to the wide photometric band of the VIS instrument, which averages the signal from distinct regions of the disk,  only a few fields (the \textit{Euclid} Deep Fields) will be imaged repeatedly by \textit{Euclid} over its survey, providing rather poorly sampled lightcurves compared to ground studies. One possible exception is represented by the SelfCal field which is imaged every few weeks for calibration purposes and may allow to probe the faint end of the AGN luminosity function discovering and monitoring many thousand of LLAGN, as done by past studies with \textit{Hubble} on much smaller sky regions (Sect.\,\ref{sec:optvar_discover}).

We finally mention that once these limitations are removed, variability may become an effective tool to, e.g., measure masses independently from spectroscopy, \citep{Kelly2013,Treiber2023} and provide a new observable to constrain the connection between SMBH and their host galaxies \citep[e.g.,][]{Sartori2019,georgakakis21}.

\clearpage
\section{X-ray variability}
\label{sec:x-rayvar}

We define, for the purpose of this review, the X-ray band as from $\sim 0.1$ to 100 kev, with an emphasis on the X-ray variability studies below 10 keV. The X-ray band is the only spectral band in which rapid (i.e. on timescales of minutes/hours) and large-amplitude variations (i.e. by more than a factor of 2 or so) have been detected in radio-quiet AGN. These variability properties suggest that the X-ray emitting region in AGN is small. Since the X-ray luminosity is a sizable fraction of the bolometric luminosity in these objects, it is reasonable to assume that the X-ray source is located close to the central BH, where most of the accretion power is released. Therefore, X-ray variability studies should provide important clues regarding the physical processes that operate in the innermost part of active galaxies. 

We have limited direct observational evidence on the size and location of the X-ray source in AGN. Perhaps the most constraining evidence comes from monitoring observations of several lensed quasars. Modelling of multi-wavelength light curves of these objects provides constraints on the extent of the optical, UV, and X-ray emission regions. The results so far indicate that the X-ray emission region is smaller in size than the optical/UV emission region in AGN. \cite{Chartas16} show that the half-light radius of the X-ray source of quasars, R$_{1/2,X}$, is as small as $\sim 10$R$_{\rm G}$. This is confirmed by \cite{Dogruel20}, who also found that R$_{1/2,X}\sim10$R$_{\rm G}$ in radio-quiet AGN, although with large errors (see, e.g., their Eq. (9) and Fig.\,20).  This analysis has been performed so far on a small number of quasars. Nevertheless, these are powerful results, which strongly suggest that the X-ray source in AGN is (very) small and is located at the centre of the active galaxies, next to the supermassive black hole. 

\subsection{A brief history of AGN X-ray observations}

The first report of significant X-ray variability of an AGN, using data from a single instrument, was \cite{elvis76} who reported an increase in the X-ray flux of NGC 4151 by a factor of 2 in 3 days. \cite{marshall81} studied 5 years long light curves of 28 AGN using data from the Sky Survey Instrument on board {\it Ariel V} and found that more than half showed significant variations on timescales 
less than a year. Later, \cite{tennant83} studied the X-ray variability of 38 AGN (mostly Seyfert 1), using light curves from the {\it HEAO1 A-2} experiment. They concluded that significant variability of X-rays on timescales less than $\sim$ half a day is rare among AGN. Similar results were reported by \cite{zamorani84}, who studied the X-ray variability of 51 quasars observed with the Einstein Observatory. They found that none of them showed variability on timescales less than a day. 

\begin{figure}
    \centering
    \includegraphics[width=0.47\linewidth, height=0.4\linewidth]{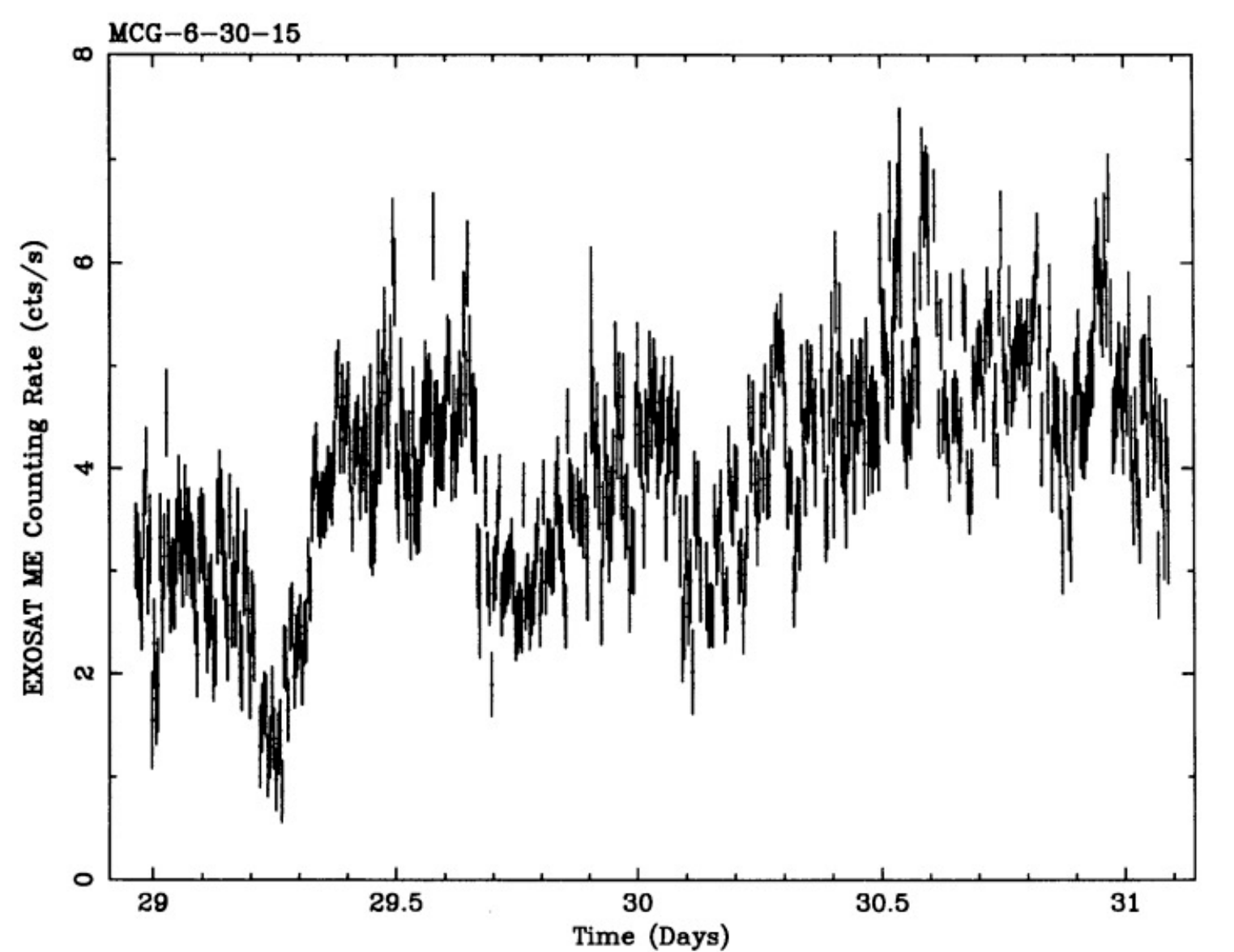},
    \includegraphics[width=0.51\linewidth,height=0.39\linewidth]{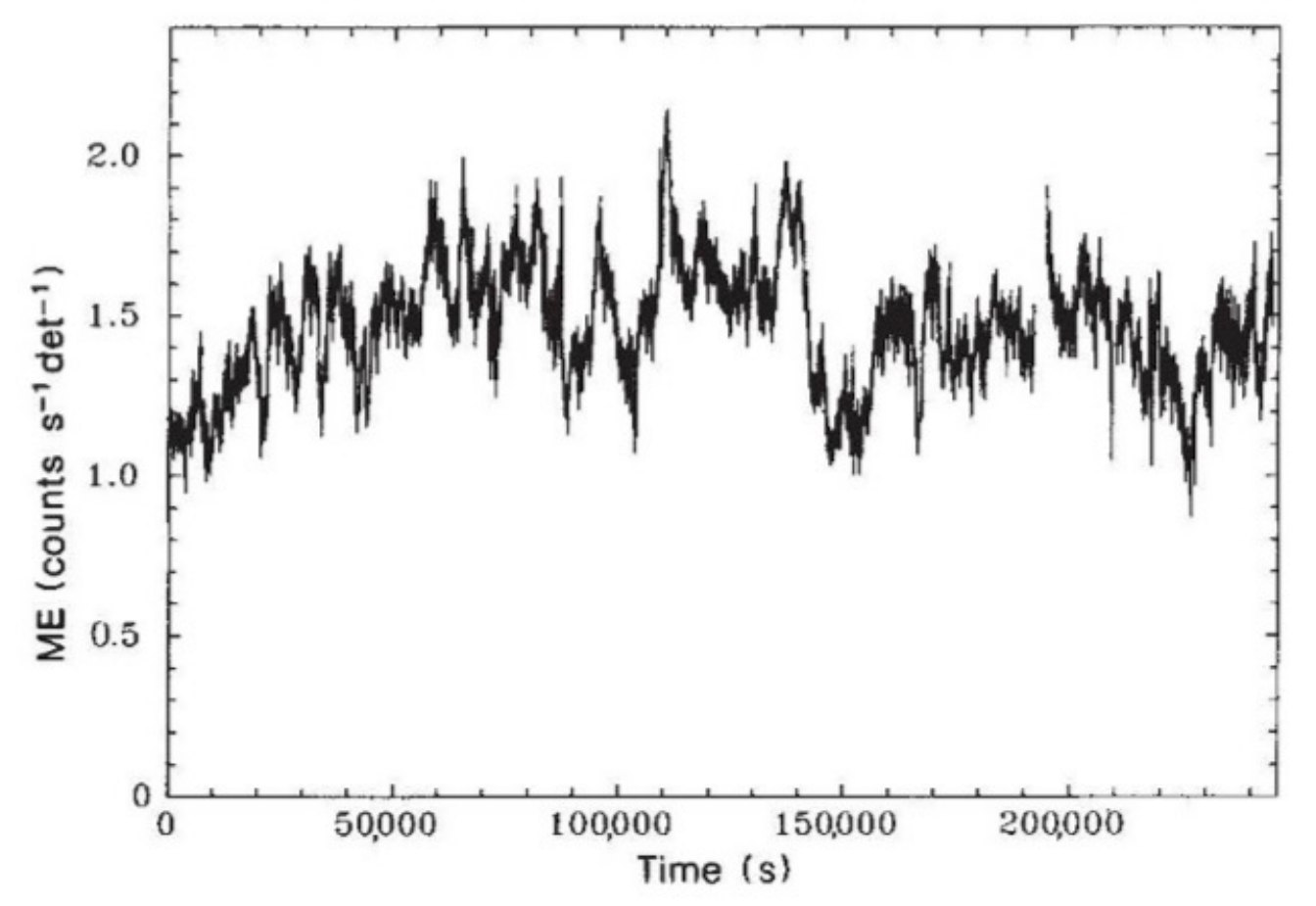}
    \caption{The ``long-look" {\it EXOSAT} ME light curves of MCG-6-30-15 and NGC 5506 (left and right panels; from Fig.\,1 of \cite{krolik93} and Fig.\,1 of \cite{mchardy87}, respectively). They clearly show that the objects are variable on all sampled time-scales.}
    \label{fig:exosatlcs}
\end{figure}

It was observations made by the Medium Energy (ME; 2-10 keV), and the Low Energy (LE; 0.05-2 keV) instruments on board the {\it EXOSAT} observatory that helped us better understand the nature of the X-ray variability of AGN. This was because these instruments were more sensitive than previous X-ray detectors and because of the unique capability of {\it EXOSAT} to continuously monitor a source for up to 3 days. Figure \ref{fig:exosatlcs} shows an example of the {\it EXOSAT} ``long-look'' light curves. AGN appeared to be variable on all the timescales sampled, from minutes to $\sim$ days. 
The long {\it EXOSAT} light curves could also be used for power spectrum analysis, for the first time in the X-ray variability studies of AGN. Since then, power spectral analysis has been the main tool for studying and characterising the X-ray variability of these objects. Thus, before continuing, 
we briefly summarise the basics of power spectrum analysis, as done in Sect.\,\ref{sec:optvar_methods} for the structure
function analysis.

\subsection{The power spectral density function of a random process.}\label{sec:x-rayvar_psd}


The power spectral density function (PSD) is one of the most important properties of a stationary random process\footnote{A random process is called stationary when the statistical properties of the process
do not change over time.} 
It is closely related to the representation of a stationary process, $X(t)$, in the frequency domain, as a Fourier-Stieltjes transform of the form 

\begin{equation}
X(t)=\int_{-\infty}^{+\infty} e^{i\omega t}dZ(\omega),
\end{equation}

\noindent where $dZ(\omega)$
is a complex random function, which determines the amplitude of the sinusoids, $e^{i\omega t}$, into which we can decompose $X(t)$ \citep[see, e.g.,][]{Priestley1982}. In fact, the amplitude of the sinusoids is proportional to $|dZ(\omega)|^2$. We therefore define the PSD of the stationary random process $X(t)$ as follows: 

\begin{equation}
{\rm PSD(\omega)}= \overline{|dZ(\omega)|^2}/d\omega,
\end{equation}

\noindent where the overbar denotes the average taken over all realisations of the process. The equation above shows that the PSD at frequency $\omega$ is proportional to the amplitude of the sinusoid of the same frequency. This property is one of the reasons that explains the usefulness of the PSD analysis: it is easier to search for periodic signals in the PSD of a light curve instead of the light curve itself in the presence of experimental errors. Poisson noise will tend to smooth out any periodic components in the light curve, especially those of low amplitude. On the other hand, when working in the frequency domain, the PSD of experimental noise will be constant with frequency (white noise), whereas the intrinsic PSD will show a peak at the frequency of the sinusoid that is responsible for the periodic signal which may be easier to detect.

It can be shown that PSD is closely related to the autocovariance function of a random process $R(\tau$)\footnote{Here we prefer this notation instead of the $cov(s_i,s_j)$ used in the optical variability part, since it is more common in X-ray variability studies.},

\begin{equation}
R(\tau)=E[\{ X(t)-\bar{X} \} \{ X(t+\tau)-\bar{X} \}],
\end{equation}
\smallskip
\noindent where $E$ is the mean operator, $\bar{X}$ is the mean of the process, and $\tau$ is usually called ``time lag" (or simply ``lag"). As discussed also in Sect.\,\ref{sec:optvar_methods}, this function is very useful, as it
quantifies the ``memory" of the system, which is one of the most important characteristics of a random process. For most of the random processes observed in astrophysics, memory is clearly present in the sense that the value of the process at time $t$ depends on its past values. Consequently $R(\tau)$ is non-zero over a wide range of lags. The broader $R(\tau)$ is, the longer the memory of the system. It can be shown that, for a stationary random process, the PSD is the Fourier transform of the autocovariance function of the process, i.e.  

\begin{equation}
{\rm PSD}(\omega)=\int_{-\infty}^{\infty} R(\tau)e^{-i\omega \tau} d\tau.
\end{equation}
\smallskip

\noindent The equation above implies that the PSD holds exactly the same information as the auto-covariance does. We can express $R(\tau)$ as the inverse Fourier transform of $PSD(\omega)$. Since $R(\tau=0)=\sigma^2$, we can then write 

\begin{equation}
    \sigma^2=\int PSD(\omega)d\omega.
\end{equation}
\smallskip

This equation highlights another useful property of the power spectral density function. The product $PSD(\omega)d\omega$ is the contribution to the total variance of the random process of sinusoids in $X(t)$ with frequencies between $\omega$ and $\omega + d\omega$. Furthermore, it turns out that PSD is a more useful tool to study a variable process due to the statistical properties of the estimators of these functions. Estimators of $R(\tau)$ use the observed points in a light curve and turn out to be highly correlated, exactly due to the intrinsic memory of the process, which one wishes to study. This issue was discussed in Sect.\,\ref{sec:optvar_methods} as well, when describing the properties of the SF. Consequently, they can be heavily biased towards the intrinsic value of $R(\tau)$, over a wide range of time-lags. Furthermore, their statistical properties, such as their variance (i.e. their error), are unknown, as they depend on the unknown statistical properties of the variable process under study. Therefore, the sampled autocovariance functions cannot be easily fitted by models. 

The most frequently used estimator of the PSD is called the ``periodogram", and its statistical properties are much superior. If computed over a certain range of frequencies, the periodogram estimates are asymptotically independent random variables with a known distribution. Their error depends on the intrinsic PSD (which is unknown), but this deficit can be overcome by considering either the logarithm of the periodogram \cite[in which case the log-periodogram error is know, i.e.][]{papadakis93}, or by ``smoothing" the periodogram estimates. The periodogram is an asymptotically unbiased estimate of the intrinsic PSD, although in the case of ``red-noise" power spectra the lowest frequency estimates can be biased (this is the so-called ``red-leakage" bias, see \citealp{papadakis93}, and references therein).  Given the superior statistical properties of the PSD estimates and the fact that it can help us easily detect characteristic timescales (either signaling periodicity or otherwise), PSD analysis has been the dominant method of studying the variability in the observed X-ray light curves of AGN. 

\subsection{Results from the PSD analysis of the X-ray light curves}

\begin{figure}
    \centering
    \includegraphics[width=0.47\linewidth,height=0.4\linewidth]{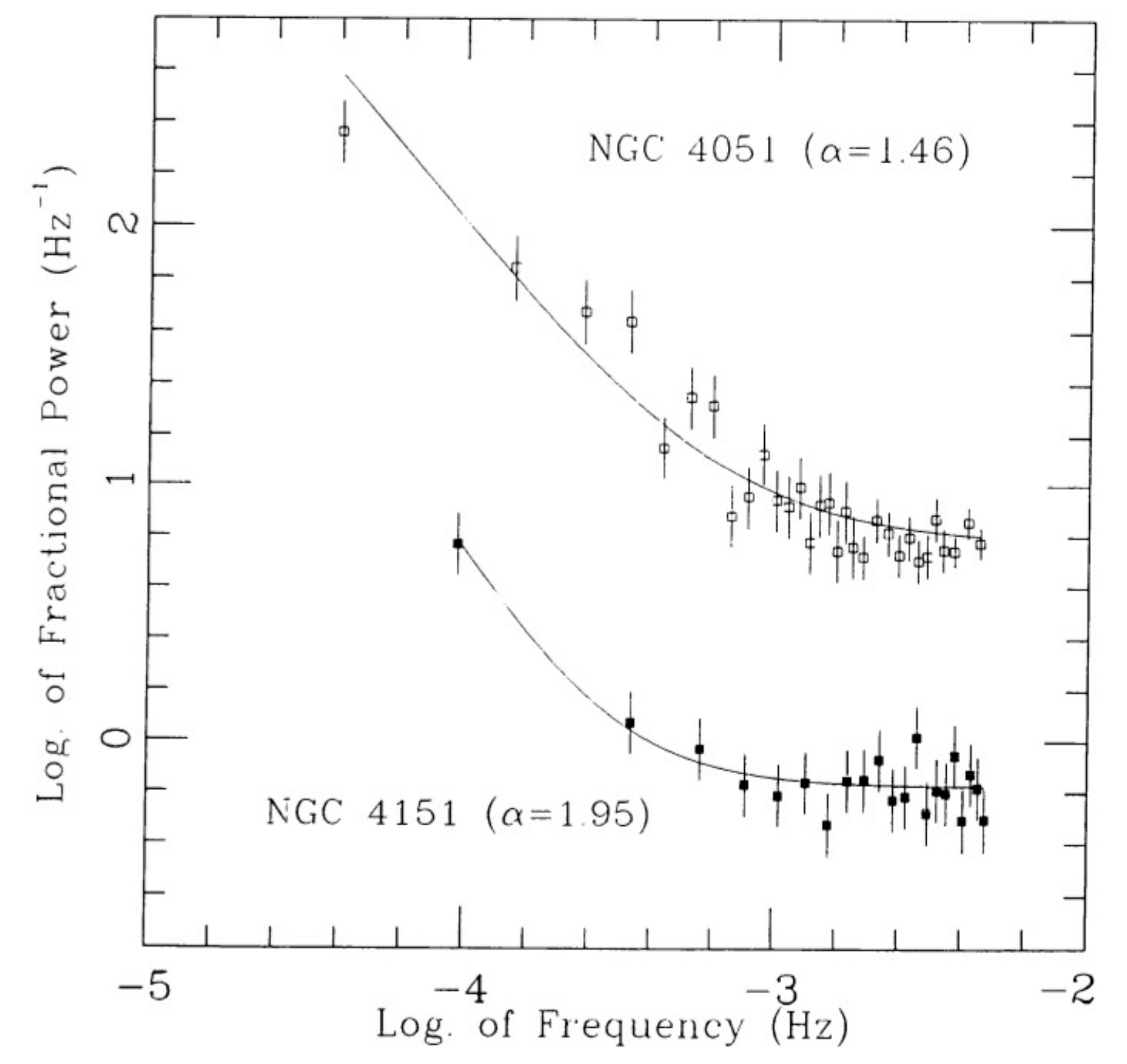}
    \includegraphics[width=0.52\linewidth,height=0.39\linewidth]{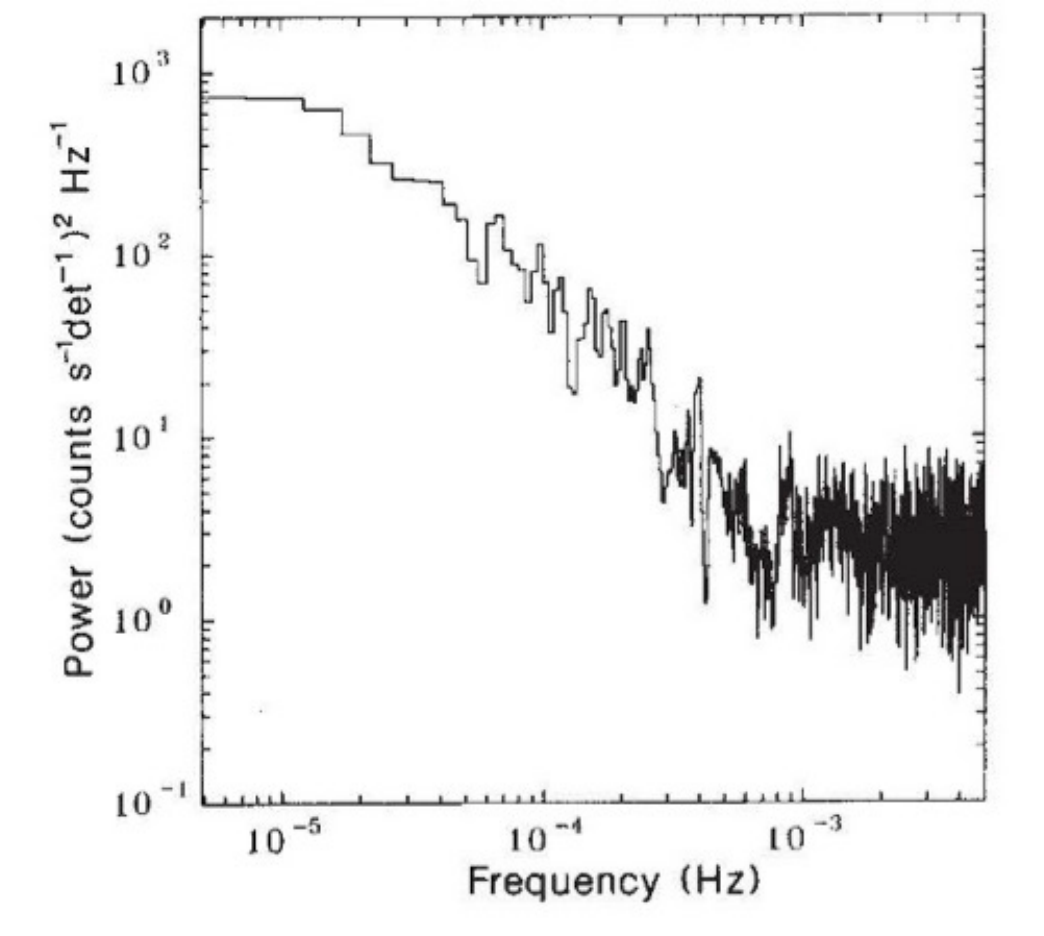}
    \caption{Examples of power spectra computed using {\it EXOSAT} ME long light curves. The \textit{left panel} shows the NGC\,4051 and NGC\,4151 PSDs (from Fig.\,1 of \citealt{lawrence93}); solid lines show the best-fit power-law model to the data, and $\alpha$ indicates the best-fit slope). The \textit{right panel} shows the NGC\,5506 power spectrum (from Fig.\,2 of \citealt{mchardy87}).}
    \label{fig:exosatpsds}
\end{figure}

The {\it EXOSAT} long observations yielded light curves that could be  used for a power spectral analysis. 
The resulting power spectra did not show any periodicities or other characteristic timescales. Instead, PSDs appeared to increase towards low frequencies with a featureless, power-law-like shape, indicating that X-ray variability in AGN is purely stochastic (see Fig.\,\ref{fig:exosatpsds}).

Soon after the first PSD results appeared, a systematic and detailed power spectral analysis of many {\it EXOSAT} light curves was carried out. The results were presented by \cite{lawrence93} and \cite{green93}. These studies confirmed that PSDs are well fitted by a simple power-law (PL) model over a broad frequency range (from $\sim 10^{-3}$ Hz to $\sim 10^{-5}$ Hz), with a slope of $\sim -1.5$. In addition, they also found a strong anticorrelation between X-ray variability amplitude and X-ray luminosity, with the more luminous objects being less variable. These results confirmed earlier findings reported by \cite{barr86}. 

Another interesting result from the power spectrum analysis of the {\it EXOSAT} light curves was that the ME PSDs were significantly flatter than the lower-energy PSDs \citep{papadakis95}. This is in contradiction with the expectations from inverse Compton scattering of low-energy photons in a single-temperature X-ray corona. Hard X-ray photons should undergo more scatterings in this case, hence high frequency variations should be washed out, and the PSD at high energies should be separately steeper, contrary to what is observed. 
Similar results were also reported for galactic X-ray black hole binaries (GBHBs; e.g. \citealt{Nowak99}). 

\subsubsection{The {\it RXTE/XMM-Newton} legacy} 
\label{sec:rxtexmmlegacy}

Power spectrum studies of AGN were revolutionised in the late 1990s, with the launch of {\it RXTE}. NASA's {\it RXTE}  was specifically designed for rapid slewing, allowing monitoring of AGN on a variety of timescales from hours to years. Several groups began monitoring AGN and the resulting light curves were ideal for high-quality, long-timescale PSDs. 

It was long proposed that the X-ray PSDs of AGN cannot continue with a steep slope to very low frequencies because the total variance would then be unphysically large.  There were indications that PSD ``breaks" to a flatter slope at frequencies below a characteristic ``break" or ``bending" frequency ($\nu_b$) in the PSD analysis of past observations of AGN \citep[e.g.][]{papadakis-mchardy-95}. However, the first clear and unambiguous detection of such a ``break timescale" was based on the PSD analysis of the {\it RXTE} light curve of NGC 3516 and was reported by \cite{Edelson1999}. Soon after, \cite{Uttley02} and \cite{Markowitz03} published the results of a detailed study of the power spectrum of a few Seyferts using {\it RXTE} light curves and reported further detections of such bending frequencies. Based on the results of these papers, a picture of a power spectrum with a slope of $\sim -1$ at low frequencies that steepens to $\sim -2$ above $\nu_b$ emerged in the 2--10 keV band. 

\begin{figure}
    \centering
    \raisebox{0.2cm}{\includegraphics[width=0.46\linewidth, height=0.38\linewidth]{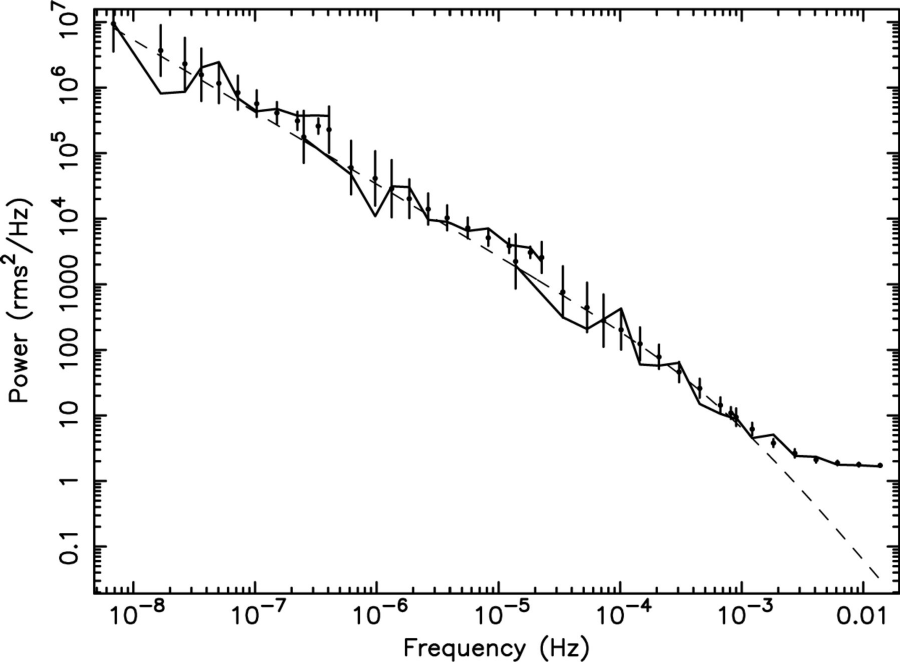}}
    \includegraphics[width=0.5\linewidth, height=0.4\linewidth]{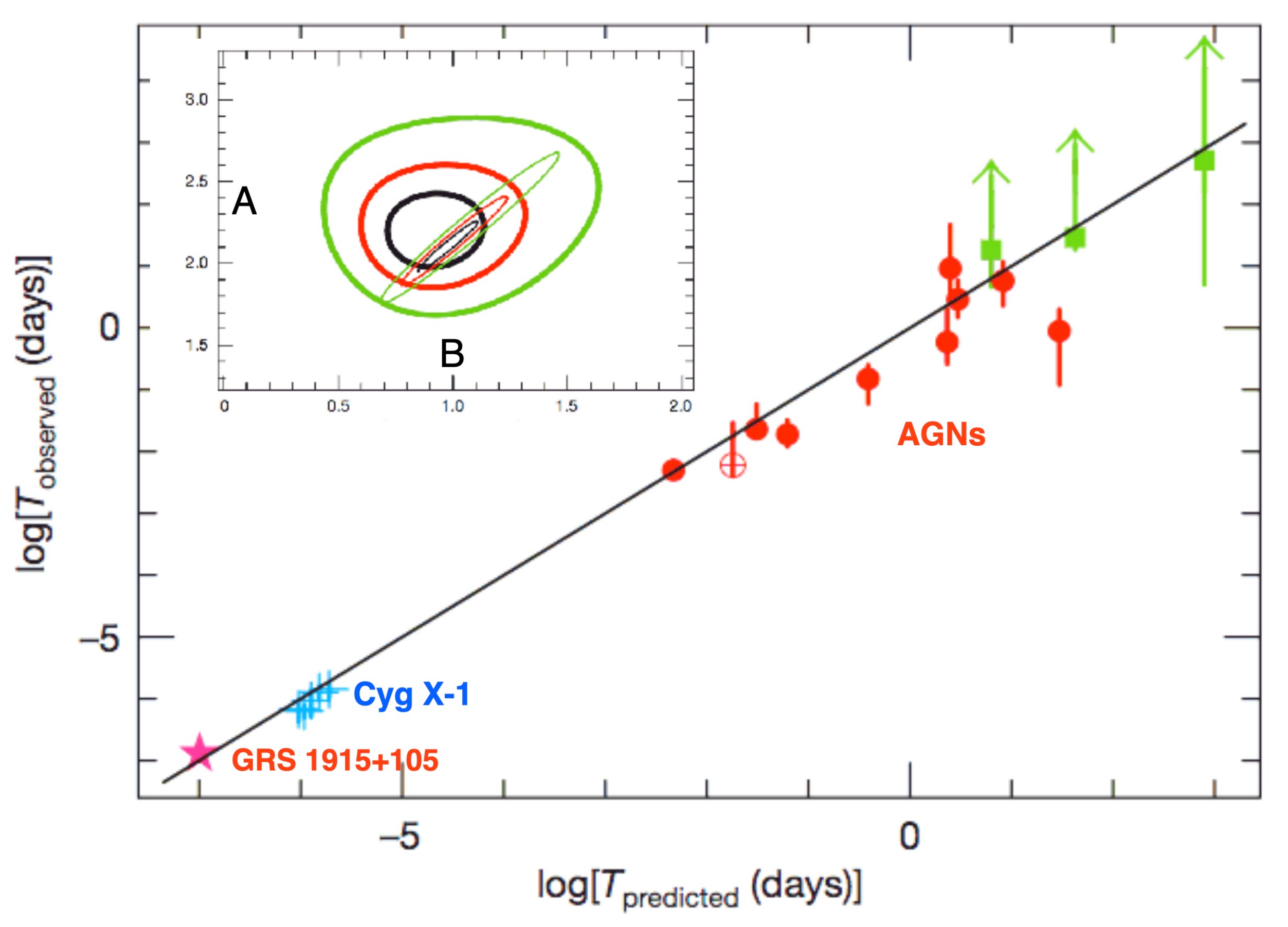}
    \caption{{\it Left panel}: Combined {\it RXTE} and {\it XMM–Newton} PSD of NGC\,4051 (from Fig.\,6 of \citealp{mchardy04}). This is arguably the most well defined PSD in AGN, covering $\sim 6.5$ decades of frequency. The thin dashed line is the underlying best-fit bending PL model. The PSD below $\nu_b$ continues to low frequencies with $\alpha_L \approx -1$ over a very broad range of frequencies. {\it Right panel:} The observed PSD break timescale, T$_{\rm B, observed}$, plotted as a function of the predicted time-scale, T$_{\rm B, predicted}$, using the best-fit relation of \cite{McHardy06} for AGN and selected GBHBs. The inset shows the confidence contours for the mass and bolometric luminosity indices A and B, with and without including the Galactic black hole system GRS 1915+105 (thin and thick lines respectively, adapted from Fig.\,1 and 2 of \citealp{McHardy06}). }
    \label{fig:4051psd}
\end{figure}

The next significant improvement in the determination of the AGN X-ray PSDs was made possible by the combination of data from {\it RXTE} and {\it XMM-Newton} in order to calculate the power spectrum over a very wide frequency range. Arguably, the best example of such a well-defined PSD 
is the power spectrum of NGC\,4051 presented by \citet[see the left panel in Fig.\,\ref{fig:4051psd}]{mchardy04}. This power spectrum is well fitted by a ``bending power law (BPL) model'' of the form: 

\begin{equation}
    {\rm PSD}(\nu)= A\nu^{-a_L}[ 1+(\nu/\nu_b)^{a_H-a_L}]^{-1},
\label{eq:psdbendmodel}
\end{equation}

\noindent where $a_L$ and $a_H$ are the PSD slopes below and above $\nu_b$, respectively, while $A$ is equal to $2$\,PSD($\nu_b)\times (\nu_b)^{a_L}$. In the case where $a_L=1$, then PSD$(\nu_\textrm{b})\times \nu_\textrm{b}=A/2$. At frequencies much lower than $\nu_\textrm{b}$, then PSD$(\nu)\times \nu=A$ and, therefore, $A$ is representative of the PSD amplitude (when $a_L=1$).

In NGC\,4051 the PSD extends with a slope of $a_L\sim 1$ over more than 4 orders of magnitude in frequency below $\nu_b$. Similar quality PSDs were soon reported for MCG -6-30-15 \citep{McHardy05}, as well as for NGC\,3227, NGC\,5506 and NGC\,3783 \citep{Uttley2005,Summons07}. These results suggest that AGN are like ``soft-state" GBHBs. When the Galactic black hole binaries are in this state, there is only one bending frequency in their PSD, and the power spectrum extends with a slope of $-1$ over a wide range of frequencies below $\nu_b$. 

\subsubsection{The scaling of the PSD break-frequency on BH mass and accretion rate.}\label{sec:X-ray_bend_dep}

By the mid 2000s, PSD analysis of the $RXTE$ and {\it XMM-Newton} light curves had resulted in the detection of the PSD bending frequency in more than 10 AGN. Using these data, \cite{McHardy06} were the first to demonstrate that the respective characteristic timescale, $T_b=1/\nu_b$, depends both on the mass of the BH and the bolometric luminosity, L$_{\rm bol}$, according to the relation: log$(T_b)$ = A log(M$_{\rm BH}$)-B log(L$_{\rm bol}$) + C, where ${\rm A}\approx 2$ and $\rm{B}\approx 1$. If $\lambda_{Edd}=$L$_{\rm bol}/$L$_{\rm Edd}$ (where L$_{\rm Edd}$ is the Eddington luminosity), then these results suggest that $T_b\propto$ M$_{\rm BH}/\lambda_{\rm Edd}$\footnote{$\lambda_{\rm Edd}$ is an estimate of \medd\ (i.e. the accretion rate of an AGN normalised to the Eddington accretion limit) Although they are not identical (for example, $\lambda_{Edd}$ will be smaller than \medd\ if the inclination angle of the accretion disc is large), in this work we use these terms interchangeably.}.

The dependence of T$_b$ on \medd\ was later questioned by \cite{GMV11}, who presented the results of a uniform analysis of the X-ray power spectra of 104 nearby ($z < 0.4$) AGN, using more than 200 {\it XMM-Newton} observations. Their work is the most comprehensive X-ray PSD analysis of AGN variability in the 0.3-10 keV band at frequencies higher than $\sim 10^{-5}$ Hz that has been performed to date. 
However, although the {\it XMM-Newton} light curves may be adequate for accurate PSD analysis for sources with a small BH mass, they may be less so for larger BH mass AGN.  For example, the \cite{GMV11} best-fit log($\nu_b$) estimates for MCG-6-10-15, NGC\,3227, NGC\,5506 and PKS0558-504 in the 2--10 keV band are $\sim -3.6, -3.6, -3.8$ and $-4.8$, respectively. These values are systematically higher than the respective estimates from the best model fits to the combined {\it RXTE} and {\it XMM-Newton} PSDs, that is $\sim -4.5, -4.6, -4.4$ and $-5.2$, determined by \cite{McHardy05}, \cite{Uttley&McHardy2005} and \cite{papadakis10}, respectively. A combined analysis of all available {\it RXTE} and {\it XMM-newton} light curves, and more accurate estimates of \mbh\ and of L$_{\rm bol}$ are necessary conditions to better determine how $\nu_{b}$ depends on \mbh\ and \medd. 

There is plenty of observational evidence on the increase in the characteristic frequencies in the GBHB X-ray power spectra with increasing X-ray flux, that is, with an increasing accretion rate in this case (e.g. \citealt{Axelsson05} for Cyg X-1, \citealt{reig13} for GRO J1655-40, \citealt{alabarta20} for MAXI J1727–203). Assuming that the X-ray variability properties of AGN and GBHBs are similar, this observational evidence supports the findings of \cite{McHardy06}. If true, this result puts interesting constraints on the physical timescale that could be responsible for $\nu_b$ in AGN. 

For example, as discussed in Sect.\,\ref{sec:optvar_model}, the dynamical and thermal timescales, which are among the basic characteristic timescales of a Keplerian, geometrically thin, and optically thick disc, do not depend on the accretion rate of the disc. Their value at each disc radius depends only on \mbh. On the other hand, the viscous timescale, $t_{visc}$, and the timescale of the propagation of sound waves in the radial direction, $t_{sound-R}$, depend on $H/R$ according to eqs.\,\ref{eq:tvisc} and \ref{eq:tsound}, which may depend on \medd\ (see eq.\,\ref{eq:hover-mdot}). Both $t_{visc}$ and $t_{sound-R}$ could be responsible for the break timescale that has been detected in the AGN X-ray PSDs, through the dependence of $H/R$ on \medd.  
If the X-ray corona is powered by the accretion process, 
the mechanism that powers the X-ray corona (and is responsible for the observed variations) could be modulated by $t_{visc}$ or $t_{sound-r}$. However, there are other possibilities as well. 

It is possible that the corona is highly magnetised and is powered by magnetic processes. For example, according to \citet[][see also \citealt{galeev79}]{haardt94} the energy that powers the X-ray corona may be stored in magnetic field structures that produce active X-ray regions on top of the disc. The energy is stored in the magnetic field on a charge timescale, $t_c$, and is released on a shorter discharge timescale, $t_d$. In this scenario, the main characteristic timescales of the X-ray corona are $t_c$ and $t_d$. The charge timescale is given by:
\begin{equation}
    t_c=530 (\frac{H}{R})^{-1} (R/R_S)^{3/2} M_{\rm BH,8} (s).
    \label{eq:tcharge}
\end{equation}

\begin{figure}
    \centering
    \includegraphics[width=0.48\linewidth, height=0.4\linewidth]{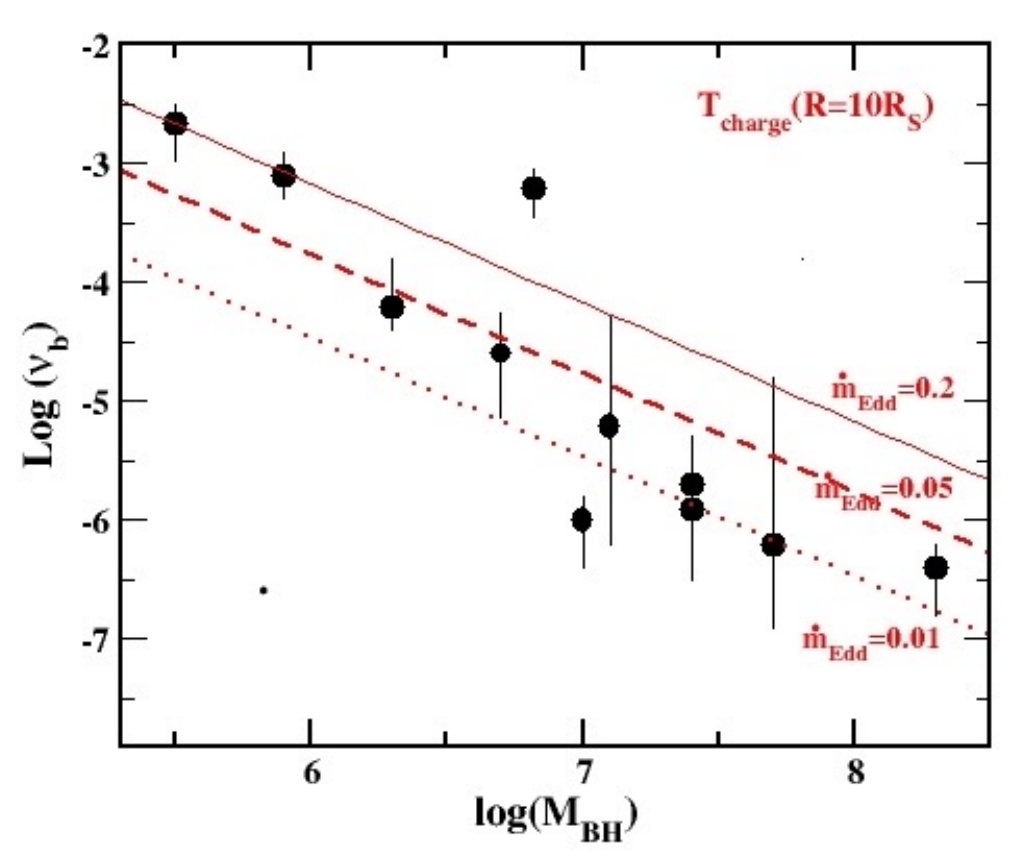}
        \includegraphics[width=0.5\linewidth, height=0.4\linewidth]{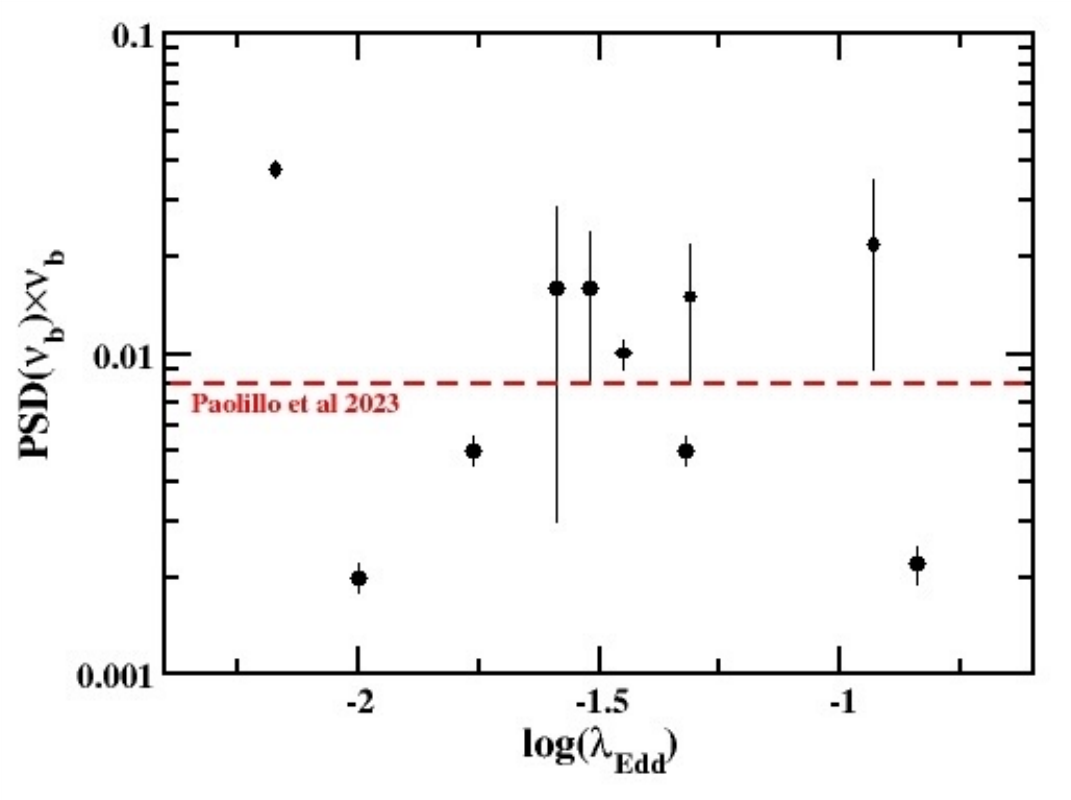}
    \caption{{\it Left panel}: The $\nu_b$ versus \mbh\ relation in AGN, using $\nu_b$ estimates mainly from joined PSD analysis of {\it RXTE} and {\it XMM-Newton} light curves. Solid, dashed and dotted lines indicate the charge time-scale, $t_c$, of the X-ray corona (eq.\ref{eq:tcharge}) for various accretion rates. {\it Right panel:} Power spectrum amplitude of Seyferts (i.e. PSD$(\nu_b)\times \nu_b)$ plotted as a function of $\lambda_{Edd}$. The horizontal dashed line is the amplitude derived by \citet{Paolillo23} from a sample of local and high-$z$ AGN. }
    \label{fig:psdres}
\end{figure}

\noindent The left panel of Fig.\,\ref{fig:psdres} shows a plot of log($\nu_b)$ versus BH mass, for a small sample of AGN. We used data from \cite{GMV11} for NGC\,4395 and \cite{markowitz07} for Mrk\,766, who studied the power spectra using only {\it XMM-Newton} light curves. For all the other objects, we used results from power spectral analysis of combined {\it RXTE} and {\it XMM-Newton} PSDs (\citealt{mchardy04} for NGC\,4051, \citealt{McHardy05} for MCG -6-30-15, \citealt{Uttley&McHardy2005} for NGC\,3227, \citealt{Markowitz10} for NGC\,7469, \citealt{Summons07} for NGC\,3783, and \citealt{Markowitz03} for NGC\,4151, NGC\,3516, NGC\,5548 and Fairall 9). For the BH mass estimates, we used the measurements listed in the ``AGN Black Hole Mass Database'', which have been calculated by spectroscopic reverberation mapping studies \cite[assuming a mean $f$ factor of 4.3,][]{bentz15}. 
Solid, dashed, and dotted lines show the theoretical $\nu_b-$\mbh\ relation assuming that $T_b=t_c(R=10R_S)$, in the case where $\lambda_{Edd}=0.2,0.05$ and 0.01, respectively. Clearly, the hypothesis that 
$\nu_b=1/t_c$ can explain most of the scatter in the observed relation between $\nu_b$ and \mbh, as long as the accretion rate decreases from the smaller to the larger BH mass objects, which is what is observed in low redshift samples of AGN, such as the BASS sample \citep[e.g.][]{koss22}. 

The theoretical lines in the case where $T_B=t_{sound-R}$ and $T_B=t_{visc}$ ($\alpha=0.1$)  have the same slopes but are $\sim 4$ and 40 times lower in normalisation than the lines plotted in the left panel of Fig.\, \ref{fig:psdres}, for the same accretion rates. The $T_B=t_{sound-R}(R=10R_S)$ solution can agree with the observed $\nu_b-$\mbh\ relation but for \medd\ four times larger than those shown in this panel.

\subsubsection{The PSD amplitude.}
Most of the work so far has focused on the study of the power spectrum break/bending frequency and on its dependence on \mbh\ and \medd. Relatively less attention has been paid to the other characteristics of the PSDs, such as the amplitude and the low- and high-frequency slopes of the PSD. 

The right panel in Fig.\,\ref{fig:psdres} shows the product $\nu_b \times$PSD($\nu_b)$ for Seyeferts where this product can be calculated using the best-fit results from the BPL model fits to the 2-10 keV PSDs. This quantity is representative of the PSD amplitude (as we explained in \S\ref{sec:rxtexmmlegacy}). 
We have used the results of \cite{mchardy04,McHardy05,Uttley&McHardy2005} for NGC\,4051, MCG-6-30-15, and NGC\,3227, respectively,  and \cite{Markowitz03} for Fairall 9, NGC\,5548 and NGC\,3516. These authors used {\it RXTE} and {\it XMM-Newton} light curves to compute the power spectrum over a wide range of frequencies. For the rest of the sources (i.e. Mrk\,335, Mrk\,766, NGC\,4395 and NGC\,6860) we used the \cite{GMV11} results. PSD amplitudes are plotted as a function of log($\lambda_{Edd})$ (we used the \citealt{gupta24} measurements of $L_{bol}$ to calculate $\lambda_{\rm Edd}$).

There is significant scatter in this plot. Some of it could be due to the fact that if we only use XMM light curves to compute the PSD, then the PSD amplitudes may be smaller than in reality (since the best-fit fit $\nu_b$ may be higher, as we have already discussed). However, some of the apparent scatter in this plot could also be real, indicating that the PSD amplitude may not be exactly the same in all AGN. Instead, there could be a distribution of PSD amplitudes. However, even if this is the case, the right plot in Fig.\,\ref{fig:psdres} shows that the mean PSD amplitude probably does not depend 
on the accretion rate. Furthermore, one can also show that there is no apparent dependence on BH mass either. 

The horizontal dashed line indicates the mean amplitude of the X-ray PSD that \cite{Paolillo23} found, using the so-called "variance frequency plot" (VFP, see Sect.\,\ref{sec:X-ray_excess_var}) to estimate the PSD of a large sample of AGN over a wide range of frequencies. The PSD amplitude of the Seyferts plotted in Fig.\,\ref{fig:psdres} are also broadly consistent with the mean PSD amplitude of $\sim 0.017$ found by \cite{Papadakis04} based on the analysis of the normalised excess variance of AGN.

\cite{Ponti12} suggested that the amplitude of PSD may decrease with increasing accretion rate as PSD$_{amp}\propto \lambda_{Edd}^{-0.8}$. Their results were based on the analysis of the normalised excess variance of many AGN, using XMM-Newton observations. This steep dependence does not appear to be supported by the data plotted in the right panel of Fig.\,\ref{fig:psdres}. Other works also report a dependence of the PSD amplitude on the accretion rate. For example, \citet{Paoillo17} suggest that some dependence of the PSD amplitude on $\lambda_{Edd}$ is necessary to explain the normalised excess variance of high-redshift AGN (see \S\ref{sec:X-ray_excess_var}). \cite{Papadakis24} also found evidence for a dependence of the PSD amplitude on $\lambda_{Edd}$, with PSD$_{amp}\propto \lambda_{Edd}^{-0.3}$. Their results were based on the PSD analysis of many Swift/BAT light curves in the 14-195 keV band, from the 157-month Swift/BAT hard X-ray survey, and are suggestive. But even if real, the dependence of the PSD amplitude on $\lambda_{\rm Edd}$ that they found is less steep than the \cite{Ponti12} results. 

The number of direct measurements of the PSD amplitude so far has been rather small, and we cannot draw any firm conclusions. The data so far indicate a distribution of PSD amplitudes with the same mean for all AGN, but further PSD analysis of combined {\it RXTE} and {\it XMM} light curves is necessary to reach conclusive results.

\subsubsection{The low frequency PSD slope.}

\begin{figure}
    \centering
    \includegraphics[width=0.46\linewidth, height=0.4\linewidth]{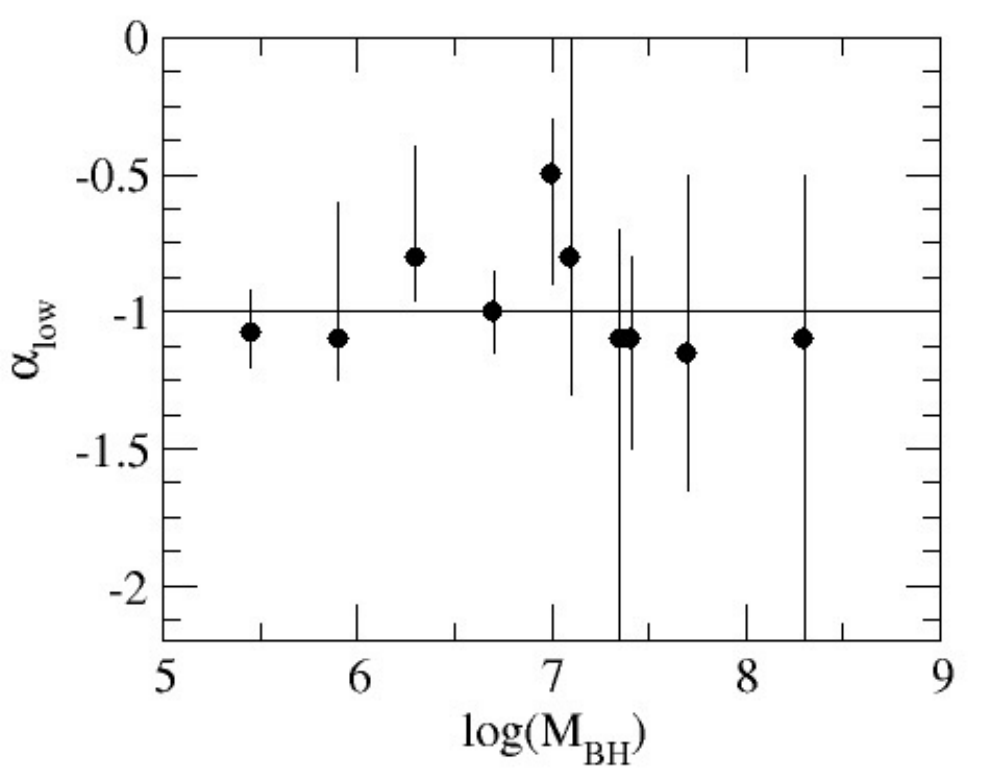}
    \includegraphics[width=0.5\linewidth, height=0.4\linewidth]{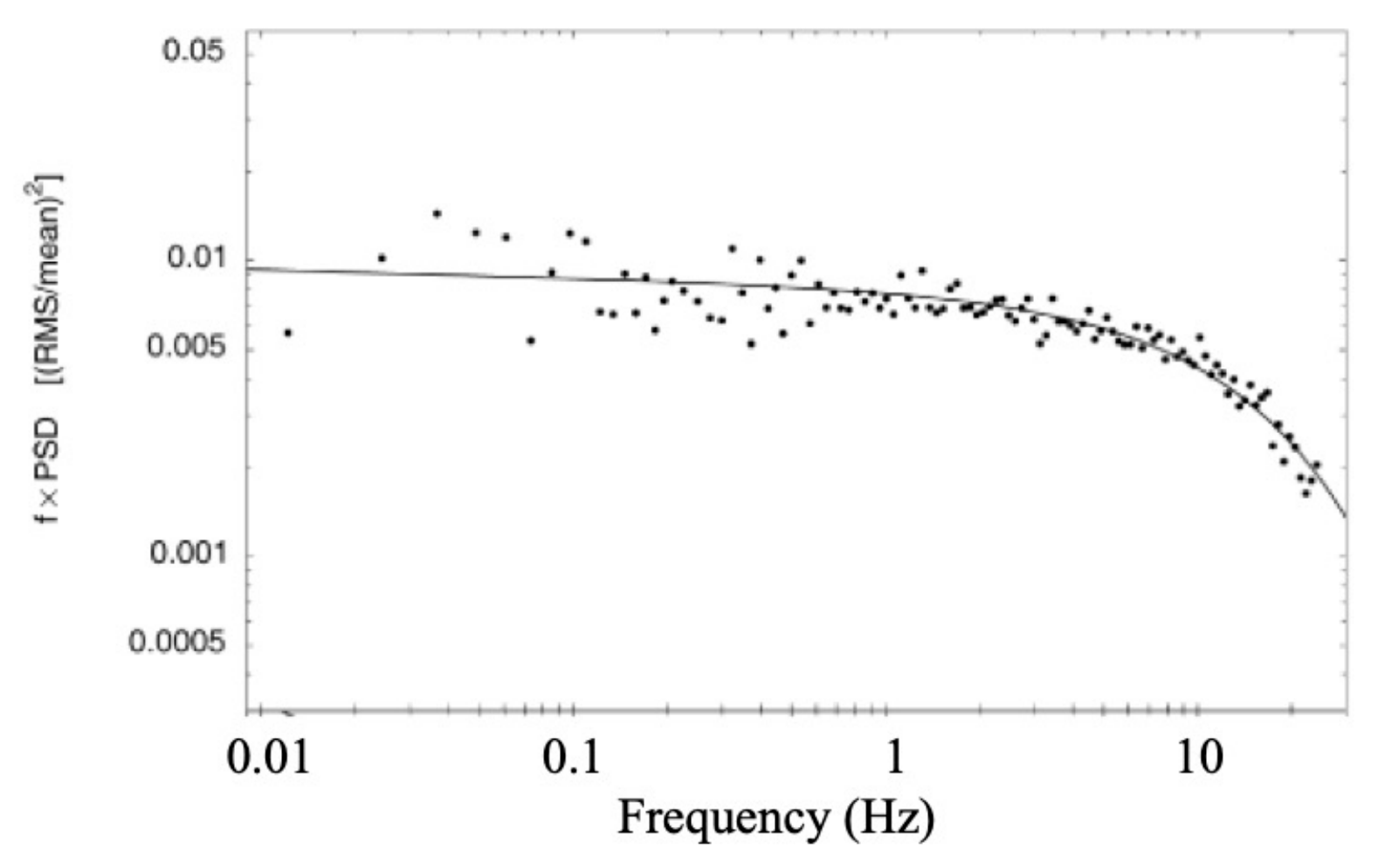}
    \caption{{\it Left panel}: The low frequency PSD slope for Seyferts where power spectral analysis has been performed using long term {\it RXTE} lightcurves. {\it Right Panel:} Filled squares show the Cyg X-1 PSD in the 2-12 keV band, plotted in PSD$(\nu)\times \nu$, when the source in its soft state and its PSD follows a cut-off PL model. 
    The fact that the PSD in this representation is flat below $\nu_b$ indicates that the low-frequency slope is $\sim -1$ over a wide  frequency range, very similar to the NGC\,4051 PSD plotted in the left panel of Fig.\,\ref{fig:4051psd}. It is also interesting that the Cyg X-1 PSD amplitude is $\sim 0.01$, entirely consistent with the mean PSD amplitude of AGN shown in Fig.\,\ref{fig:psdres}. Right panel graph from Fig.\,6 of \cite{Axelsson05}.}
    \label{fig:psdres2}
\end{figure}

The PSD slope at frequencies below $\nu_b$ is generally believed to be very close to $-1$. This is a rather well established fact, based on the results of the bending PL model fits to PSDs computed using long-term {\it RXTE} light curves. The left panel of Fig.\,\ref{fig:psdres2} shows the best fit $a_L$ for NGC\,4051 \citep{mchardy04}, MCG-6-30-15 \citep{McHardy05},  Fairall 9, NGC5548, NGC3516 and NGC4151 \citep{Markowitz03}, as well as NGC3227 \citep{Uttley&McHardy2005}, NGC3783 \citep{Summons07} and NGC7469 \citep{Markowitz10}. The solid horizontal line indicates the value of $-1$. Clearly, all the low-frequency PSD slope estimates are consistent with this value (within the error). The PSDs of NGC4051 and MCG-6-30-15  (\citealp{mchardy04} and \citealp{McHardy05}, respectively) are particularly important in this respect, as they show that the slope of -1 extends over $\sim 5$ and $3.5$ orders of magnitude below $\nu_b$. This is typical of the XRB PSDs when they are in their so-called ``soft state". 

The fact that the low-frequency PSD slope is $\sim -1$ in AGN is further supported by other studies of larger AGN samples. For example, \cite{Paolillo23} studied a large number of AGN, with a wide range of BH masses and accretion rates, on a wide range of timescales. They used excess variance measurements, and they showed that the ``variance-frequency" plot of AGN is fully consistent with a PSD slope of -1 below the PSD break frequency.

\subsection{Comparison between AGN  and Galactic black hole binaries}

\cite{Uttley02} were among the first to notice that the Galactic black hole X-ray binaries (GBHBs) and the AGN PSDs are very similar in both shape and amplitude (see Section 6.4 and Figure 2 in their paper). They also noticed that the characteristic break/bending frequency scales with the BH mass. Similar comments were also made by \cite{Markowitz03}, \cite{McHardy05}, and \cite{Uttley&McHardy2005}. \cite{McHardy06} further showed that the ``$\nu_b$ - BH mass" relation is the same for both AGN and GBHBs (see Fig.\,\ref{fig:4051psd}). 

The comparison between AGN and GBHB PSDs is complicated by the fact that the GBHB PSDs depend on the ``state" in which they operate \cite[e.g.][]{Belloni10, Remillard06}. The right panel in Fig.\,\ref{fig:psdres2} shows the power spectrum of Cyg X-1 in its ``canonical soft state", taken from \cite{Axelsson05}. As we discussed above, this PSD is very similar to the AGN PSDs (both in shape and in amplitude). 
However, as \cite{Axelsson05} notice, only 35\% of the observations made during the soft state of Cyg X-1 allowed a fit using a simple cutoff PL model (indicated by the solid line in this plot). 

Furthermore, there is a significant difference between the PSD of other GBHBs and Cyg X-1 in the soft state. For example, Fig.\,2c in \cite{Heil15} shows a number of GBHB power spectra in various states. The PSD of XTE\,J1550-564 in its soft state 
appears to have a single break frequency and a low-frequency PSD slope of $\sim -1$. However, its amplitude is significantly lower than the Cyg X-1 PSD amplitude shown in Fig\,\ref{fig:psdres2}. On the other hand, some GBHBs in their hard state also show PSDs with a cutoff PL-like shape with a low-frequency slope of $-1$, as shown, for example, in the bottom blocks of plots in the appendix of \cite{Heil15} (which appears only in the astro-ph version of this paper). The GX\,339-4 and the GS\,1354-64 PSDs shown in these graphs are very similar to the Cyg X-1 PSD shown in Fig.\,\ref{fig:psdres2} (although their amplitude is slightly larger than the amplitude of the Cyg X-1 PSD). It seems that only a small number of GBHB observations (including Cyg X-1) show PSDs that are similar to the AGN PSD. 

There are reports of AGN whose PSD is different from the Cyg X-1 PSD in its soft state. The most studied case is Ark,564, a highly variable X-ray bright narrow-line Seyfert 1. \cite{Papadakis02} were the first to show that the Ark\,564 X-ray PSD shows two breaks. It has a slope of $\sim -2$ at high frequencies, which flattens to a slope of $-1$ below $\nu_b$, and then flattens even more and becomes constant at even lower frequencies, below a second low frequency break. They suggested that this PSD shape is similar to the PSD of Cyg X-1 in its hard state. Later, \cite{McHardy07} confirmed that the Ark\,564 PSD is well described by a two-break power law, which is different from the power spectra of almost all other AGN observed so far. Based on the apparently high accretion rate of the source, \cite{McHardy07} suggested that this source operates in the so-called intermediate/very high state. A second AGN that may also operate in this state is IRAS 13224-3809, a highly variable bright X-ray Seyfert galaxy \citep{alston19}. \cite{Markuttley05} on the other hand found that the PSD of the low-luminosity AGN NGC\,4258 resembles the PSD of GBHB in their hard state, based on the fact that $\nu_b$ in this object appears to be much smaller than expected based on its BH mass and the \mbh-$\nu_b$ relation of AGN and GBHBs. 

However, these sources appear to be the exception among the AGN that have been studied so far. The results so far suggest that the X-ray power spectrum of most AGN is similar to the Cyg X-1 PSD (and possibly other GBHB PSDs as well) in a particular state during their evolution from one state to another,  which is rather rare. This appears strange, as one would expect that the AGN power spectrum would be more similar to the PSD of GBHBs in their most common state. Perhaps this is a selection effect: AGN in this state are both luminous and highly variable, and this is the reason why it is easier to study their PSD. Detailed PSD analysis of more sources is necessary in order to better understand the correspondence between the GBHB and AGN PSDs. This would be helpful in better understanding the physical properties that operate in AGN. 

\subsection{Quasi-periodic oscillations (QPOs) in AGN}

QPOs are commonly observed in the X-ray light curves of Galactic X-ray binaries, both in black hole and in neutron star systems. QPOs are detected using Fourier techniques. They appear in the power spectrum as narrow peaks. By definition, their width is (at least) less than half of their centroid frequency. In GBHBs, QPOs are generally grouped into low-frequency (LF) QPOs, with centroid frequency $\leq 30$ Hz, and high-frequency (HF) QPOs, with centroid frequency $\geq 60$ Hz \cite[e.g.][]{Belloni10}. LF QPOs in GBHBs have been classified into three types: A, B and C. In general, Type C QPOs are observed during the start of the outburst when GBHBs are in their hard state (but also appear in later stages), whereas Type A and Type B QPOs are only observed during the transition to the soft state \citep{Ingramqpo}. 

QPOs in AGN have been detected with the use of {\it XMM-newton} light curves. Since the maximum length of individual {\it XMM-Newton} observations is less than $\sim 120$ ksec, all QPOs that have been reported in AGN are at frequencies higher than $\sim 10^{-5}$ Hz. If QPO detections in AGN are real and if characteristic frequencies scale with the BH mass in AGN and GBHB then assuming a typical BH mass of $\sim 10^{6-7}$ M$_{\odot}$ for the AGN where QPO detections have been reported, these QPOs should be analogues of QPOs with frequencies higher than $\sim 10^{2-3}$ Hz in GBHBs, assuming a typical BH mass of $\sim 10$ M$_{\odot}$ in these systems. This simple reasoning suggests that AGN QPOs should be analogues to HF QPOs in GBHBs. 

HF QPOs are rare in GBHBs \citep[e.g.][]{Belloni12}. They appear only in high-flux observations, with a fairly specific hardness ratio \citep{Ingramqpo}. According to \cite{Remillard06}, the HF QPOs appear during the ``Very High" or ``Steep Power law" state of GBHBs, where the PSD amplitude is $\sim 10^{-3}$ (in the PSD$(\nu)\times \nu$ representation). This is significantly lower than the mean PSD amplitude of AGN. 
Therefore, it is not surprising that QPOs are very rare in AGN as well. For example, \cite{GMV11} find no strong evidence for QPOs in their analysis of archival {\it XMM-Newton} light curves, except for the well-known case of RE J1034+396. The QPO in this galaxy was first reported by \cite{Gierlinski08}, and has been reconfirmed many times since then \cite[e.g.][]{qpoalston,qpojin,qpozhang,qpoxia}. 

Detections of similar QPOs in other AGN \cite[e.g.][]{Alstonqpo,Panqpo,Zhangqpo-mrk766,Guptaqpo} are less convincing. The main issue with these detections is the transient nature of these QPOs. They have been detected in only one {\it XMM-Newton} observation in all cases, casting some doubt on the validity of the detections. For example, a period of $\sim 10^5$ s in AGN (this is the typical duration of an {\it XXM-Newton} observation) will correspond to a period of $\sim 0.1-1$ s in GBHBs (if we assume a BH mass of $\sim 10^{6-7}$ M$_{\odot}$ in AGN and of $\sim 10$ M$_{\odot}$ in GBHBs). If real, the QPOs in AGN should correspond to QPOs that would appear only for a period of time not more than $\sim 0.1-1$ s in GBHBs (and then they would disappear). Such QPOs have never been observed (although, even if they existed, we would not be able to detect them in the GBHB light curves). 

\subsection{The energy dependence of X-ray variability in AGN}

\begin{figure}
    \centering
    \includegraphics[width=0.42\linewidth, height=0.4\linewidth]{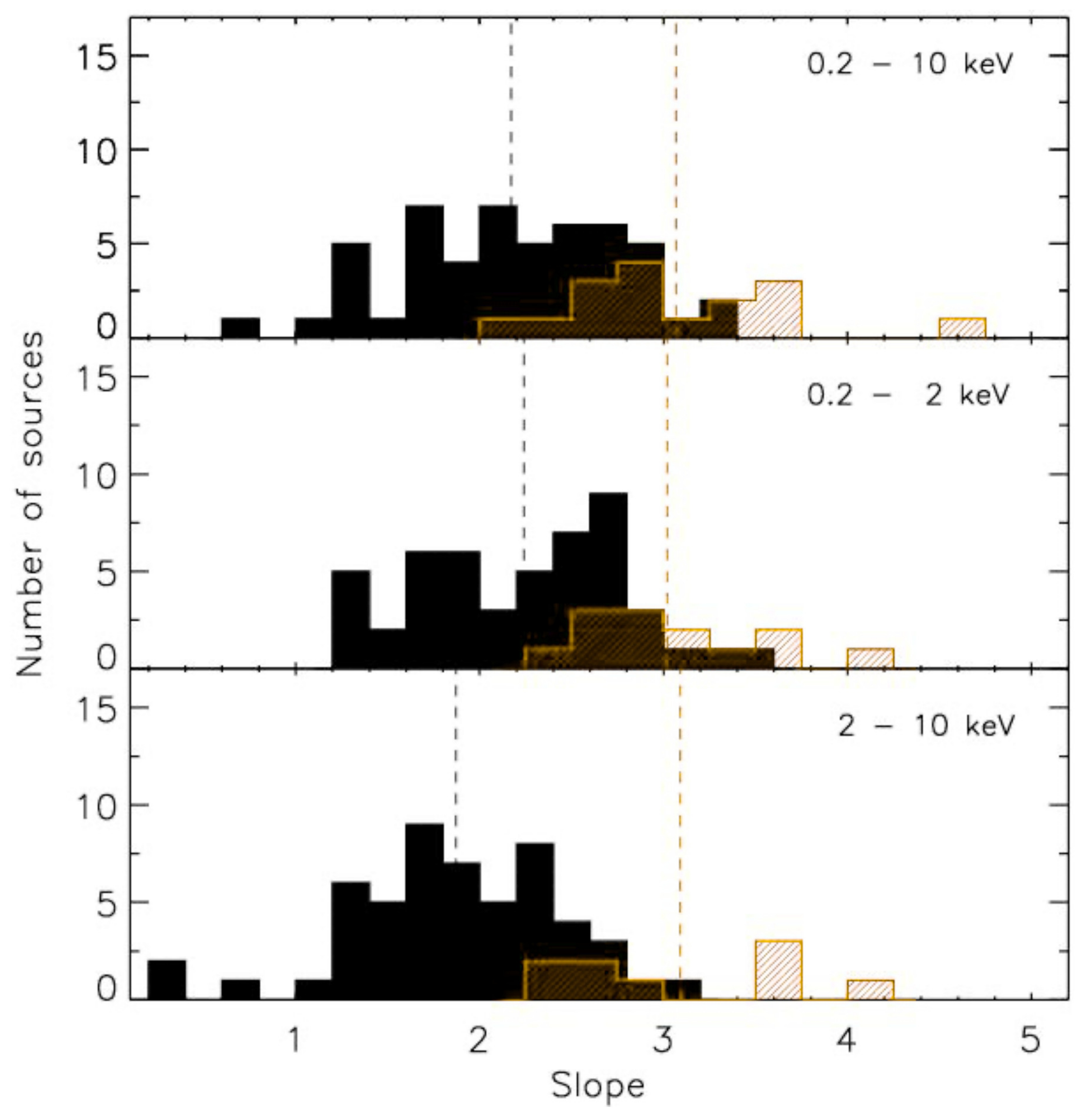}
    \includegraphics[width=0.54\linewidth, height=0.4\linewidth]{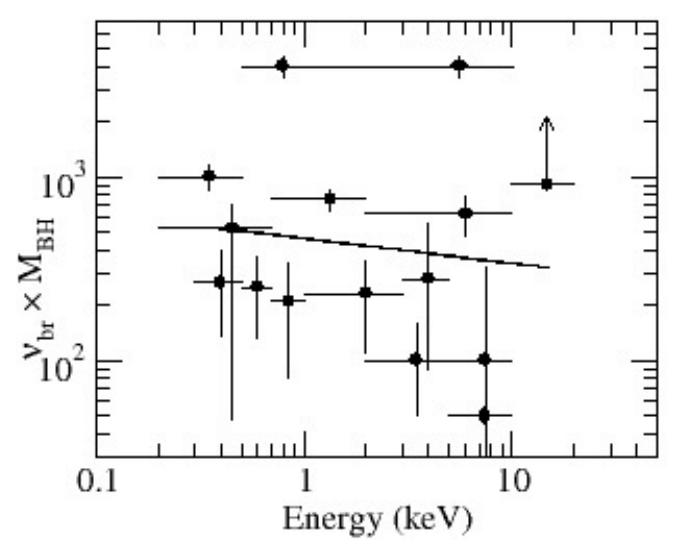}
    \caption{{\it Left panel}: Histogram of the PSD slopes in the case when the power spectrum is well fitted by a PL and a BPL model (black filled and red dashed histograms, respectively; from Fig.\,2 of \cite{GMV11}).  {\it Right panel}: Plot of the PSD bending frequency as a function of energy for a few AGN ($\nu_b$ is multiplied by \mbh\ in order to remove the dependence on BH mass).}
    \label{fig:psdres-energydep}
\end{figure}

\subsubsection{The high frequency PSD slope}

As discussed above, the low-frequency PSD slope is $\sim -1$ below the PSD bending frequency in the 2--10 keV band. This result also holds at higher energies. \cite{Papadakis24} found that the low-frequency slope of AGN PSDs in the 15-190 keV band is also close to $-1$. However, this is not the case for the PSD slope at frequencies higher than $\nu_b$. \cite{papadakis95} were the first to notice that the slope of the PSD becomes flatter (that is, the PSD ``hardens") with increasing energy, when studying the X-ray PSD of NGC 4051 below and above 2 keV, using {\it EXOSAT} light curves. This effect has been observed in many more AGN and is similar to what has also been observed in GBHBs \cite[e.g.][for Cyg X-1]{Nowak99}. 

The left panel in Fig.\,\ref{fig:psdres-energydep} shows the distribution of the best-fit PSD slopes from \cite{GMV11}. The black dashed vertical lines indicate the mean PSD slope when the power spectra are well fitted by a PL model. On average, the mean slope of the 2--10 keV band PSDs is flatter than the one in the 0.2-2 keV band (middle and bottom panels in the same Figure). Interestingly, this is not the case for the sources where the BPL model fits the power spectrum well. 
The high-frequency slope in this case is steeper than the slope in the case of the PL models, and it does not appear to depend on energy (see the red dashed vertical lines in the same figure). It is not clear whether this result depends on the fact that, as we have already commented, it may not be easy to accurately determine the PSD properties by using {\it XMM} light curves only, especially in narrow energy bands, when the Poisson noise level will increase. On the other hand, the analysis of high-$z$ samples by \citet{Paolillo23} suggests that the high-frequency PSD slope in the rest frame $2-10$ keV band is $-2.7$, in excellent agreement with \cite{GMV11}. A detailed study of the X-ray variability properties of NGC\,4051 using {\it NICER} light curves showed that the slope of the high-frequency PSD is $\sim -2.4$ at all energies from $\sim 0.3$ to 3 keV. This is an order of magnitude in frequencies, and the high signal-to-noise {\it NICER} light curves allow the determination of the PSD at very high frequencies. So, it is not clear yet whether the high-frequency PSD slope depends on energy or not.

Nevertheless, we can put strong constraints on the nature of the X-ray corona in AGN even if the PSD slope remains constant with energy.  As \cite{papadakis95} showed, if X-rays are produced by Inverse Compton scattering off hot electrons, then we should expect a strong steepening of the PSD with increasing energy in the case of a single corona with a uniform temperature (see their Figure 16). The loss of high-frequency variability is due to the fact that photons must be scattered more times to attain higher energies, and this process smooths out fast variations. Therefore, the observed dependence of the high frequency PSD slope on energy rules out the hypothesis of an X-ray corona with a single-electron temperature. The X-ray emitting region should consist of multiple active regions with different temperatures, or it may be a single medium but with a stratified temperature. 

\subsubsection{The energy dependence of the bending frequency} 
\label{sec:X-ray_enerdydep_nu}
The dependence of the bending frequency on energy has not been studied in detail so far, although PSD analysis has been performed in various energy bands for many AGN. The plot in the right panel of Fig.\,\ref{fig:psdres-energydep} shows $\nu_b$ as a function of energy, using the results of the energy-dependent PSD analysis of NGC4051 \citep{mchardy04}, NGC\,4395 \citep{vaughan05}, Mrk\,766 \citep{markowitz07}, IC4329A \citep{Markowitz09}, and NGC\,7469 \citep{Markowitz10}. 
The best-fit bending frequencies are multiplied by the BH mass to account for the BH mass dependence of this timescale (Sect.\,\ref{sec:X-ray_bend_dep}). There is considerable scatter in this graph, probably due to the fact that $\nu_b$ also depends on the accretion rate. The solid line indicates the best-fit straight line to the data. This line indicates the energy dependence of $\nu_b$, and suggests that $\nu_b$ may decrease with increasing energy, although this trend is not statistically significant. Even if this is not the case, 
the data plotted in Fig.\,\ref{fig:psdres-energydep} suggest that $\nu_b$ stays roughly constant over a wide range of energy of the order of 30 or so. \cite{Rani25} provided extra evidence for the the fact that $\nu_b$ does not depend on energy in the case of NGC\,4051, using {\it NICER} light curves. 

As noted in \cite{Rani25}, the (non)dependence of $\nu_b$ on energy can put constraints on various theoretical models. 
For example, according to the propagating fluctuations model (see Sect.\,\ref{sec:optvar_model}) X-rays are emitted from a flat corona, 
whose emission is modulated by variations in the accretion flow. 
The bending frequency in each energy band should correspond to the viscous timescale at the radius where most of the photons in this band are emitted. Because higher-energy photons are emitted closer to the central BH, we would expect the bending frequency to increase with increasing energy, which does not appear to be the case.

\subsubsection{The PSD Amplitude}

The results of the PSD analysis of many Seyferts using {\it RXTE} light curves suggest that, on average, PSD$(\nu)\times \nu$ is roughly flat and equal to $\sim 0.01-0.02$ at frequencies below $\nu_b$ \cite[see also][]{Paolillo23}. \cite{Papadakis24} found that PSD$(\nu)\times \nu$ is also flat and equal to $\sim 0.014$ at low frequencies in the 14-195 keV energy band. The PSD amplitude does not appear to significantly depend on energy (if at all). 

This result has interesting implications on the nature of the X-ray variability in AGN. Power spectra in AGN are (almost) always divided by the square of the mean of the light curve. As a result, the PSD does not have units of flux$^2$/Hz but units of Hz$^{-1}$. In this way, we can compare PSDs that are computed using light curves from different X-ray instruments. 
However, the PSD amplitude is affected by this normalisation. If there are spectral components that are constant in an energy band, then the PSD amplitude will be reduced. Furthermore, if the observed variations are not due to the X-ray continuum variations only (e.g. variable absorption could affect the observed variations at low energies), then the PSD amplitude will increase. The fact that the mean PSD amplitude of the X-ray bright AGN is very similar in the 2--10 and 14-195 keV bands indicates that the main driver of the observed variations over almost two decades in energy is the intrinsic X-ray continuum variations, whose amplitude does not depend on energy.

\subsection{The energy-dependent X-ray time-lags in AGN}
\label{sec:xraytimelags}

\begin{figure}
    \centering
    \raisebox{-0.1cm}{\includegraphics[width=0.5\linewidth, height=0.4\linewidth]{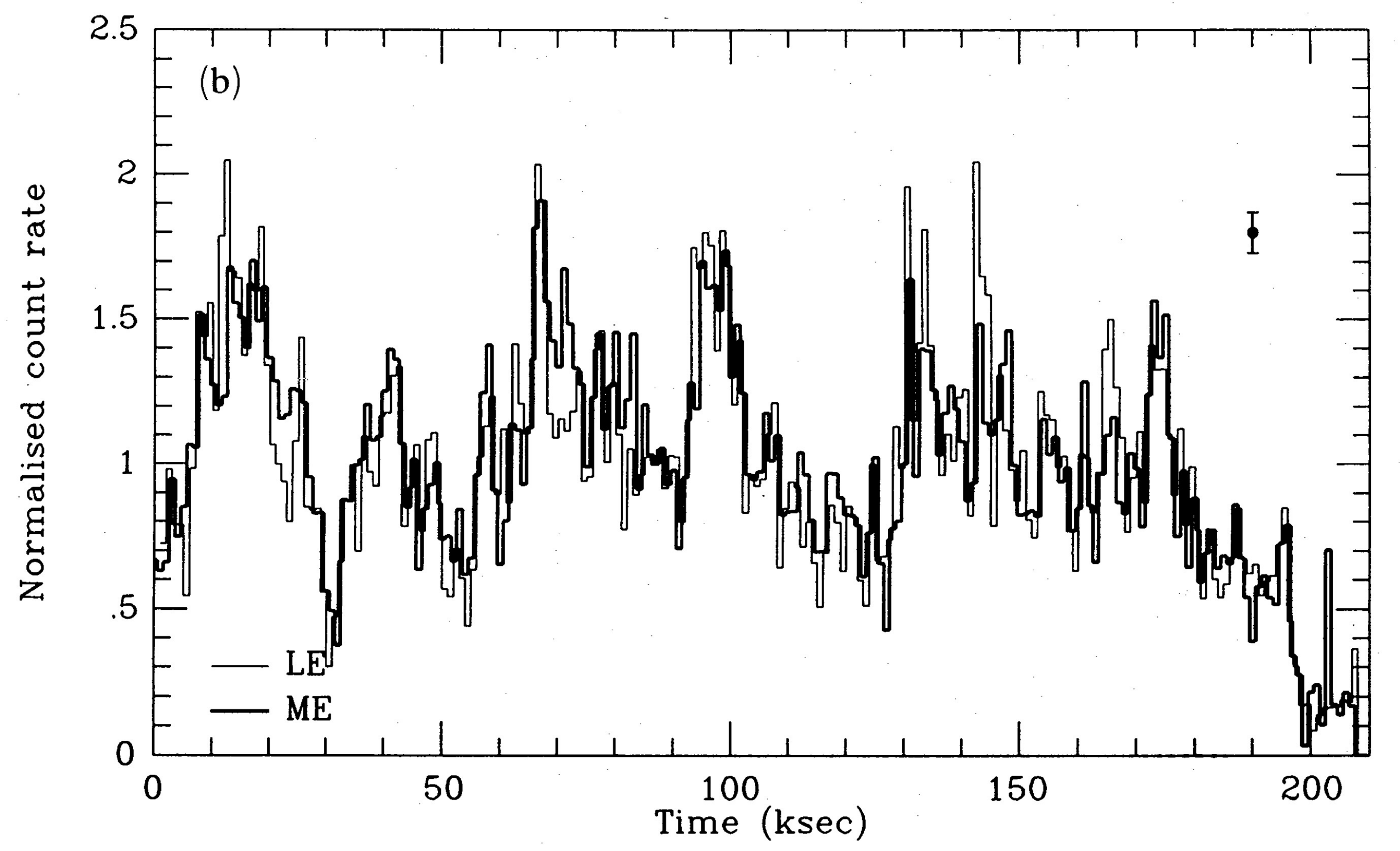}}
    \includegraphics[width=0.49\linewidth, height=0.39\linewidth]{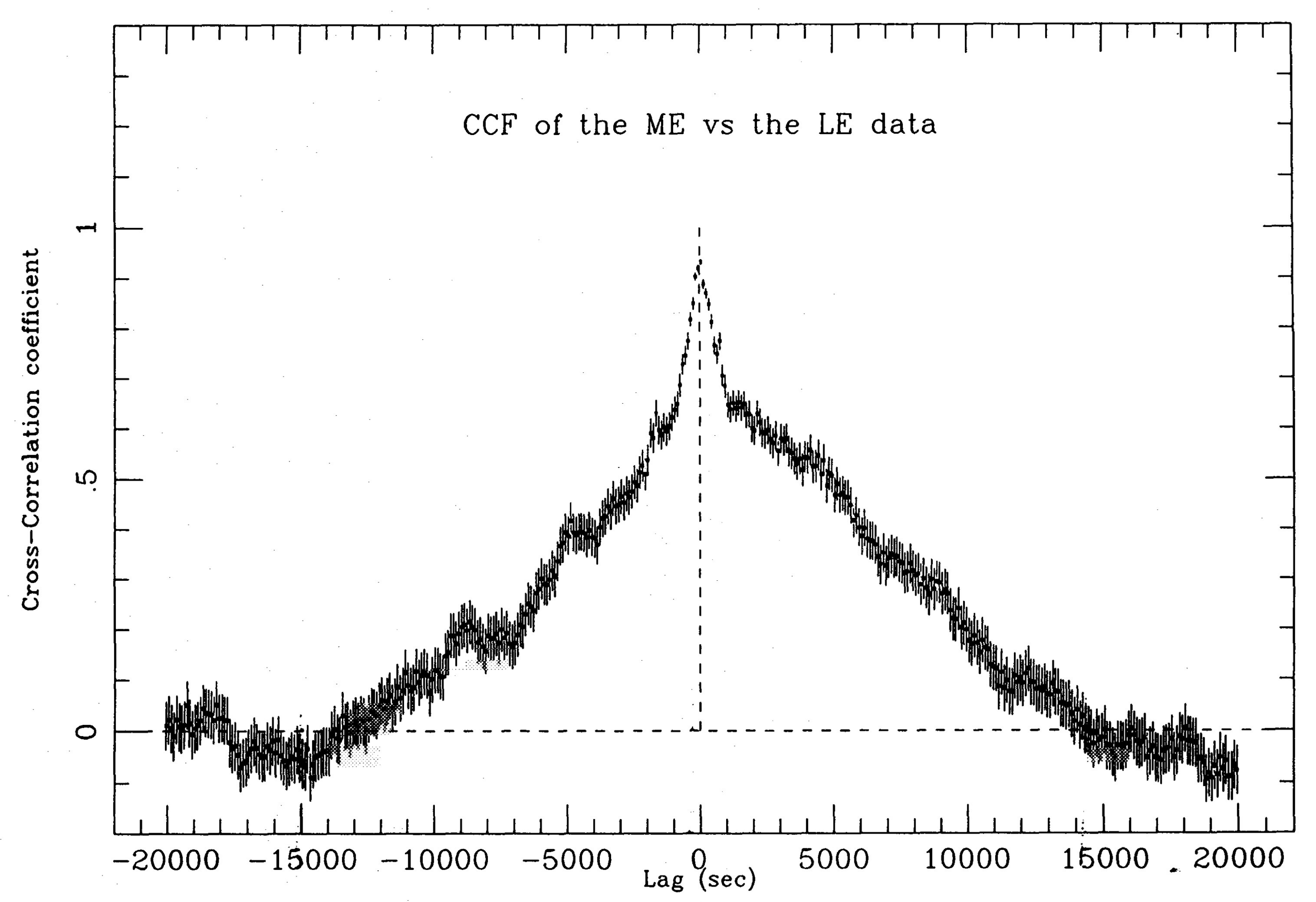}
    \caption{{\it Left and Right panels}: The {\it EXOSAT} ME and LE light curves of NGC\,4051 (in bins of 1000 s) and their cross-correlation, respectively (from Fig.\,1 and 14 of \cite{papadakis95}).}
    \label{fig:4051-corr}
\end{figure}

The X-ray variations in various energy bands are well correlated. The left panel in Fig.\,\ref{fig:4051-corr} shows the {\it EXOSAT} long-look light curves of NGC\,4051. The grey and black lines show the LE and ME light curves, respectively. Both bands show similar variations on all sampled scales. By eye, the variations happen at (almost) the same time in both bands. The right panel of the same figure shows a plot of the cross-correlation function between the LE and ME light curves. A strong peak appears at lag zero, indicating the good correlation between the two light curves. However, the cross-correlation function is clearly skewed toward positive lags, indicating that the ME variations are delayed with respect to the LE variations. This is expected in the case of thermal Comptonization: higher-energy photons are scattered more than the lower-energy photons, hence the delay between the variations in the high- and lower-energy bands. 

Studying delays in the time domain is not easy. The points in the sampled cross-correlation function are heavily correlated (because the light curve points are correlated themselves). Therefore,  cross-correlation functions cannot easily be used to fit theoretical models (see Sect.\,\ref{sec:x-rayvar_psd}). The average time delay between light curves in different energy bands can be determined by various Monte Carlo techniques (such as ICCF, for example; \citealp{peterson04}). However, the delays between X-ray variations in various energy bands are more complicated than just a constant delay that increases with energy separation. It turns out that if we decompose the light curves into sinusoids, we can show that there is a difference between the phase of these sinusoids at each frequency that increases with decreasing frequency. When divided by 2$\pi$, this phase shift gives us the time delay between each of these variability components in two energy bands.

\begin{figure}
    \centering
    \includegraphics[width=0.45\linewidth, height=0.4\linewidth]{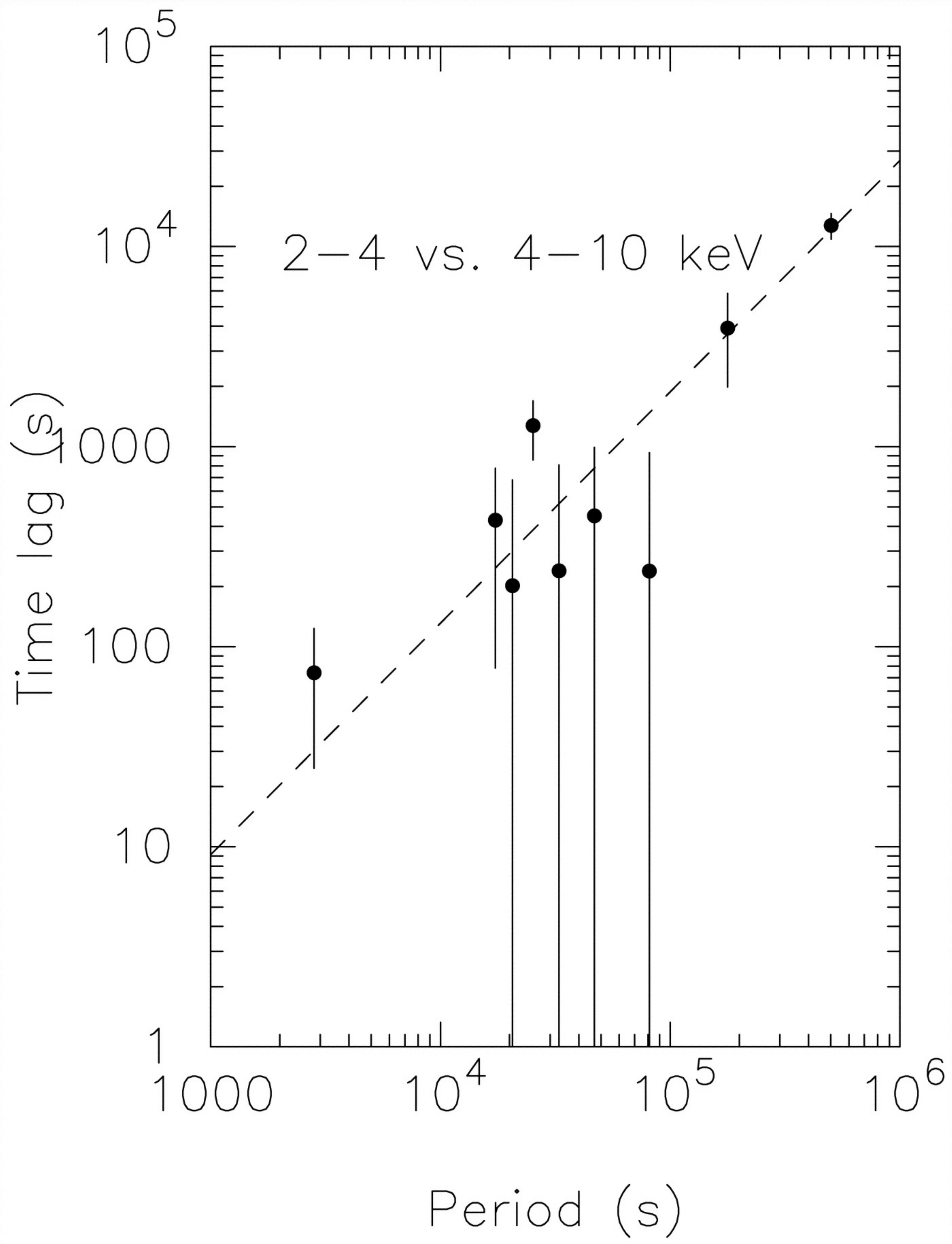}
    \includegraphics[width=0.49\linewidth, height=0.4\linewidth]{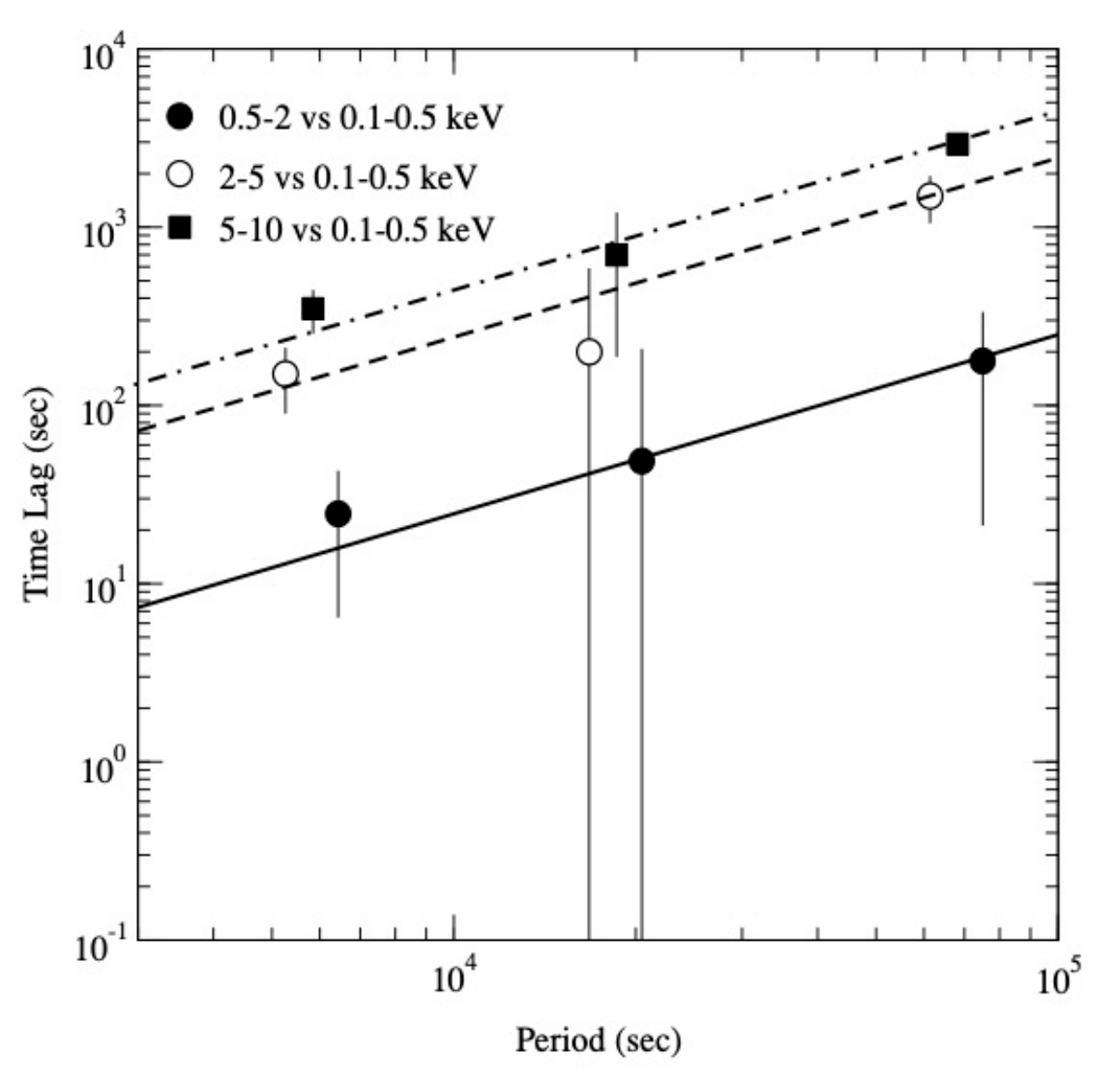}
    \caption{{\it Left panel}: Time lags vs. Fourier period for the 
2–4 keV vs. 4–10 keV band variations in NGC\,7469. The dashed lines show the best-fit power-law model (from Fig.\,1 of \cite{Papadakis01}). {\it Rigth panel:} Time lag versus Fourier period for the 0.1–0.5 keV versus the 0.5–2, 2–5, and 5–10 keV bands cross-spectrum for NGC\,4051.
The solid, dashed, and dot-dashed lines show the best-fitting PL model to the respective time lag plots (from Fig.\,16 of  \cite{mchardy04}). For both sources, the observed lags show that the higher energy band variations are delayed with respect to the variations in the lower energy bands}
    \label{fig:timelags}
\end{figure}

The first detection of these frequency-dependent time delays in AGN (the so-called ``hard" time lags) was presented by \cite{Papadakis01}. They studied {\it RXTE} light curves of NGC\,7469 and found that the delay between the variations detected in the 4--10 and 2--10 keV bands is not constant. Instead, it increases with increasing Fourier period (left panel of Fig.\,{\ref{fig:timelags}). The dashed line in this plot shows the best fit power-law model to the data (with slope of $-1$). The delays are on the order of $\sim 1-2$\% of the Fourier period. These results are similar to what has also been detected in Galactic black hole and neutron star binaries.  
Hard X-rays delayed with respect to softer X-rays, with a time delay approximately proportional to the Fourier period, have also been detected in many other AGN since then \cite[e.g.][]{vaughan-mcg, markowitz-timelags,McHardy07,markowitz07,Arevalo-timelags,sriram-timelags,walton-timelags,alsto-timelags,loban-tlags-1,loban-tlags-2,mastroserio-tlags,loban-tlags-3}.

The right panel of Fig.\,\ref{fig:timelags} shows time lags versus the Fourier period for NGC\,4051 \citep{mchardy04}. Time-lags in this source have been studied extensively. \cite{alston-tlags-4051} found that the delays may depend on the source flux, while \cite{papadakis-4051tlags} found that the time delays between the hard and soft band variations may extend to very long periods (of the order of $\sim$ hundreds of days). A comparison between the graph on the right panel of Fig.\,\ref{fig:timelags} and the cross-correlation plot in the right panel of Fig.\,\ref{fig:4051-corr} reveals the power of studying the delays in the frequency domain instead of the time domain. The cross-correlation function indicates a constant delay between the hard- and soft-band variations, whereas the time lags reveal that the delays between the sinusoids that are
responsible for the observed variations are not constant. Instead, the delays increase with increasing Fourier period, and for any given period, the time lag increases with the energy separation of the bands. As we mentioned, this behaviour can be explained by the Comptonization hypothesis, but in this case, we would expect that the delays between various energy bands should increase by the same amount for all Fourier periods, which is not what is observed \cite[e.g.][]{arevalo-pap}.

\cite{epitropakis} presented the results from a systematic analysis of the X-ray continuum time-lags (and intrinsic coherence) between the 2-4 keV and various energy bands in the 0.3-10 keV range, for 10 X-ray bright and highly variable AGN. They used {\it XMM-Newton} light curves and estimated the time lags following the statistical methods of \cite{epitropakis-papadakis}. They were able to demonstrate that time lags have a power-law dependence on frequency, with a slope of -1. Their amplitude scales with the logarithm of the mean-energy ratio of the light curves, and it increases with the square root of the X-ray Eddington ratio. Furthermore, they found that the intrinsic coherence is approximately constant at low frequencies and then decreases exponentially at high frequencies. Both the constant value of the coherence at low frequencies and the frequency above which it decays have a logarithmic dependence on the light curve mean energy ratio.

The continuum, hard time-lags in AGN can also be explained in many ways. \cite{Arevalo-Uttley} showed that the hard time lags can be explained if the X-ray corona is accreting, its temperature increases inward, and there are inward-propagating mass accretion fluctuations (within the accreting X-ray corona), which are produced at a wide range of radii. The fluctuations modulate the X-ray emitting region as they move inward and can produce the observed temporal-frequency-dependent lags between various energy bands. 

On the other hand, a Comptonization origin of the lags is possible if the Comptonizing medium has a radially dependent electron density, as proposed by \cite{kazanas}. \cite{mastroserio-tlags} argued that the observed hard lags can be explained by fluctuations in the slope and strength of the X-ray energy spectrum.  Finally, \cite{wenda} studied the Fourier time lags due to Comptonization of disc-emitted photons in a spherical and uniform X-ray corona, which is located on the rotational axis of the black hole. They used a general relativistic Monte Carlo radiative transfer code \citep{monk} to calculate the Compton scattering of photons emitted by a thin disc with a Novikov–Thorne temperature profile. They provided equations that can be used to determine the time lags for a wide range of values for the corona radius, temperature, optical depth, height, and for various accretion rates and black hole masses. They showed that the results of \cite{epitropakis} are consistent with thermal Comptonization of disc photons in the case of a dynamic corona, with a variable location, size, and lifetime. 


\subsection{Excess variance measurements and high-redshift AGN.}
\label{sec:X-ray_excess_var}

\begin{figure}
    \centering
    \includegraphics[width=0.47\linewidth, height=0.4\linewidth]{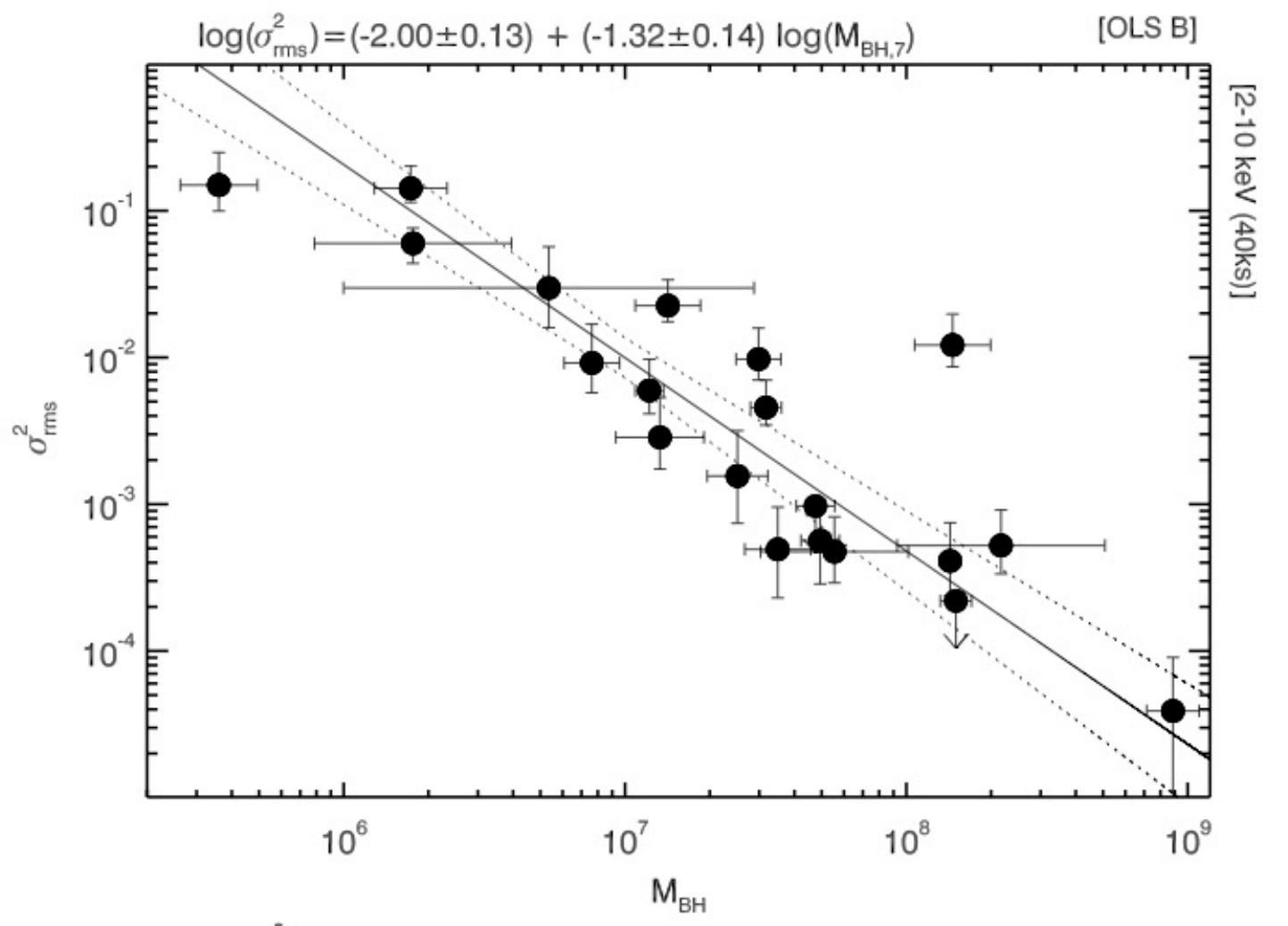}
    \includegraphics[width=0.52\linewidth, height=0.38\linewidth]{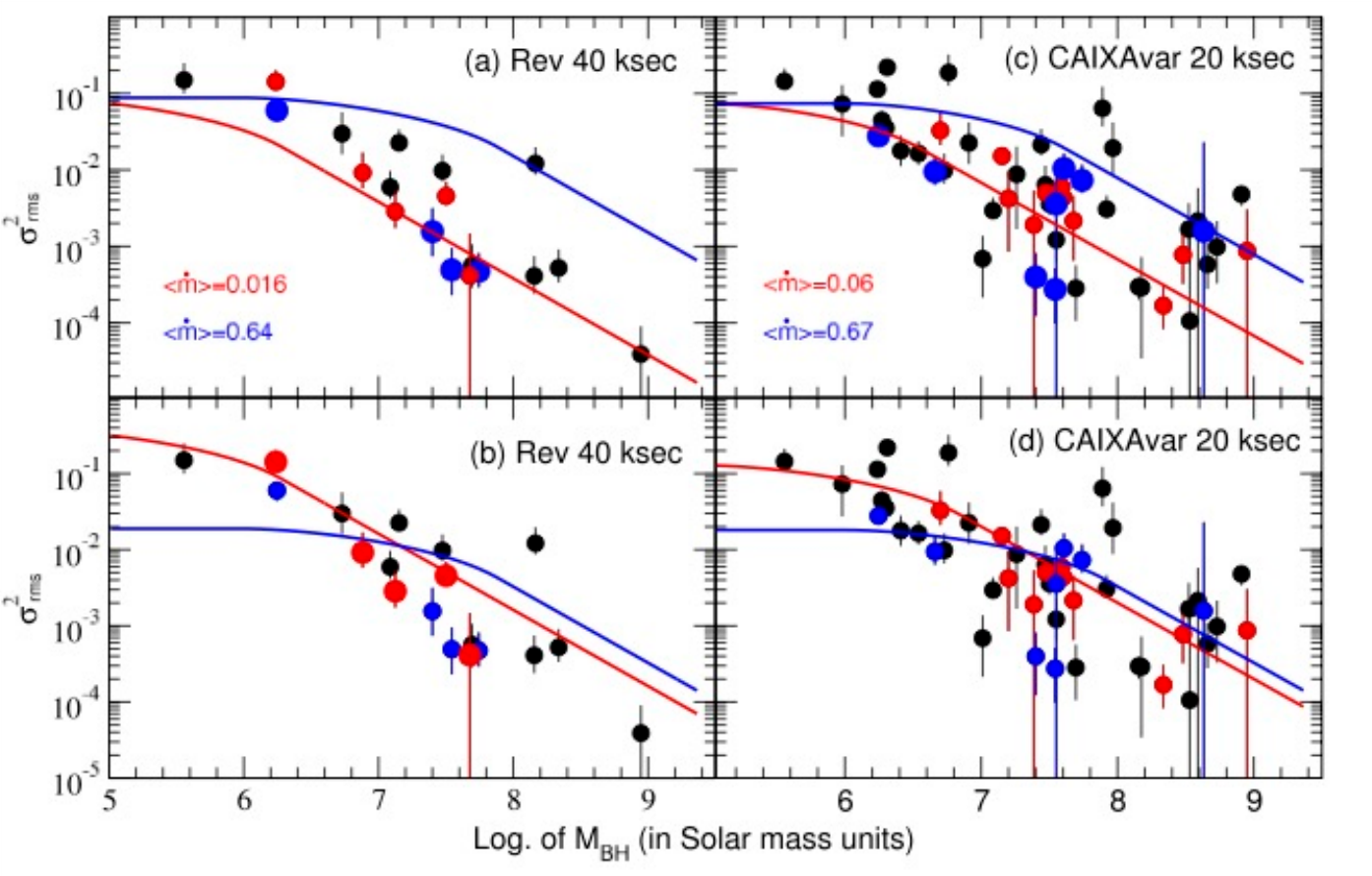}
    \caption{{\it Left panel:} The \nxv\ vs \mbh\ relation for AGN in the ``Reverberation" sample of \cite{Ponti12}. Excess variance in this graph is measured using light curve segments of 40 ksec duration. Solid and dotted lines indicate the best-fit line and the 1$\sigma$ error (from Fig.\, 3 of \cite{Ponti12}). {\it Right panel:} The left and right graphs in this panel also show the \nxv\ vs \mbh\ relation of AGN in the ``Reverberation" sample of \cite{Ponti12}, when \nxv\ is measured using light curves of 40 and 20 ksec duration, respectively. The solid and blue lines indicate the predicted relations for various models (from Fig.\,6 of \cite{Ponti12}).}
    \label{fig:nxvponti}
\end{figure}

Power spectrum analysis is a powerful tool to study
the variability properties of AGN. However, it requires long, uninterrupted, and evenly spaced observations. 
Short, evenly sampled, high signal-to-noise X-ray observations, as well as longer, sparsely sampled light curves for hundreds of objects, are available in the archive of X-ray observatories. They may not be ideal for power spectrum analysis; however, these light curves offer us the possibility of studying the variability of X-rays in various energy bands using the so-called 'normalised excess variance', \nxv \citep{nandra97}. 

This statistic is an estimate of the integral of the power spectrum from the shortest to the longest frequency sampled by the light curve. It is a biased estimator of this integral (even in the case of evenly sampled light curves), because of the red noise character of the intrinsic X-ray variations, and the bias depends on the PSD slope and the sampling pattern of the light curve. Its statistical properties have been studied by \citet{vaughan-var}, in the case of evenly sampled light curves, and by \cite{Allevato}, in all cases (also see \citealt{almaini2000} and \citealt{Buchner2022} for alternative approaches in the low-count-rate regime). In general, a single measurement of the normalised excess variance is not very useful. Even when corrected for bias, single \nxv\ measurements follow a very asymmetric and broad distribution of unknown variance (that is, their error is unknown), especially in the case of sparse sampling and low signal-to-noise light curves. If we can calculate excess variance using light curves from many individual sources, then 'ensemble' estimates of \nxv\ can be used, as they have a known error and distributions similar to a Gaussian \citep{Allevato}. The same procedure can be followed with excess variance measurements using segments of long light curves of a single object. However, the individual \nxv\ values in this case are not independent (due to the strong correlation between the points of the light curve over very long periods), therefore it will be difficult to determine the statistical properties of the resulting ensemble estimates.

The close relationship to PSD and the availability of large samples of objects for which excess variance can be easily measured can explain the popularity of the use of \nxv\ in the study of AGN variability. It has been used many times in the past to study the X-ray variability of AGN and to investigate the dependence of various PSD parameters on the physical parameters of the system. Almost all of the work that has been done so far focusses on the analysis of light curves in the (rest frame) 2-10 or  $0.5-2$ keV bands. 

Early studies of local AGN using {\it ASCA} light curves showed that the excess variance is anticorrelated with AGN luminosity, as already observed in the optical/UV (Sect.\,\ref{sec:optvar_scaling}), and that the observed excess variance appeared to be larger for NLS1 galaxies \citep{nandra97,Turner1999}. \cite{luyu,bianzhao} and \cite{Papadakis04} further showed that \nxv\ is inversely proportional to \mbh. In fact, \cite{Papadakis04} used the fact that the variance is a measure of the PSD integral to demonstrate that the PSD amplitude is probably the same in all AGN while the PSD bend frequency decreases with increasing BH mass even before \cite{McHardy06}. Later, \cite{oneill,miniutti} and \cite{Ponti12} confirmed the dependence of \nxv on the BH mass (see left panel of Fig.\,\ref{fig:nxvponti}). 

\cite{Ponti12} computed \nxv\ for a large sample of (nearby) AGN using {\it XMM-newton} light curves, to study the X-ray variability of AGN in various energy bands and on timescales shorter than $\sim$ a day. They were able to show that their results could only be explained if the PSD bending frequency depends on both the BH mass and the accretion rate as shown by \citet[][see the right panel of Fig.\,\ref{fig:nxvponti}]{McHardy06}. They also suggested that the PSD amplitude should increase with decreasing accretion rate in AGN. Although they predicted a rather steep dependence of the PSD amplitude on \medd, which is not supported by other observations, their results showed that \nxv\ analysis can provide powerful insights into the properties of AGN X-ray variability. In fact, since \nxv\ can be used to characterise the X-ray variability of large samples of sources, the results of the \nxv\ studies can impose constraints on models for fundamental relations in AGN. For instance, \cite{georgakakis21} linked the demographics of AGN to their ensemble X-ray variability properties in order to investigate the black hole mass versus stellar mass relation in these objects (also see \citealp{Sartori2018} for a similar application using PSD analysis).

Given the tight correlation between \nxv\ and \mbh, it was soon suggested that \nxv\ could be used to estimate the BH mass in AGN \cite[e.g.][]{nikolajuk1,Nikolajuk2,Nikolajuk3}. \cite{Ponti12} provided BH mass estimates for tens of AGN based on their \nxv\ measurements, and \cite{zhou-bhm} provided a calibration for the correlation between \nxv\ and \mbh\ in AGN. \cite{akylas} critically assessed \nxv\ as a tool to measure \mbh\ in AGN. They showed that it is possible to accurately measure \mbh\ using \nxv\ measurements in the 3–10 and the 10–20 keV bands; however, the signal-to-noise ratio and the duration of the light curves should be larger than $\sim 3$ and $\sim 80$ ksec, respectively. 

All the works mentioned above used light curves that are shorter than $\sim$ day. Excess-variance measurements have perhaps played an even more important role in the study of the X-ray variability of high-$z$ AGN on longer timescales. In most cases, such objects are observed only a few times, in a highly irregular pattern. Consequently, the resulting light curves are not appropriate for the application of Fourier techniques. Therefore, \nxv\ has been widely used in the study of their X-ray variability properties. The lack of BH mass estimates was another obstacle in investigating the correlations between the amplitude of the X-ray variability and the physical parameters of AGN (i.e. \mbh\ and \medd). As a result, most studies originally tried to investigate the correlation between variability amplitude and source luminosity, as well as redshift \cite[e.g.][]{almaini2000,manners,paolillo04,papadakis08,mateos,Young2012,shemmer14,Yang2016,Paoillo17,Zheng2017}, confirming the anticorrelation with luminosity observed at lower redshift. This finding was also confirmed by studies based on SF analysis \citep{Vagnetti2016}. \citet{almaini2000,manners,paolillo04} also observed a tentative evolution of X-ray variability with redshift, which was however disproved by later studies using better data and more complete samples.

\cite{lanzuisi} were the first to show that the anti-correlation between \nxv\ and \mbh\ also holds for very luminous AGN at high redshifts. \cite{zhang-longpsd} also studied the relationship between \nxv\ and BH mass. They used All Sky Monitor (on board {\it RXTE}) light curves to calculate the normalised excess variance of many AGN on timescales from 1 to 14 years. They found that the observed variance appears to be independent of \mbh\ and bolometric luminosity on such long timescales. They convincingly argued that AGN are scaled-up GBHBs in the high/soft accretion state. 

\begin{figure}
    \centering
    \includegraphics[width=0.49\linewidth]{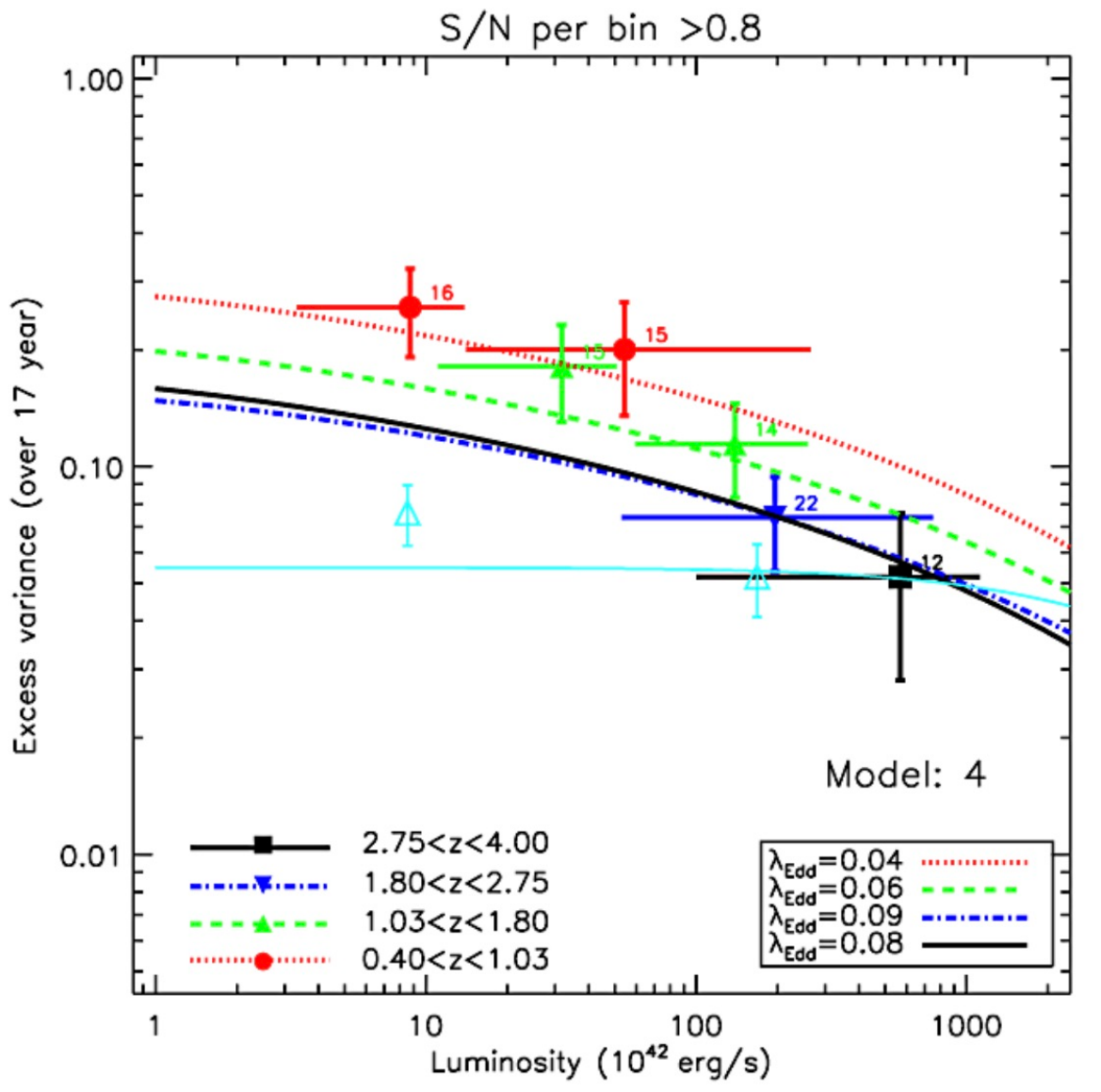}
    \includegraphics[width=0.49\linewidth]{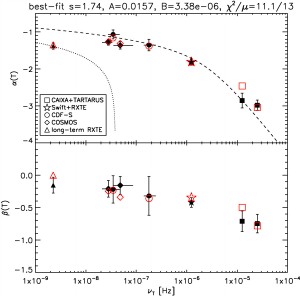}
    \caption{{\it Left panel}: Points show the excess variance measurements of AGN in various redshift bins, computed using light curves from 17 years long CDF-S observations. The lines indicate the model predictions, for each redshift bin, assuming a particular PSD model (from Fig.\,8 of \citealp{Paoillo17}).  {\it Right panel:} The amplitude and slope of  the \nxv-\mbh\ relation of a large sample of AGN with various redshifts (top and bottom graphs, respectively), computed using light curves of various durations. The dashed line on the top shows the predictions of a model which assumes the same PSD for all AGN up to redshift of $\sim 3$ (see Fig.\,8 of \citet{Paolillo23} for details).}
    \label{fig:nxvpaolillo}
\end{figure}

The relation between X-ray variability and \mbh\ and \medd\ was also studied indirectly by \cite{papadakis08} and \cite{Paoillo17}. They took advantage of the fact that \nxv\ is an estimate of the PSD integral over the sampled timescales in a light curve, and they considered various PSD models to fit their \nxv -- X-ray luminosity plots. \cite{Paoillo17} studied the X-ray variability properties of AGN up to $z\sim 4$ in the Chandra Deep Field-South region over a period of 17 years. They were able to produce \nxv\ -- Luminosity plots in various redshift bins and for different timescales (see the left panel of Fig.\,\ref{fig:nxvpaolillo}). They fitted their data 
assuming the PSD scaling relations with mass and accretion rate of \cite{McHardy06} and \cite{GMV11}. They found that the X-ray variability of distant AGN is consistent with the hypothesis that $\nu_b$ depends on both \mbh\ and \medd, while \medd\ may also affect the PSD amplitude. They also considered the evolution of the mean accretion rate of AGN as a function of redshift. They found an indication for the increase of accretion rate with increasing redshift (see the middle panel in their Figure 9), although due to the large error bars the hypothesis of a constant mean accretion rate was also acceptable. 

Finally, \cite{Paolillo23} presented a novel way of using \nxv\ measurements to construct a ``variance-frequency" plot (VFP) for AGN, which is directly analogous to the average PSD of AGN. They used \nxv\ measurements computed from short and long light curves ($\sim 20$ ksec - 14 years), for many objects over a wide range of redshifts (up to $\sim$ 3), luminosity and black hole mass. The top plot in the right panel in Fig.\,\ref{fig:nxvpaolillo} show the excess variance (for a 10$^8$ solar mass AGN) plotted as a function of frequency. The dashed line shows the expected VFP in the case of a PSD with a low- and high-frequency slope of $-1$ and $\sim -2.7$, respectively, amplitude of $\sim 0.01$ and $\nu_b$ that scales like the \cite{McHardy06} results. The \cite{Paolillo23} results show that the PSD results, which are based on the study of a small number of nearby Seyferts, hold for the general AGN population as well, over a wide range of BH masses, luminosity and redshift. 


\subsection{Stationarity}

A stochastic process is considered to be stationary if its statistical properties (like the mean, variance, autocovariance, and power spectral density function) remain constant and do not change over time. The stationarity of the variability process is implicitly assumed whenever a PSD analysis of long light curves is performed in AGN.  Otherwise, if the intrinsic PSD evolves with time on short timescales, then one has to consider the use of the PSDs that are computed using {\it RXTE} light curves that are years long. 

It is well-known that the X-ray variability process in AGN is not strictly stationary because the variance of the lightcurves increases linearly with source flux. This ``rms-flux" relation was first noticed by \cite{uttley-rms}, and it is well established in AGN as well as in GBHBs \citep[e.g.][]{heil12}. This is an important property and can place constraints on various models that have been proposed to explain the observed variations. For example, \cite{Uttley2005} showed that if the rms-flux relation holds for variations on all timescales, then the variability process in AGN should be multiplicative in nature: variations should be coupled together on all timescales. Therefore, the observed variability cannot be due to a shot-noise process or the result from completely independent variations in separate emitting regions. 

The rms-flux relation in AGN does not impose a strong constraint on the PSD analysis of long light curves, as long as the PSDs vary only in amplitude as the source flux changes and not in shape \citep{uttley-rms}. In this case, the observed PSDs should remain constant over time if we normalise the light curves to their mean, as is (almost) always the case when we calculate power spectra with the AGN light curves. When normalised to the light curve mean squared, the AGN PSDs are generally thought to be constant over long timescales. 

\cite{alston19} found that the PSD changes in amplitude and shape (to a lesser extent) even when the light curves are normalised to their mean in the case of IRAS 13224--3809. This is a highly variable source, which has been extensively observed by {\it XMM-Newton}. Such subtle, but significant, PSD changes imply violation of the assumption of stationarity in the X-ray variability process in AGN. However, this is not the case for other AGN. 
For example, \cite{Rani25} did not find a significant change between the {\it NICER} and {\it XMM-Newton} PSDs of NGC\,4051, despite the fact that the observations were taken quite a few years apart. As \cite{alston19} discuss,  the significant PSD changes observed in IRAS 13224-3809 on timescales of $\sim$ days are equivalent to PSD changes on a timescale of $\sim$ 10 s in a 10 M$_{\odot}$  GBHB. PSD variations on the GBHB PSDs on such short timescales have not been reported. Perhaps such a significant change in the variability process may rarely occur in accreting compact objects. 

\subsection{Constrains on theoretical models}

X-ray variability studies have provided various clues regarding the X-ray emission mechanism in AGN. It is generally accepted that X-rays are produced by inverse Compton scattering of optical/UV photons emitted by the disc from hot electrons in the corona. As we have already mentioned, the fact that the PSD slope flattens with increasing energy already suggests that the electron density and temperature cannot be uniform in the corona. If that were the case, then, due to the larger number of scatterings, the PSD slope should steepen (significantly) with increasing energy. 

The propagating mass accretion fluctuations model of \cite{Kotov} and \cite{Arevalo-Uttley} can explain the steepening of the power spectrum, as the temperature of the corona is not homogeneous. According to the model, the X-ray corona is accreting and the temperature increases inwards. Therefore, higher-energy X-ray photons are emitted from regions closer to the BH. Furthermore, the mass-accretion fluctuations propagate from the outer corona to the inner corona. At each radius, the corona is variable at all time scales up to the characteristic viscous timescale, $t_{visc,corona}(R)$ at that radius (which decreases with smaller $R$). Therefore, the variability of high-energy photons should increase at high frequencies as well, since $1/t_{visc,corona}$ increases inward. The same model can also explain the continuum hard time lags we described in
Sect.\,\ref{sec:xraytimelags}, for the same reason. It takes time for the accretion rate fluctuations to propagate from the outer to the inner corona, therefore, the variations at high energies should be delayed with respect to the variations at lower energies. 

The model predicts that $\nu_b$ should increase with increasing energy because higher energy photons are produced at the hotter regions of the corona, which are closer to the BH, where $t_{visc,corona}(R)$ is shorter. However, this does not appear to be the case with the energy-dependent PSDs in AGN (see Sect. \ref{sec:X-ray_enerdydep_nu}). The bending frequency remains roughly the same throughout the observed frequency band from $\sim 0.3-10$ keV, which is against the predictions of the model. Furthermore, according to the model, the power spectra should extend with a slope of $-1$ over a frequency range determined by $1/t_{visc,corona}(R_{out})$ up to 1/$t_{visc,corona}(R_{min})$ (where the PSD break should appear).  However, $R_{out}$ should be very large in order to explain power spectra such as the NGC 4051 PSD (left panel in Fig.\,\ref{fig:4051psd}). This power spectrum extends with a slope of $-1$ over more than five orders in frequency down to $\sim 5\times 10^{-9}$ Hz. See also the high-quality X-ray PSD of NGC 4395 in \cite{beard25}, which extends with a slope of $-1$ below $\nu_b$ over four and a half orders of magnitudes in frequency. In order to explain this, the hot corona must extend to very large radii. For example, \cite{rapisarda17} fitted the Cyg X-1 PSD in its soft state with a fluctuations propagation model and found that the corona (which is located on top and below the disc) should have a very large radius of $\sim 2500$ $R_g$. This is clearly much larger than the size of the X-ray corona, as measured by monitoring observations of lensed quasars (see references at the beginning of Sect.\,\ref{sec:x-rayvar}). Apart from the difference between the model and the observed corona size, it is also difficult to understand how such a corona can be formed at such large radii from the BH, whether it will be stable or how it may evolve over time. Currently, there are no physical models that could be used to determine the accretion rate of the corona, its temperature, and the density radial distribution.

Note that the requirement of an extended corona does not come from the need to explain the observed long-term variations since these variations could be produced by accretion rate fluctuations which could propagate in the accretion disc and then within the corona as well). 
However, an accreting X-ray corona, with a large radius, is necessary to explain the long-term, continuum hard X-ray time-lags in AGN, especially if they really extend to very low frequencies \cite[e.g.][]{papadakis-4051tlags}. We also stress the similarity between the AGN PSDs and that of Cyg X-1 in its (own) soft state. In this state, a strong black-body-like component appears at low energies in the energy spectrum of the source. It is generally accepted that this component is due to thermal emission from the accretion disc, which extends down to the innermost stable circular orbit. If the similarity of the timing properties implies similar physical conditions as well, then it is possible that the accretion disc in the AGN also extends to its innermost radius, leaving not much space for a radially extended corona.

Regarding extended coronae, one may also consider the models of \cite{kazanas} and \cite{reig03}. In the first case, the corona has a uniform temperature, but the electron density decreases as $\propto r^{-1}$ so that the corona has the same optical depth per decade of radius and extends for approximately 3 decades of radius. In the second case, the corona is outflowing with a speed of 0.5-0.8$c$. Both models have been developed mainly to explain the X-ray variability of GBHBs and can explain many of the observations but, since the phenomenology of GBHBs is very similar to that of AGN, these models could also be applicable to AGN. 

There are also a few models that study the variability properties of magnetic flares in accreting compact objects (i.e GBHBs and AGN). \cite{poutanen99} assumed that X-rays are produced in compact magnetic flares at short distances from the BH. They showed that the tendency for magnetic loops to inflate and detach from the underlying accretion disc causes the time delays between hard and soft photons. They considered the case of an avalanche model in which flares can, with some probability, trigger other flares. The resulting power spectra, time lags and coherence functions are very similar to what is observed in AGN and GBHBs. \cite{merloni01} considered a similar model of magnetic flares above the accretion disc which produce the X-ray spectrum via inverse Compton scattering of soft photons. The typical size of the flaring events is comparable to the thickness of the accretion disc, and they are triggered at a height at least an order of magnitude larger than their size. Furthermore, they assumed that the spatial and temporal distribution of the flares is not random. Instead, they assumed that heating of the corona proceeds in correlated trains of flares in an avalanche fashion. The amplitude of the avalanches obeys a power-law distribution and determines the size of the flares. The model can explain many of the observed variability properties in GBHB and AGN. 

\begin{figure}
    \centering
    \includegraphics[width=0.6\linewidth]{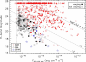}
\raisebox{0.7cm}{\includegraphics[width=0.35\linewidth]{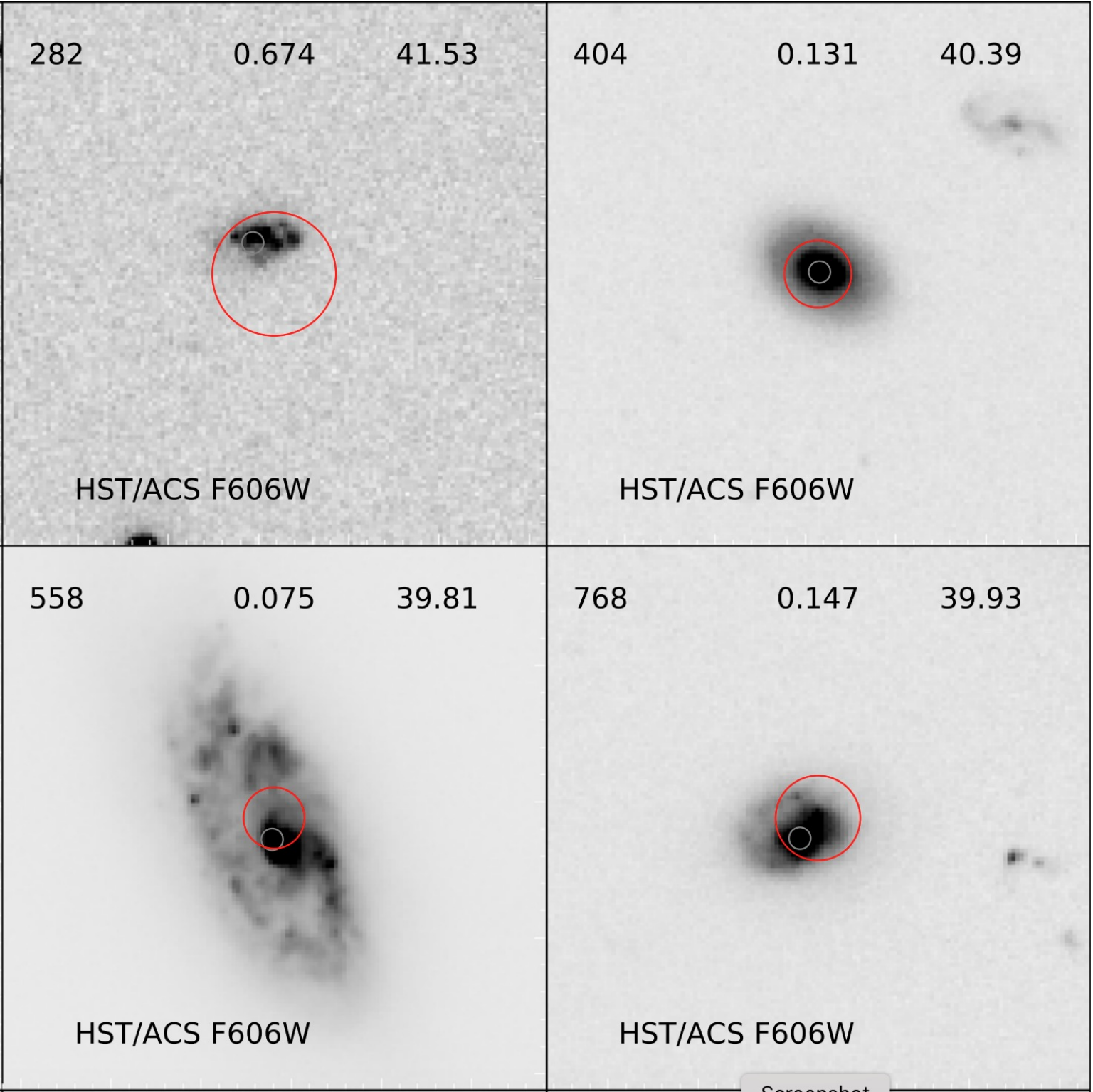}}
    \caption{\textit{Left:} Classical optical/X-ray ratio diagnostic for AGN.
    Constant flux ratios are marked with diagonal lines: bright AGN are usually detected as sources with $\log(f_x/f_R) > -1$.
    Several sources with significant variability (filled symbols) lie in the region dominated by normal galaxies. \textit{Right:} Images of LLAGN candidates detected from variability. From Fig.4 and 5 of \cite{Young2012,Ding2018} respectively.}
    \label{fig:XO_ratio}
\end{figure}

In general, there is no model that can explain all the variability properties of AGN. As we mentioned in \S\ref{sec:optvar_model}, X-ray reverberation appears to be able to explain many of the optical/UV variability properties of AGN. In most of these models, the X-ray source is located on the spin axis of the BH (and presumably also of the accretion disc). However,  there is no explanation as to how this model could reproduce the X-ray PSD (and in particular the Long-term part of the observed PSDs) or the variance-mean relation that is observed in some AGN. As we noted, \cite{wenda} showed that this model could explain the time lags in AGN, however, more detailed work should be done to investigate whether this model is consistent with X-ray variability properties of AGN. Perhaps such a study could be done within the framework of the ``aborted jet" model \citep{ghisellini04}. This model aims to unify radio-quiet and radio-loud AGN, and proposes a reasonable physical framework within which ``lamp-post" models can be studied. The model has been shown to be consistent with the spectral variability of AGN, but a more detailed study of its timing properties is necessary. 

\subsection{Detecting AGN through X-ray variability}

Since X-ray emission is one of the most prominent AGN features, it has been pivotal for the discovery and study of the population of SMBH in the Universe \citep{Brandt2005}. This is because normal galaxies emit little radiation at these energies. Thus, the presence of X-ray emission, in excess of the one expected from the X-ray binaries and SNe in a galaxy, is among the most reliable methods to discover accreting SMBHs. In addition to the detection of significant X-ray flux from a distant source, variability can also play a significant role in its identification. This applies in particular to the elusive population of low-luminosity/low-accretion SMBH which are numerically more abundant than bright AGN, especially in the local universe \citep[e.g.][]{Alexander2011, Aird2018, Torbaniuk2021, Torbaniuk2024}.

The use of persistent variability to detect faint AGN has not been exploited in the X-ray band as much as in the optical band (Sect.\,\ref{sec:optvar_discover}). Nevertheless, it has been shown to be effective in pushing the limit of AGN detection in deep, multi-epoch X-ray surveys. In particular, the works of \cite{Young2012, Ding2018} have exploited the repeated observations of the CDF-S spanning up to 17 years 
to identify LLAGN which escape classical selection criteria. As shown in Fig.\,\ref{fig:XO_ratio}, 10\%-20\% of sources with optical-to-X-ray flux ratio lower than 0.1 ( the lower limit usually adopted to confirm the presence of an AGN, \citealt{Maccacaro1988}) are detected to be significantly variable. Such objects possess a luminosity $< 10^{41}$ erg/s and a (stacked) spectral power-law index typical of AGN. However, their BH mass and accretion rate are $\sim 2$ and $\sim 20$ times smaller than their bright counterparts, respectively. We point out that any deep multi-epoch survey has the potential of applying the same method to identify LLAGN if the survey cadence is properly planned. In fact, based on these examples, similar results can be expected from eROSITA or the future \textit{Athena} mission.

Obviously, the scenario changes entirely if we concentrate instead on quiescent non-accreting SMBH, where episodic events such as tidal disruption of stars can produce sudden bursts of optical/UV/X-ray emission. However, this topic, although very active today, is outside the scope of this review.

\subsection{Variability studies with current and next generation X-ray missions.}

There are 23 AGN listed in \cite{GMV11} for which high-frequency bending frequencies have been detected in their PSDs. Such frequencies have been detected using only {\it XMM} light curves in 11 AGN. A combination of light curves from {\it XMM} and other satellites (mainly {\it RXTE}) have been used for the detection of break timescales in the remaining 12 AGN. Some additional detections of bending/break timescales have been reported since then, and we also have information from many more objects using excess variance analysis. However, the reality is that our knowledge of the X-ray power spectra in AGN is based on the study of only two dozens of objects with good-quality PSDs so far (for some of them we do not even have an accurate estimate of their BH mass). 
These are nearby, X-ray bright, and very variable objects. It may well be that our view of the X-ray variability properties of active galaxies is biased. It is important to study as many AGN as possible to significantly improve our understanding of their X-ray variability. 

There are many data in the archive of satellites like {\it XMM-Newton} and {\it RXTE} which have not been studied yet.
There are also archival {\it NICER} and {\it NUSTAR} data that could further help us better understand the energy dependence of the X-ray variability in AGN. We also note that archival data from past and current missions can be combined to deliver long-term light curves. This approach has been used by, for example, \citet{Middei2017} who merged {\it ROSAT} and {\it XMM-Newton} data, or \citet{Georgakakis2024, Prokhorenko2024} who used {\it XMM-Newton} and eROSITA observations. We believe it will be beneficial if the X-ray community tries to exploit the full power of the public data archives to probe X-ray variability on timescales of decades, as is being done by the optical community in preparation for LSST.

To make substantial further progress, we would like to accurately determine the PSD of at least 200 sources or so (i.e. an order of magnitude more than the current number of AGN with well-determined PSD). To do so, we would need densely sampled light curves (with a bin size well below one day for sources with BH masses between 10$^6-10^7$M$_{\odot}$),  with a duration longer than $\sim 2000$ days, in order to sample the low-frequency part of the PSD well, and hence determine $\nu_b$ accurately even for a BH mass of $\sim 10^9$M$_{\odot}$. Moreover, it would be important to obtain high signal-to-noise light curves over narrow energy bands, so that we could study the dependence of the PSD properties on energy as well. 

{\it Swift}/BAT provides continuous monitoring of many AGN but is not sufficiently sensitive to create densely sampled light curves, on short timescales and in various energy bands. The available light curves can be used for a study of the average properties of the X-ray PSD at low frequencies. Analysis of the long term {\it Swift/}BAT light curves of the brightest AGN so far has confirmed that the slope of the power spectrum in AGN is $\sim -1$ below $\nu_b$, and the PSD amplitude at energies above $\sim 15$ keV is similar to the PSD amplitude in the 2-10 keV band \citep{shimizu13,Papadakis24}. However, the light curve errors are large even with a binning of one month, therefore it is not possible to investigate how the PSD properties (slopes, amplitude, $\nu_b$) depend on \mbh\ and \medd, as a function of energy. 

The situation will improve with the Wide Field Monitor on board {\it eXTP}, which is a future satellite led by the Chinese Academy of Sciences and universities in China, with partnerships with various European institutions, and is planned to launch in 2027. The WFM will monitor hundreds of sources daily, but the number of AGN among these sources will not be very large. According to \cite{extp19}, the maximum sensitivity of the Wide Field Monitor will be $\sim 2.1$~mCrab in 50 ksec. If this is over the 2--50 keV energy range, this could imply a $\sim 1$~mCrab flux in the 2-10 keV band for power-law-like spectra with a slope of $\Gamma=-2$ (the flux limit increases/decreases with steeper/flatter PL slopes). There are $\sim 70$ unobscured AGN in the nearby Universe with flux higher than $\sim 1$~mCrab \citep{gupta24}. So, WFM will help us to determine low-frequency PSD slopes, as well as amplitudes, for a sizable number of sources. We should be able to investigate their dependence on BH mass and/or accretion rate as well, although we will probably not be able to investigate PSD variations with energy and we will not be able to study the PSD bending frequency in detail. However, as before with data from other satellites, we will be able to use excess variance and VFP analysis (see Sect.\,\ref{sec:X-ray_excess_var}) on sparsely sampled lightcurves of several hundred sources, especially in the regions of the sky observed repeatedly by the scanning strategy of the mission (see Fig.\,14 of \citealp{intZand2019}). This will give us the opportunity to investigate the variability properties of a much larger sample of objects, especially at higher redshifts \cite[e.g.][]{lanzuisi,Paolillo23}.

{\it Einstein Probe} is another X-ray mission which is currently monitoring the sky in (soft) X-rays. It is a mission of the Chinese Academy of Sciences in collaboration with the European Space Agency and the Max-Planck Institute for Extraterrestrial Physics. It was launched in January 2024, and one of its objectives is to provide light curves with a uniform sampling and long duration (on the order of a few years) for a large number of AGN. The detection sensitivity at 5 sigma of the Wide Field X-ray Telescope on board {\it Eisntein probe} is less than 1 mCrab within an exposure of 1000 s (see Figure 5 in the ``Science Capability" page at the central web page of the mission, i.e. https://ep.bao.ac.cn/ep/). The light curves will be in the range of 0.5-4 keV, where the contribution of variable absorption to the observed variations is the strongest in AGN. However, if the number of AGN with high-qulity light curves is large, the {\it Einstein probe} can contribute significantly to our understanding of AGN variability. 

eROSITA (extended ROentgen Survey with an Imaging Telescope Array) is the primary instrument on-board the Russian-German "Spectrum-Roentgen-Gamma" (SRG) mission.  eROSITA was developed by the Max-Planck Institute for Extraterrestrial Physics (MPE) and was launched in 2019. Since December 2019, eROSITA has been conducting an all-sky survey in the X-ray band up to 10 keV, in which the entire celestial sphere will be mapped once every six months. Up to 2025, eROSITA has completed four of its eight intended all-sky survey passes. Consequently, eROSITA will provide light curves with a maximum of four points over two years for faint sources over most of the sky in the 0.2-8 keV band (rest frame). It is difficult to perform PSD analysis when there are few points in the light curves. Furthermore, the analysis of eROSITA data is intrinsically complex due to the changing PSF and sensitivity during the scans \citep{Buchner2022,Bogensberger2024a}. However, significantly better sampled light curves will be available for sources near the poles, where eROSITA will observe AGN from tens to hundreds of times (see Fig.5.7.3 of \citealp{eROSITA_wb} and updates on the mission web page\footnote{\hyperref[https://erosita.mpe.mpg.de]{https://erosita.mpe.mpg.de}}). Consequently, a detailed PSD analysis may be possible for these AGN. 
The big advantage of eROSITA when it comes to variability studies is the enormous number of AGN that will be detected during the sky survey. The very large number of AGN should allow for a detailed characterisation of X-ray variability in AGN through the use of statistics such as \nxv or any other statistic that could be an estimate of the integral of the PSD over the sampled timescales. 
eROSITA will benefit from redshift and BH mass estimates from several ancillary spectroscopic and photometric surveys, including LSST, which should allow us to better constrain the ensemble variability dependence on AGN properties, as demonstrated by the first published results \citep{Bogensberger2024b,Georgakakis2024,Prokhorenko2024}.}

Finally, we point out that any other future sensitive/wide-field mission may allow ensemble variability studies of thousands of AGNs, if the proper observing strategy is adopted. As proof of concept, we refer to the {\it Wide Field X-ray Telescope} (WFXT) mission \citep{Murray2010,Rosati2010} which was envisioned as a potential successor to eROSITA. WFXT, with a uniform PSF ($FWHM< 5^"$) across the 1 deg$^2$ field-of-view and a large sensitivity ($1~m^2$ at 1 keV), could monitor tens of thousand AGN in the 0.5-2 keV band (observed frame) for over five years \citep{Paolillo12} thus delivering results similar to those obtained by {\it Chandra} and {\it XMM-Newton} in their deepest surveys \citep{lanzuisi,Paoillo17,Zheng2017} but for samples at least two order of magnitude larger. 

\bmhead{Acknowledgements}

We thank D. De Cicco, V. Petrecca and E. Kammoun for valuable discussions and contributions. I. Papadakis thanks the University of Naples Federico II for support through its ``Visiting professor'' program. M. Paolillo acknowledges support from the Italian PRIN - MIUR 2022 ``SUNRISE'' and the INAF grant TIMEDOMES. This research has been supported in part by the EU HORIZON MSCA Doctorate Network \href{https://www.star.bris.ac.uk/TALES/}{``TALES''}.

\section*{Declarations}
\begin{itemize}
\item \textit{Funding}: I. Papadakis thanks the University of Naples Federico II for support through its ``Visiting professor'' program. M. Paolillo acknowledges support from the Italian PRIN - MIUR 2022 ``SUNRISE'' and the INAF grant TIMEDOMES. This research has been supported in part by the EU HORIZON MSCA Doctorate Network \href{https://www.star.bris.ac.uk/TALES/}{``TALES''}.
\item \textit{Conflict of interest/Competing interests}: the authors have no relevant financial or non-financial interests to disclose.
\item \textit{Ethics approval and consent to participate}: not applicable.
\item \textit{Consent for publication}: all the non-original material reproduced in this review is published under explicit consent of the original authors, journals or according to the individual journal reproduction permission policies.
\item \textit{Data availability}: for access to the data cited in this review the reader is referred to the original publications.
\item \textit{Materials availability}: not applicable.  
\item \textit{Code availability}: not applicable. 
\item \textit{Author contribution}: all authors have contributed equally to the preparation of this manuscript.
\end{itemize}

\newpage

\bibliography{main.bib}

\end{document}